\documentclass[12pt,preprint]{aastex}

\slugcomment{Accepted for publication in ApJ}

\shorttitle{$L_{\rm X}-T$ AND PROPERTIES OF GALAXY CLUSTERS}
\shortauthors{Ota et al.}

\begin{document}

\title{$L_{\rm X}-T$ RELATION AND RELATED PROPERTIES OF GALAXY CLUSTERS}

\author{Naomi Ota} 
\affil{Cosmic Radiation Laboratory, RIKEN (The Institute of Physical
  and Chemical Research), 2-1 Hirosawa, Wako, Saitama 351-0198, Japan
  \email{ota@crab.riken.jp}}

\author{Tetsu Kitayama}
\affil{Department of Physics, Toho University, 2-2-1 Miyama,
  Funabashi, Chiba 274-8510, Japan}

\author{Kuniaki Masai} 
\affil{Department of Physics, Tokyo Metropolitan University, 1-1
  Minami-osawa, Hachioji, Tokyo 192-0397, Japan}

\and

\author{Kazuhisa Mitsuda}
\affil{Institute of Space and Astronautical Science, Japan Aerospace
  Exploration Agency, 3-1-1 Yoshinodai, Sagamihara, Kanagawa 229-8510,
  Japan}

\begin{abstract}
  An observational approach is presented to constrain the global
  structure and evolution of the intracluster medium based on the {\it
    ROSAT} and {\it ASCA} distant cluster sample. From statistical
  analysis of the gas density profile and the connection to the
  $L_{\rm X}-T$ relation under the $\beta$-model, the scaled gas
  profile is found to be nearly universal for the outer region and the
  luminosity evaluated outside $0.2r_{500}$ is tightly related to the
  temperature through $\propto T^{\sim 3}$ rather than $T^{2}$. On the
  other hand, a large density scatter exists in the core region and
  there is clearly a deviation from the self-similar scaling for
  clusters with a small core size. A direct link between the core size
  and the radiative cooling timescale, $t_{\rm cool}$ and the analysis
  of X-ray fundamental plane suggest that $t_{\rm cool}$ is a
  parameter to control the gas structure and the appearance of small
  cores in regular clusters may be much connected with the thermal
  evolution. We derive the luminosity-`ambient temperature' relation
  ($L_{\rm X}-T'$), assuming the universal temperature profile for the
  clusters with short cooling time and adopting a correction
  $T'=1.3T$, and find the dispersion around the relation significantly
  decreases in comparison to the case of the $L_{\rm X}-T$ and the
  slope becomes less steep from $3.01^{+0.49}_{-0.44}$ to
  $2.80^{+0.28}_{-0.24}$. $L_{\rm 1keV}$, which is defined as a
  normalization factor for each cluster, can be regarded as constant
  for a wide range of the cooling time.  We further examined the
  $L_{\rm X}-T\beta$ and $L_{\rm X}-T'\beta$ relations and showed a
  trend that merging clusters segregate from the regular clusters on
  the planes. A good correlation between the cooling time and the
  X-ray morphology on the $L_{\rm 1keV}-(t_{\rm cool}/t_{\rm age})$
  plane leads us to define three phases according to the different
  level of cooling, and draw a phenomenological picture: after a
  cluster collapses and $t_{\rm cool}$ falls below the age of the
  universe, the core cools radiatively with quasi-hydrostatic
  balancing in the gravitational potential, and the central density
  gradually becomes higher to evolve from an outer-core-dominant
  cluster, which marginally follows the self-similarity, to
  inner-core-dominant cluster.
\end{abstract}


\section{INTRODUCTION}

The X-ray luminosity-temperature ($L_{\rm X}-T$) relation of galaxy
clusters is one of the most fundamental parameter correlations,
established from previous X-ray observations
\citep[e.g.][]{Edge_etal_1990}. Since the X-ray luminosity reflects
temperature and density profiles of hot intracluster medium (ICM), the
$L_{\rm X}-T$ relation should contain information on physical status
and evolution of the ICM.

Observationally, the correlation is well approximated with a power-law
function: $L_{\rm X} \propto T^{\alpha}$, with $\alpha\sim3$
\citep[e.g.][]{Edge_etal_1990,David_etal_1993, Markevitch_1998,
  Arnaud_Evrard_1999}.  In addition, it shows a significant scatter
around the mean power-law relation and little redshift evolution. On
the other hand, the self-similar model \citep[e.g.][]{Kaiser_1986}
predicts $\alpha=2$. Thus the inconsistency between the observations
and the simple theoretical model has been debated for many years and
various possibilities including non-gravitational heating
\citep[e.g.][]{Evrard_Henry_1991, Cavaliere_etal_1997} and dependence
of gas mass or gas-mass fraction on the temperature have been proposed
\citep[e.g.][]{David_etal_1993,Neumann_Arnaud_2001}.  Recently,
hydrodynamical simulations have indicated that the effect of radiative
cooling plays an important role to reproduce the observed $L_{\rm
  X}-T$ relation \citep[e.g.][]{Muanwong_etal_2002}.  They also
suggested a further requirement of significant non-gravitational
heating mechanism so as to account for the observed gas-mass fraction,
however, the physical origin of the additional heating is yet to be
understood.

Under the virial theorem and the isothermal $\beta$-model
\citep{Cavaliere_Fusco_1976}, an ICM temperature $T_{\rm gas}$ is
proportional to a virial temperature, $T_{\rm vir}$, and differs by a
factor of $\beta$, $T_{\rm vir}\sim T_{\rm gas}\beta$.  However, since
$T_{\rm gas}$ is usually an emission-weighted temperature measured
from X-ray spectroscopy and strongly reflects the temperature of the
cluster core region, a temperature decrease due to radiative cooling
may have much influence on the $T_{\rm gas}$ measurement.  The {\it
  ASCA} spectroscopy of central region of clusters particularly with
cD galaxy provided important results that cool emission from the core
is systematically less luminous than that estimated in the past mainly
on the basis of imaging observations
\citep[e.g.][]{Makishima_etal_2001}.  After the advent of the {\it
  XMM-Newton} and {\it Chandra} X-ray satellites, more detailed
temperature profiles have been measured for nearby `cooling-flow'
clusters, which typically show central temperature decrement by a
factor of $\sim 2$ \citep[e.g.][]{Kaastra_etal_2004}.  These facts
gave a piece of evidence against the standard cooling-flow model
\citep{Fabian_1994}, which triggered explorations of a variety of new
scenarios for gas heating: heat conduction from outer hot layers
\citep[e.g.][]{Narayan_Medvedev_2001}, AGN heating
\citep[e.g.][]{Bohringer_etal_2002}, magnetic reconnections
\citep{Makishima_1997}, Tsunami \citep{Fujita_etal_2004} etc. Recently
\cite{Masai_Kitayama_2004} proposed a quasi-hydrostatic model for
cluster gas under radiative cooling, which predicts a characteristic
temperature profile with an asymptotic temperature for the central
region being $\sim1/3$ of the non-cooling outer region. Their
calculation agrees well with that observed in the cooling-flow
clusters.  If the `universal' temperature profile emerges in all or a
fraction of clusters with short cooling timescales, the effect of
temperature drop should be properly taken into account in the
discussion of the $L_{\rm X}-T$ relation.

Another point is due to a connection of the $L_{\rm X}-T$ relation to
the cluster core sizes.  \cite{Ota_Mitsuda_2002} found based on the
X-ray data analysis of 79 distant clusters with {\it ROSAT} and {\it
  ASCA} that the distribution of the cluster core radius shows two
distinct peaks at $\sim 50$~kpc and $\sim 200$~kpc for $H_0=70$. When
dividing the sample into two subgroups corresponding to the two peaks,
they show a significant difference in the normalization factor of the
$L_{\rm X}-T$ relation. Some possibilities have been discussed to
understand the origin of the double-peaked distribution
\citep{Ota_Mitsuda_2002, Ota_Mitsuda_2004}, however, it is yet to be
clarified. As long as one rely on the hydrostatic equilibrium and the
isothermal $\beta$-model, their result may indicate that underlying
dark matter distribution is likely to have two preferable scales of 50
kpc and 200 kpc.  On the other hand, it is also found that the core
radius is tightly related to the radiative cooling timescale 
\citep[e.g.][]{Ota_Mitsuda_2004, Akahori_Masai_2005}.
Thus the core structure might largely be a result of the thermal
evolution of the ICM.

In this paper, using a large number of {\it ROSAT} and {\it ASCA}
clusters compiled by \cite{Ota_2001,Ota_Mitsuda_2004}, we investigate
correlations between the fundamental cluster parameters.  Since {\it
ASCA} has a high sensitivity to measure the X-ray spectrum in the wide
energy band while {\it ROSAT} is good at imaging in the soft X-ray
band, the two observatories are an excellent combination to study
properties of the intracluster medium.  At present the {\it
XMM-Newton} and the {\it Chandra} satellites are in orbit and generate
much cluster data with higher sensitivities. However, the data set
used in the present paper will be one of the best existing to
construct the largest sample of distant clusters and study global
X-ray structures.  While \cite{Ota_Mitsuda_2002} mainly focused on the
discovery of the typical two core sizes based on the analysis of 79
distant clusters with {\it ROSAT} and {\it ASCA},
\cite{Ota_Mitsuda_2004} described the uniform analysis, and showed the
X-ray parameters for the individual clusters and the results on
scaling relations. The present paper addresses 
the $L_{\rm X}-T$ relation in more detail, which is organized as below. In
\S~\ref{sec:sample}, the sample and parameter estimation are
described. In \S~\ref{sec:lx-t_density}, we show gas density profiles
of the sample and how they are related to a scatter about the $L_{\rm
X}-T$ relation. Since redshift evolution is not clearly seen in the
present distant sample and our analysis is based on a conventional
$\beta$-model, a study of redshift evolution in the $L_{\rm X}-T$
relation is beyond the scope of the present paper.  In
\S~\ref{sec:lx-t-rc} and \S~\ref{sec:lx-t-tcool}, we mainly focus on
the $L_{\rm X}-T$ relation and show its connection with the cluster
core radius and the cooling time.  In \S~\ref{sec:lx-tbeta} we also
derive a $L_{\rm X}-T\beta$ relation and show a trend that possible
merging clusters segregate from regular clusters on the plane.
Finally in \S~\ref{sec:discussion}, major results are summarized and
the evolution of the ICM are discussed particularly in the light of
radiative cooling.

We use $\Omega_{\rm M} = 0.3$, $\Omega_{\Lambda}=0.7$ and
$h_{70}\equiv H_0/(70~{\rm km\,s^{-1}Mpc^{-1}})=1$. The quoted
parameter errors are the 90\% confidence range throughout the paper
unless otherwise noted.  The $1\sigma$ error bars are plotted in all
figures.

\section{SAMPLE AND PARAMETER ESTIMATIONS}
\label{sec:sample}

\subsection{The Sample}

The sample is selected from an X-ray catalogue of the {\it ROSAT} and
{\it ASCA} distant clusters presented in \cite{Ota_Mitsuda_2004}.
Though the original catalogue includes 79 clusters in the redshift
range of $0.1 < z < 0.82$, the final sample used in this work
comprises 69 clusters in $0.10 \le z \le 0.56$ in order to reduce
possible selection effect and measurement uncertainties which are more
serious at higher redshifts, as detailed below.
  
Because the sample was collected from the archival data of the
pointing observations, in order to reduce the selection effect that is
not negligible at higher redshift, we exclude 5 high-redshift clusters
with large measurement uncertainties from the \cite{Ota_Mitsuda_2004}
sample: they are \#75 \object{MS2053.7-0049}, \#76 \object{3C220.1},
\#77 \object{MS1137.5+6625}, \#78 \object{RXJ1716.6+6708}, \#79
\object{MS1054.5-0321} (see \cite{Ota_Mitsuda_2004} for details). We
exclude \#71 \object{3C295} and \#76 \object{3C220.1} due to possible
contaminations of X-ray emission from the central radio sources as
reported by {\it Chandra} observations (see \cite{Allen_etal_2001a}
for \object{3C295} and \cite{Worrall_etal_2001} for \object{3C220.1}),
and \#56 \object{CL0500-24}, \#61 \object{CL2244-02}, and \#66
\object{CL0024+17} due to uncertainties in the spectroscopy caused by
contaminations from point sources.  We did not use \#43 A1758 and \#79
MS1054.5-0321 because their $\beta$-model parameters were not well
constrained.

\subsection{Brief Summary of X-ray Data Analyses}    
The data analyses consisted of two major steps: 1) the spatial
analysis with the {\it ROSAT} HRI and 2) the spectral analysis with
the {\it ASCA} GIS and SIS.

For 1), the radially-averaged X-ray surface brightness distribution
was fitted with the isothermal $\beta$-model to determine the slope
parameter $\beta$, the core radius $r_c$ and the central surface
brightness.  The model gave an reasonably good fit to the observed
surface brightness of many clusters: it resulted in a statistically
acceptable fit for about 2/3 of the sample. For some clusters,
however, it has left residual emission in the central region and a
significant improvement was found when including an additional
$\beta$-model component particularly for 9 clusters (termed as
``double-$\beta$'' clusters).  We have further found that adoption of
the single $\beta$-model for such double-$\beta$ clusters will return
parameters consistent with one of the two $\beta$-model components, as
shown in Fig. 9 of \cite{Ota_Mitsuda_2004}, which means the single
$\beta$-model tends to pick up a component that dominates the
emission.  Thus the present work basically stands on the single
$\beta$-model for the reason that we are mainly interested in a
systematic study on the global structure in a large number of distant
clusters. Note that possible systematic errors of physical parameters
estimated under the $\beta$-model will be discussed in the next
subsection. Furthermore, the X-ray morphology was measured with using
`centroid variations' of the cluster image and the sample was
classified into ``regular'' and ``irregular'' clusters (see \S 3.2 of
\cite{Ota_Mitsuda_2004} for details).

For 2), the emission-weighted temperature and bolometric luminosity
were measured from the {\it ASCA} spectroscopic data under the
Raymond-Smith model \citep{Raymond_Smith_1977}, utilizing the XSPEC
software version 9.0.  $kT$ is the emission-weighted temperature
determined from the simultaneous fitting to the {\it ASCA} GIS and SIS
spectra extracted from $r<6\arcmin$ and $r<3\arcmin$ circular regions,
respectively, and the bolometric luminosity, $L_{\rm X}$ is calculated
within a virial radius $r_{500}$.  Here $r_{500}$ is defined as the
radius within which the average matter density is equal to
$\Delta_c=500$ times the critical density of the Universe at the
cluster redshift.

In what follows, we use the X-ray parameters derived under the single
$\beta$-model; the slope parameter $\beta$, the core radius $r_c$, the
central electron density $n_{e0}$, and the radiative cooling time
$t_{\rm cool}$ as well as spectroscopically measured $T$ and $L_{\rm
X}$. All quantities are listed in Tables 2, 4, and 5 of
\cite{Ota_Mitsuda_2004}.

\subsection{Parameter Estimation under the $\beta$-model and systematic errors} 
To perform a reliable estimation of the $\beta$-model parameters, we
have carefully checked the effect of the {\it ROSAT}/HRI PSF on the
spatial analysis. As detailed in \S~3.4 of \cite{Ota_Mitsuda_2004}, we
confirmed that the effect is negligibly small and the parameters are
well reproduced as long as $r_c$ is larger than $5\arcsec$. We have
also checked the consistency with the previously published values from
the {\it Chandra} observations in regard to the $\beta$-model
parameters ($\beta$ and $r_c$) and the temperature, and found a good
agreement (see \S~3.4 and \S~4.2 of \cite{Ota_Mitsuda_2004}).  Due to
the large {\it ASCA} PSF, it is not easy to measure the
temperature profile accurately for the distant clusters. We then use the
spectroscopically determined mean temperature in the parameter
estimation.  We will further evaluate possible systematic errors in
estimating physical parameters under the $\beta$-model, $L_{\rm X}$
and $t_{\rm cool}$, which are important quantities in the present
study.

\begin{itemize}
\item $L_{\rm X}$\\ To estimate a systematic error of $L_{\rm X}$, we
  first calculate the bolometric luminosity within $r< 6\arcmin$ in
  two different ways: 1) convert the 2--10 keV luminosity measured
  with the GIS spectrum into the bolometric luminosity using the
  emissivity of the Raymond-Smith plasma model and the {\it ASCA}
  temperature, and 2) integrate the $\beta$-model surface brightness
  distribution determined with the {\it ROSAT} HRI within the same
  integration area and multiply it by the band correction factor.  We
  found that two estimations are consistent within about 10--20\% in
  all three different temperature ranges of 2--5, 5--7, $>7$~keV or
  four different redshift ranges, 0.1--0.2, 0.2--0.3, 0.3--0.4,
  $>0.4$.  This also assures that the current assumption of the
  isothermal $\beta$-model gives reasonably a good approximation to
  the real clusters in estimating $L_{\rm X}$.  Then we finally
  determine the bolometric luminosity within $r_{500}$, $L_{\rm X}$
  from 1) the bolometric luminosity that was spectroscopically
  measured within $6\arcmin$ under the Raymond-Smith model and
  corrected for the emission from outside $6\arcmin$ by assuming the
  $\beta$-model surface brightness distribution.  In Fig.~\ref{fig1}a
  we show a comparison of the bolometric luminosities measured with
  GIS and HRI.

\item $t_{\rm cool}$ \\ 
In the temperature range concerned, the cooling rate of the gas is
dominated by free-free emission and then approximately proportional to
$\sqrt{T}$. Therefore, with the cooling rate in the form $q_{\rm
  ff}\sqrt{T}n_{e0}^2$, we define the cooling time for a cluster as
$t_{\rm cool} = 3k\sqrt{T}/q_{\rm ff}n_{e0}$, where $n_{e0}$ is the
gas density at the cluster center.  On the other hand, recent
higher-resolution observations of clusters with a cooling core pointed
that the conventional $\beta$-model does not sufficiently fit overall
profiles \citep[e.g.][]{Lewis_etal_2003, Hicks_etal_2002}.  In
addition, if the central temperature drop by a factor of $\sim 2$
exists like what observed in many cooling-flow clusters, it might
cause an overestimation of $t_{\rm cool}$.  We then make a comparison
with recent results from \cite{Bauer_etal_2005} as they analyzed {\it
  Chandra} data of X-ray luminous clusters at $0.15<z<0.4$ and derived
the cooling time profiles using spatially resolved spectral
information. From Fig.~\ref{fig1}b, we find that there is an agreement
between the two analyses: for 11 out of 21 distant clusters $t_{\rm
  cool}$ is consistent with the {\it Chandra} results within the
measurement errors.  For the rest of them, it is systematically larger
(smaller) for $t_{\rm cool}\lesssim 2$~Gyr ($t_{\rm
  cool}\gtrsim10$~Gyr).  As the clusters with $t_{\rm
  cool}\gtrsim10$~Gyr are mostly irregular ones, the underestimation
might be due to their complex internal structure and different
definitions of the centroid position. From temperature maps derived by
\cite{Bauer_etal_2005}, we found that clusters with systematically
smaller $t_{\rm cool}$ tend to have hot spots at the centroids or gas
hotter than the surrounding regions. This would then bias the
temperature toward lower value (as we use the emission-weighted
temperature instead of the spatially-resolved central temperature) and
then higher central density (to account for the observed emission
measure), resulting in shorter $t_{\rm cool}$.  For clusters with
$t_{\rm cool}\lesssim 2$~Gyr, the overestimation may be attributed to
possible temperature decrease at the center: assuming that the central
temperature is 1/2 -- 1/3 of ambient temperature, $T'$, and the
emission-weighted temperature is related through $T\sim T'/1.3$
(Eq.~\ref{eq:temp_correction} in \S~\ref{subsec:lx-t-tcool}), $t_{\rm
  cool}$ is possibly overestimated by a factor of $\sim 1.3 - 1.8$.
Thus this may reconcile the factor of $\sim 2$ difference seen in
Fig.~\ref{fig1}b. However, because the above estimation depends on the
assumed temperature profile, we use $t_{\rm cool}$ obtained from the
$\beta$-model analysis of the {\it ROSAT} and {\it ASCA} data.  Note
that the two analyses compared in Fig.~\ref{fig1}b are consistent
within at most a factor of 2 and our discussion is not so much
affected by the absolute value of $t_{\rm cool}$ as this paper aims at
a study of a global trend of the ICM evolution, which is suggested to
proceed along the ``principal axis'' (\S~\ref{subsec:funda_plane})
over two orders of magnitudes of cooling time.
\end{itemize}

\section{$L_{\rm X}-T$ RELATION AND GAS DENSITY PROFILES}
\label{sec:lx-t_density}

\subsection{Core Radius Distribution and $r_{500}-r_c$ Relation}
\label{subsec:r500-rc}

Since the $L_{\rm X}-T$ relation is sensitive to the structure of the
cluster core region, we first examine the core radius.  As noted in
\cite{Ota_Mitsuda_2002}, together with 45 nearby clusters analyzed in
\cite{Mohr_etal_1999}, the $r_c$ distribution of 121 clusters exhibits
high concentrations around $50~h_{70}^{-1}$kpc and
$200~h_{70}^{-1}$kpc . We show the core radius distribution of the
present sample in Fig.~\ref{fig2}b. Though the $r_c$ has a large
cluster-to-cluster dispersion, we find that about 60\% of clusters are
included in narrow ranges corresponding to the two peaks. Thus the
large dispersion is partly due to the double peak structure of the
$r_c$ distribution.
 
Under the self-similar model, the internal structure of the gas should
be scaled by the virial radius, and then $r_{500}/r_c$ should be
constant for all clusters. However, as is seen from Fig.~\ref{fig2}a,
$r_c$ does not simply scaled by $r_{500}$.  As mentioned in
\cite{Ota_Mitsuda_2004}, we should be careful about the results of the
$r_{500}-r_c$ relation because it depends on the choice of the $r_c$
range.  When we fit all the sample at the same time, we obtain a weak
correlation, $r_{500}\propto r_c^{0.15\pm0.04}$ ($\chi^2/{\rm
  d.o.f.}=953.2/67$).  Since $r_{500}/r_c \gg 1$ in general, $r_{500}$
is mostly determined by the temperature.  Actually $r_c$ and $T$ do
not show a strong correlation.  If we divide the sample into two $r_c$
groups relative to $r_c=100$~kpc, $r_{500}\propto
r_c^{0.37^{+0.12}_{-0.11}}$ for 34 clusters with $r_c>100$~kpc
($\chi^2/{\rm d.o.f.}=107.7/32$) while no meaningful correlation was
found for 35 clusters with $r_c<100$~kpc (the correlation coefficient
is 0.21).  We further restrict the large core group to a narrow range,
for example $100<r_c<200$~kpc, $r_{500}\propto
r_c^{0.71^{+0.52}_{-0.27}}$ ($\chi^2/{\rm d.o.f.}=21.0/16$).  Thus we
consider that the large core clusters may marginally satisfy the
self-similar condition (see also \cite{Akahori_Masai_2005}).  On the
other hand, since the small core group clearly shows a deviation from
the $r_{500}-r_c$ relation, the formation of the small cores is
suggested to be determined from some physical process other than the
self-similar collapse. Furthermore, for 6 clusters with a very large
core radius of $r_c>400$~kpc, which forms the third peak around
500~kpc in the $r_c$ histogram (Fig.~\ref{fig2}b), there is no
meaningful $r_{500}-r_c$ correlation (the correlation coefficient is
0.21). Thus they do not satisfy the self-similarity. This deviation
may be attributed to the past or on-going merging events, which we
will discuss in more detail in \S~\ref{subsec:lx-tbeta}.

\subsection{Scaled Gas Density Profiles}\label{subsec:gas_profile}

To investigate the global gas density structure, we plot the best-fit
$\beta$-model profiles for 69 clusters in Fig.~\ref{fig3}, where the
radius was normalized by $r_{500}$.  A significant scatter of the
density exists within $\sim 0.1r_{500}$ region (typically $0.1
r_{500}\sim 100$~kpc) and the smaller core clusters show a
systematically higher density in comparison to the large core
clusters.  On the other hand, the profiles are fairly universal
outside that radius.  We consider that this is consistent with the
similarity found in outer regions of scaled profiles for nearby
clusters \citep{Neumann_Arnaud_1999}.

In the next two subsections we will treat the core region and the
envelope separately in order to directly show how the density profile
is related to the scatter around the $L_{\rm X}-T$ relation.  We found
that the $r_{500}-r_c$ relation and the density profile do not show a
clear evolution against the observed redshift.  However it is
important to test whether the $L_{\rm X}-T$ relation would evolve with
redshift or it is more closely related with other physical parameters.
Then in \S~\ref{subsec:lx-t_redshift} we will investigate the redshift
dependence of the $L_{\rm X}-T$ relation since the present sample
covers the redshift range of $0.1<z<0.56$.

\subsection{$L_{\rm X}-T$ outside $0.2r_{500}$}\label{subsection:lx-t_outside}

We calculate the luminosities outside a radius of $0.2r_{500}$ and
plot them against the temperature in Fig.~\ref{fig4}b.  The radius was
chosen so as to be a typical $r_c$ of the large core clusters, $\sim
200$~kpc and thus the emission from the core region can be removed
from the overall emission.  As a result, we obtained
\begin{equation}
L_{\rm X}(>0.2r_{500}) 
= 4.95^{+3.45}_{-2.13}\times10^{42} (kT)^{2.85^{+0.30}_{-0.27}}.
\end{equation}
We found that there is no significant difference in the relation
between the two $r_c$ groups (see also Table~\ref{tab1}) and the data
scatter around the mean relation is surprisingly small.  Despite the
similarity of the outer profiles shown in \S\ref{subsec:gas_profile},
however, the measured power-law index is yet close to 3 rather than
the self-similar expectation of 2.  This is also true for 20 clusters
with $100<r_c<200$~kpc, which marginally satisfy the self-similar
relation as noted in \S~\ref{subsec:r500-rc}; the power-law index
is $3.13^{+0.58}_{-0.51}$.

Note that the $L_{\rm X}-T$ relation is highly sensitive to the way
the central emission is treated.  \cite{Allen_Fabian_1998} pointed
out that the slope of the $L_{\rm X}-T$ relation would be flattened
from $T^{3}$ to $T^{2}$ once corrected the central emission based on
the standard cooling flow model. We find that the slope is still
steeper than the prediction of the self-similar model by simply
excluding the central component in a model-independent manner.

\subsection{$L_{\rm X}-T$ inside $0.2r_{500}$}
In Fig.~\ref{fig4}a, we show the $L_{\rm X}-T$ relation calculated
for $r<0.2r_{500}$.  The power-law fitting resulted in
\begin{equation}
L_{\rm X}(<0.2r_{500}) 
= 3.02^{+45.22}_{-2.83}\times10^{41} (kT)^{4.37^{+1.54}_{-1.50}}.
\end{equation}
There is a significant data scatter in the plot and the luminosity is
systematically higher for the small core group for a fixed
temperature. If assuming the best-fit slope of 4.37, the
  relations for the small and large core groups are obtained as:
\begin{eqnarray}
L_{\rm X}(<0.2r_{500}) &=&5.82^{+1.91}_{-1.12}\times10^{41} (kT)^{4.37}
	~{\rm for}~r_c\le 0.1~{\rm Mpc}, \\
L_{\rm X}(<0.2r_{500}) &=&8.32^{+1.58}_{-1.57}\times10^{40} (kT)^{4.37}
	~{\rm for}~r_c> 0.1~{\rm Mpc}
\end{eqnarray}
(see also Table~\ref{tab1}). This significant difference in the
normalization factor can be attributed to a negative correlation
between the central density and the core radius as mentioned in
\S\ref{subsec:gas_profile}.

We then see from Figs.~\ref{fig3} and \ref{fig4} that the density
scatter inside $0.2r_{500}$ is indeed a source of scatter seen in the
$L_{\rm X}-T$ relation.  In comparison to the result for
$r>0.2r_{500}$, the large variety of the luminosity for $r<0.2r_{500}$
and the significant deviation from the self-similarity for the small
core clusters may imply that the dispersion around the $L_{\rm X}-T$
is closely related to the evolution of the ICM in the central region
after the collapse.

\subsection{Redshift Dependence of $L_{\rm X}-T$ and Comparison to
  Previous Results}\label{subsec:lx-t_redshift}

We investigate the redshift dependence of the $L_{\rm X}-T$ relation
by evaluating the relations for the following three subsets of data:
18 clusters in $0.1\le z<0.2$, 27 clusters in $0.2\le z<0.3$, and 24
clusters in $0.3\le z < 0.56$ (only 7 out of 24 have $z>0.4$) and
compare with the previous results on high-redshift ($z>0.4$) and
low-redshift ($z<0.1$) clusters.  To constrain the redshift dependence
in the core and outer region separately, we also derive the $L_{\rm
X}-T$ relation for $r<0.2r_{500}$ and $r>0.2r_{500}$.  We did not
apply the evolution correction to $L_{\rm X}$ (i.e. multiplying
$L_{\rm X}$ by $E(z)^{-1}$ since it does not significantly change the
relation for $\Delta_c=500$ as shown in Table~9 of
\cite{Ota_Mitsuda_2004}.)

\begin{itemize}
\item $L_{\rm X}-T$ relation\\
  The power-law fitting of the $L_{\rm X}(<r_{500})-T$ relations for
  three redshift subgroups resulted in
\begin{eqnarray}
L_{\rm X} &=& 1.23^{+2.62}_{-0.95}\times10^{43} (kT)^{2.75^{+0.85}_{-0.71}}
	~{\rm for}~0.1\le z<0.2, \label{eq1}\\
L_{\rm X} &=& 5.13^{+19.0}_{-4.23}\times10^{42} (kT)^{3.18^{+0.92}_{-0.84}}
	~{\rm for}~0.2\le z<0.3, \label{eq2}\\
L_{\rm X} &=& 2.95^{+108.5}_{-2.92}\times10^{41} (kT)^{4.65^{+2.33}_{-1.86}}
	~{\rm for}~0.3\le z<0.56 \label{eq3}
\end{eqnarray}
(Fig.~\ref{fig5}a). There is a marginal trend to obtain a steeper
slope and a smaller normalization factor for the higher redshfit
clusters.  However, because of the large uncertainties seen in
  the slope parameter, if fixed at the best-fit value of the total
  sample, 3.01, the normalization factors overlap with each other
  within the errors as shown in Eqs.~\ref{eq4}--\ref{eq6} and
  Fig.~\ref{fig5}b, and thus the redshift dependence is not clear in
  the observed redshift range.
\begin{eqnarray}
L_{\rm X} &=& 7.94^{+1.37}_{-1.17}\times10^{42} (kT)^{3.01}
	~{\rm for}~0.1\le z<0.2, \label{eq4}\\
L_{\rm X} &=& 7.08^{+0.71}_{-1.23}\times10^{42} (kT)^{3.01}
	~{\rm for}~0.2\le z<0.3, \label{eq5}\\
L_{\rm X} &=& 6.92^{+1.32}_{-1.60}\times10^{42} (kT)^{3.01}
	~{\rm for}~0.3\le z<0.56.\label{eq6}
\end{eqnarray}

\cite{Ettori_etal_2004} reported a steep slope of $3.72\pm0.47$
($1\sigma$ error) for high redshift clusters ($0.4<z<1.3$) based on
the {\it Chandra} data analysis and suggested a negative evolution.
Eqs. \ref{eq2} and \ref{eq3} agree with their relation within the
errors. Thus there might be a weak redshift evolution in the $L_{\rm
  X}-T$ relation. However, due to the large uncertainties associated
with the relations for the higher redshifts, it is not statistically
significant.

\item $L_{\rm X}(<0.2r_{500})-T$ and $L_{\rm X}(>0.2r_{500})-T$ relations\\
   We obtain the $L_{\rm X}(<0.2r_{500})-T$ relation in
    the three redshift ranges as follows (see also Fig.~\ref{fig5}c).
\begin{eqnarray}
L_{\rm X}(<0.2r_{500})  &=& 3.89^{+2.53}_{-1.53}\times10^{41} (kT)^{4.37}
	~{\rm for}~0.1\le z<0.2\\
L_{\rm X}(<0.2r_{500})  &=& 2.39^{+1.17}_{-0.79}\times10^{41} (kT)^{4.37}
	~{\rm for}~0.2\le z<0.3\\
L_{\rm X}(<0.2r_{500})  &=& 1.88^{+1.34}_{-0.78}\times10^{41} (kT)^{4.37}
	~{\rm for}~0.3\le z<0.56.
\end{eqnarray}
There is only a weak signature of negative evolution regarding the
$L_{\rm X}-T$ within central $0.2r_{500}$ region: this change in the
normalization factor against redshift is not statistically significant
and smaller than that of two subgroups with different core sizes in
the above analysis (Fig.~\ref{fig4}a).  Furthermore, as shown in
Fig.~\ref{fig5}d, if the central $0.2r_{500}$ region is excluded, the
normalization factor of the $L_{\rm X}(>0.2r_{500})-T$ relation has
little redshift dependence.
    
In order to further compare our results to the previous results, we
followed the \cite{Markevitch_1998}'s analysis as much as possible:
the X-ray luminosity was computed within a fixed radius of 2~Mpc and
the possible cool emission inside 100~kpc was corrected by multiplying
the luminosity for $0.1 < r < 2$~Mpc by 1.06 for clusters with short
cooling time, $\log{t_{\rm cool}/t_{\rm age}}\leq -0.5$, while their
ambient temperature, $T'$ was estimated by using
Eq.~\ref{eq:temp_correction}.  As a result, there is no significant
difference in the $L_{\rm X}-T$ among the three subsets of data and
they are consistent with the local relation (Fig.~\ref{fig5}e).
Therefore the redshift dependence is not clearly seen in the present
sample.
    
There are recent {\it Chandra} and {\it XMM-Newton} studies of
high-redshift ($z\gtrsim0.4$) clusters which support a positive
evolution \citep{Vikhlinin_etal_2002, Kotov_Vikhlinin_2005,
  Maughan_etal_2005, Lumb_etal_2004}.  We should note that the current
result does not rule out the scaling of $L_{\rm X}(1+z)^{-1.5}$ found
by \cite{Vikhlinin_etal_2002}: we found that the normalization factors
of the $L_{\rm X}(1+z)^{-1.5}-T$ relations are again consistent with
that of local relation within their errors (Fig.~\ref{fig5}f). In his
recent review, \cite{Voit_2005} pointed that there are actually some
systematic uncertainties in the evolution measurement and it is not
precise enough to distinguish the possibilities of positive or
negative evolution.
\end{itemize}    

Based on the above analyses, we may consider that the redshift
dependence is likely to be a minor effect in discussing the source of
scatter around the $L_{\rm X}-T$ relation. Although the marginal,
negative evolution of the normalization factor seen in the $L_{\rm
  X}(<0.2r_{500})-T$ relation might imply a progress of gas
concentration as redshift, the study of redshift evolution requires
higher-accuracy determination of parameters and thus is beyond the
scope of the paper.  The scatter seems to link more strongly to the
other parameters describing the core structure, which we will try to
show more directly in \S~\ref{sec:lx-t-rc} and
\S~\ref{sec:lx-t-tcool}.

Besides, \cite{Ota_Mitsuda_2004} showed that there is not strong
redshift evolution in the observed X-ray properties particularly at
$z\lesssim 0.5$, including the central gas density, the core radius,
and temperature.  As we will show in \S~\ref{subsec:funda_plane}, the
current sample is also found to have an X-ray fundamental plane
similar to that found in nearby clusters \citep{Fujita_Takahara_1999}.
We will then carry out statistical studies regardless of their
redshifts in the paper.

\section{$L_{\rm X}-T$ RELATION AND $r_c$}\label{sec:lx-t-rc}

\subsection{Observed $L_{\rm X}-T$ Relation}\label{subsec:observed_lx-t}
In Fig.~\ref{fig6}a, we show the $L_{\rm X}-T$ relation of 69 clusters
where we again divide the sample into two subgroups according to two
different $r_c$ ranges. As pointed out in \cite{Ota_Mitsuda_2002}, the
distributions of the two subgroups are {systematically shifted} in the
figure; the luminosity of a small core cluster is on average higher
than that of a large core cluster for a fixed temperature. If we fit
all the 69 clusters at the same time, we obtain a power-law index of
$3.01^{+0.49}_{-0.44}$ (90\% error). Since we found that there is not
significant difference in the power-law indices between the two
subgroups (Table~\ref{tab1}), we obtain the best-fit normalization
factors of the relation to be $1.05^{+0.12}_{-0.11}\times10^{43}~ {\rm
  erg\,s^{-1}}$ and $4.52^{+0.37}_{-0.34}\times10^{42}~ {\rm
  erg\,s^{-1}}$ by fitting them separately but fixing the indices at
the mean value of 3.01. Thus the difference is significant: the
cluster with a smaller core radius is in general brighter for a given
temperature. The fitting results are summarized in Table~\ref{tab1}.

\subsection{$L_{\rm X}-T-r_c$ Relation}
The above observational results show that the dispersion around the
mean $L_{\rm X}-T$ relation is connected with the cluster core size.
Assuming the best-fit power-law index of $\alpha=3.01$, we scale the
luminosity of each cluster to the luminosity for temperature of 1 keV
as $L_{\rm 1keV}=L_{\rm X}(kT/1{\rm keV})^{-\alpha}$. In
Fig.~\ref{fig6}b, we plot $L_{\rm 1keV}$ as a function of $r_c$. We
found that there is a negative correlation approximated as $L_{\rm
1keV}\propto r_c^{-1.01^{+0.17}_{-0.21}}$ from the fitting.  This
suggests that $L_{\rm X}$ depends not only on $T$ but on $r_c$ as
\begin{equation}
L_{\rm X}\propto T^{3.0}{r_c}^{-1}. \label{eq10}
\end{equation}
The result is consistent with that derived by \cite{Fabian_etal_1994},
although they interpreted the dependence on $r_c$ as the cooling-flow
rate. Moreover, Eq.~\ref{eq10} is also consistent with the X-ray
fundamental plane of nearby clusters noted by
\cite{Fujita_Takahara_1999}.  In the next subsection, we will derive a
simple model for the $L_{\rm X}-T$ relation under the assumption of
$\beta$-model, incorporating the two characteristic core radii, and
compare with the observation.

\subsection{Comparison to a $L_{\rm X}-T-r_c$ Model}
We obtained the following observational facts so far: 1) $r_c$
distribution shows high concentration at 50 and 200 kpc, which result
in a large dispersion of $r_c$, 2) $r_c$ does not show strong
correlation with the virial radius, and 3) $L_{\rm X}$ depends on both
$T$ and $r_c$. These results suggest that $r_c$ is determined by some
physical process in the core region of the cluster, independent of the
cluster virial mass. We thus modify the assumption of self-similarity
and consider its consequence on the $L_{\rm X}-T$ relation.

In order to evaluate the $L_{\rm X}-T$ relation quantitatively, we
assume the isothermal $\beta$-model for the gas density and a constant
gas-mass fraction. The details of the calculation of this simple
``$L_{\rm X}-T-r_c$ model'' will be mentioned in
Appendix~\ref{appendix:model}.  For a typical value of $\beta=0.7$, we
obtain the model in the form of:
\begin{equation} 
L_{\rm X} \propto f_{\rm gas}^2 (\Delta_c\rho_{\rm crit})^{-0.05}
T_{\rm gas}^{2.6}r_c^{-1.1}. \label{eq11}
\end{equation}
This is close to the observed relation (and the equation for the X-ray
fundamental plane, Eq.~\ref{eq10}), however, the index for $T_{\rm
  gas}$ is smaller than the observation. In Fig.~\ref{fig7}, we show
the model curves for three different values of $r_c$.  Thus the
$L_{\rm X}-T-r_c$ model based on the isothermal $\beta$ model may
roughly fit the observed trends of the $L_{\rm X}-T$ relation, though
not sufficient to reproduce the data quantitatively.

\subsection{X-ray Fundamental Plane Analysis}\label{subsec:funda_plane}
To look further for an essential parameter to understand the
dispersion around the $L_{\rm X}-T$ relation, we investigate the X-ray
fundamental plane in the $(\log n_{e0}, \log T, \log r_c)$ space for
the distant sample. We utilize a method similar to
\cite{Fujita_Takahara_1999} but carry out a $\chi^2$ fit to take into
account the statistical errors.  As a result, we obtained three
orthogonal parameters as follows:
\begin{eqnarray}
X &\propto & n_{e0}^{0.44} r_c^{0.65}T^{-0.62}, \label{eq:X}\\
Y & \propto & n_{e0}^{0.45} r_c^{0.44}T^{0.78}, \label{eq:Y}\\
Z & \propto & n_{e0}^{0.78} r_c^{-0.62}T^{-0.10}. \label{eq:Z}
\end{eqnarray}
In Fig.~\ref{fig8}a, the $\log{X}-\log{Z}$ plane for the distant
clusters is shown.  We found that the above equations 
agree with those obtained for the nearby sample by
\cite{Fujita_Takahara_1999} and then the $z>0.1$ and $z<0.1$ clusters
are distributed approximately on the same fundamental plane.  In
Fig.~\ref{fig8}b we also show the distribution of nearby clusters
projected onto the same $\log{X}-\log{Z}$ plane, 
whose $\log{Z}$ range coincides well with that of distant clusters.

As $\chi^2/{\rm d.o.f.} = 492.1/66$ for the first fit
(Eq.~\ref{eq:X}), however, the constancy of $\log{X}$ i.e. the
planarity of the cluster distribution in the 3D space is rejected in
the strict sense.  Besides the data scatter, this is likely due to the
presence of a ``break'' of the fundamental plane that is seen in
Fig.~\ref{fig10}b.  We will go back to this problem in
\S~\ref{subsec:lx-t-tcool}.

By setting $X \sim \mbox{const.}$, Eq.~\ref{eq:Z} yields
\begin{equation}
Z \propto  r_c^{-1.78}T^{1.00} \propto  n_{e0}^{1.20}T^{-0.69}. \label{eq12}
\end{equation}
Thus the principal axis of the fundamental plane, $Z$ is suggested to
be closely related to $r_c$. Since $t_{\rm cool}\propto
\sqrt{T}/n_{e0}$, Eq.~\ref{eq12} can be rewritten as
\begin{equation}
Z\propto t_{\rm cool}^{-1.2}.  
\end{equation}
Since the $Z$-axis represents a direction along which the dispersion
of the data points becomes the largest in the parameter correlations,
$t_{\rm cool }$ is considered to be a key parameter to control the ICM
evolution. Hence we will focus on the effect of gas cooling in the
next section.

\section{$L_{\rm X}-T$ RELATION AND $t_{\rm cool}$}
\label{sec:lx-t-tcool}

\subsection{Radiative Cooling and Temperature Decrease}
As already mentioned in the previous section, there is a large
dispersion of the gas density inside $0.1r_{500}$ region which is
strongly related with $r_c$ (Fig.~\ref{fig3}).
\cite{Ota_Mitsuda_2004} derived the $n_{e0}-r_c$ relation to be
$n_{e0}\propto r_c^{-1.3}$ and noted that the correlation tends to
become even steeper for small core clusters: $n_{e0}\propto
r_c^{-1.9}$ for $r_c< 100$~kpc.  For clusters with such high density
cores, radiative cooling may work considerably.  It is also suggestive
that $0.1r_{500}$ is roughly equal to the typical cooling radius of
$r_{\rm cool}\sim100$~kpc.

In Fig.~\ref{fig9}a we show the radiative cooling time 
against the core radius.  We confirmed that for all
small core clusters, it is shorter than the Hubble time ($t_{\rm
H}=13.4$~Gyr; \cite{Spergel_etal_2003}), $t_{\rm cool} < t_{\rm H}$.
Even if we use the age of the Universe at the cluster redshift $t_{\rm
age}$ instead of $t_{\rm H}$, which ranges $8.0 < t_{\rm age}<
12.2$~Gyr for the present sample, that hardly alter the result.
Furthermore considering the strong dependence on the core radius,
$t_{\rm cool}\propto r_c^{1.7}$ for $r_c< 100$~kpc
\citep{Ota_Mitsuda_2004}, these results seem to indicate a central
concentration of the gas according to the progress of radiative
cooling. This is supportive of an idea of the standard cooling-flow
model in this regard. However, as seen from Fig.~\ref{fig9}b, there is
no clear difference in the temperature range between the two subgroups
of different core sizes. The average temperature (and the standard
deviation) is 5.4 keV (1.8 keV) for the clusters with $r_c<100$~kpc
and 7.0 keV (1.8 keV) for $r_c>100$~kpc, and the difference is only
30\%.  Thus there is not a strong temperature drop as predicted by the
standard cooling-flow model. Our result confirms the lack of strongly
cooled gas based on the large sample of {\it ROSAT} and {\it ASCA}
clusters.

\subsection{$L_{\rm X}-T-t_{\rm cool}$ 
and $L_{\rm X}-T'-t_{\rm cool}$ Relations}\label{subsec:lx-t-tcool}
To investigate the nature of the dispersion around the $L_{\rm X}-T$
relation in more detail, we show $L_{\rm 1keV}$ as a function of the
cooling time normalized by the age of the universe at the cluster
redshift, $(t_{\rm cool}/t_{\rm age})$ in Fig.~\ref{fig10}b. At $\log
(t_{\rm cool}/t_{\rm age})> -0.5$ the distribution is nearly constant
within the data scatter. On the other hand, at the shorter cooling
time of $\log (t_{\rm cool}/t_{\rm age})\lesssim -0.5$, a significant
deviation from the mean $L_{\rm X}-T$ relation (which is denoted with
the horizontal solid line in the figure) is clearly visible.  We found
that the result depends little on the redshift, as shown in
Fig.~\ref{fig11}.  Although this was interpreted as central excess
emission accompanied by the radiative cooling within the framework of
the standard cooling-flow model, we want to suggest another
possibility that we might underestimate the temperature for the
clusters with short cooling time because of mild temperature decrease
at the center.

We then evaluate the emission weighted temperature of the clusters
with short cooling time utilizing the universal temperature profile
found in many nearby cooling-flow clusters.  Assuming a projected
temperature profile, $T(r) \propto
0.40+0.61[(x/x_c)^{1.9}/(1+(x/x_c)^{1.9}]$ ($x=r/r_{2500}$) for
relaxed clusters \citep{Allen_etal_2001c} and the $\beta$ profile,
$S(r) \propto (1+(r/r_c)^2)^{-3\beta+1/2}$, $T = \int^{r_{\rm
max}}_{0} T(r) S(r) 2\pi r dr/\int^{r_{\rm max}}_{0} S(r) 2\pi r dr$ .
Since the temperature profile is nearly constant outside the typical
cooling radius of $\sim 100$~kpc, we obtain a temperature of the
outer, non-cooling region, $T'$ to be
\begin{equation}
         T' \sim1.3T,  \label{eq:temp_correction}
\end{equation}
for the typical values of $r_c=50$~kpc, $\beta=0.6$, $r_{2500}=
600$~kpc and $r_{\rm max} = 1$~Mpc.  If the temperature starts to
decrease again at a larger radius, $r\sim 0.15 r_{180}$ as observed in
nearby relaxed clusters \citep[e.g.][]{Vikhlinin_etal_2005,
Markevitch_etal_1998}, it would affect the above equation by only
$\lesssim10$\% due to a rapid decrease of the surface brightness at
such large radii.  Therefore $T$ is likely to underestimate the
ambient temperature by about 30\% for $t_{\rm cool} \ll t_{\rm
age}$. Notice that this estimation is also consistent with the
difference of 30\% in the average temperature between two subgroups
shown in the previous subsection.  Thus we consider that it is worth
examining the luminosity - ``ambient temperature'' relation.

By simply adopting the correction $T'=1.3T$ for 26 clusters with $\log
(t_{\rm cool}/t_{\rm age})\le -0.5$ otherwise $T'=T$ for 43 clusters,
we plot the $L_{\rm X}-T'$ relation in Fig.~\ref{fig10}c. As listed in
Table~\ref{tab1}, the difference in the normalization factors between
$r \lessgtr100$~kpc mostly disappeared in this case. The slope of the
power-law function is $2.80^{+0.28}_{-0.24}$ and thus marginally less
steep compared with that of $L_{\rm X}-T$.  \cite{Fukazawa_etal_2004}
recently derived scaling relations using the {\it ASCA} archival data
on $\sim300$ objects over a wide range of mass scales from elliptical
galaxies to clusters. They found a smaller index of the $L_{\rm X}-T$
relation, $\alpha=2.34\pm0.29$ at $kT>4$~keV. Their analysis is
different from the present one in that they utilized an integration
area extending to 2--3 times $r_{500}$ for luminosity and the area
excluding the central $1' - 3'$ region for temperature. Their
temperature should therefore be closer to $T'$ than $T$. In fact, the
index of the $L_{\rm X}-T'$ relation mentioned above is slightly
larger than but consistent within error bars with the result of
\cite{Fukazawa_etal_2004}.

In Fig.~\ref{fig10}d we show $L_{\rm 1keV}(=L_{\rm X}/T'^{2.80})$
versus $(t_{\rm cool}/t_{\rm age})$.  To quantify the dispersions and
demonstrate their changes, we plot histograms of $L_{\rm 1keV}$ for
the two $r_c$ groups and those with the above correction in
Fig.~\ref{fig12}. Since we are interested in the statistical
significance of the change of the width, we performed the Gaussian
fitting with the $\chi^2$ method to include the Poisson error of each
bin. The resultant Gaussian parameters (the mean, $\mu$ and the width,
$\sigma$) and the $1\sigma$ errors are listed in Table~\ref{tab2} and
also shown in each panel of Fig.~\ref{fig12}. We thus found that by
correcting the temperature with Eq.~\ref{eq:temp_correction}, the
width significantly decreased from
$\sigma/\mu=(7.0\pm1.2)\times10^{-3}$ to $(3.3\pm 0.9)\times10^{-3}$
($1\sigma$ error) for the smaller $r_c$ group (see Fig.~\ref{fig12}a,
c) and the difference of the widths between two $r_c$ groups
disappeared: $\sigma/\mu=(3.3\pm0.9)\times10^{-3}$ and
$(3.5\pm0.5)\times10^{-3}$ (Fig.~\ref{fig12}c, d).  Note that the
$1\sigma$ error of $\sigma/\mu$ is dominated by that of $\sigma$ and
thus calculated by simply propagating the errors of $\sigma$ and
$\mu$.

It is also remarkable that after the correction, there is no
significant difference in the dispersions of the data points between
$t_{\rm cool}\lessgtr t_{\rm age}$ and $L_{\rm 1keV}$ can be regarded
as constant within the data scatter over two orders of magnitudes of
$t_{\rm cool}$ (see also Table~\ref{tab2}).  Thus we consider that
this also gives an evidence against the standard cooling-flow model
because it predicts cooling that is increasingly enhanced with
time. Our result indicates that the luminosity, that is the rate of
thermal energy loss by the radiative cooling, does not depend on time
very much, and some steady-state is attained for the gas of many
small-core clusters. We will discuss about this in more detail in
\S~\ref{sec:discussion}.

\section{$L_{\rm X}-T\beta$ RELATION AND X-RAY MORPHOLOGY}
\label{sec:lx-tbeta}

\subsection{$L_{\rm X}-T\beta$ Relation}\label{subsec:lx-tbeta}
In the case that the intracluster gas follows the isothermal
$\beta$-model, the virial temperature should be represented by $T_{\rm
  vir} = \beta T_{\rm gas} x^2(1+x^2)^{-1} \sim \beta T_{\rm gas}$,
where $x\equiv r_{\rm vir}/r_c$ and typically $x \gg 1$
\citep{Ota_2001, Akahori_Masai_2005}. We then replace $T$ with $\beta
T$ (here $\beta$ is the slope parameter determined from the {\it
  ROSAT} HRI radial profile fitting) and derive the $L_{\rm X}-T\beta$
correlation (Fig.~\ref{fig13}a).

As a result, the power-law index of the $L_{\rm X}-T\beta$ relation is
determined to be $2.67\pm0.44$ , which is marginally smaller in
comparison to that of the $L_{\rm X}-T$. As for the data scatter
around the mean relation, it becomes smaller for the small-core
clusters: the width changes from $\sigma/\mu=(7.0\pm1.2)\times10^{-3}$
to $(4.6\pm0.7)\times10^{-3}$ from $L_{\rm X}-T$ to $L_{\rm
  X}-T\beta$, while it becomes marginally larger for the large-core
clusters: $(4.2\pm0.7)\times10^{-3}$ to $(10.8\pm7.8)\times10^{-3}$.
The results are also summarized in Tables~\ref{tab1} and \ref{tab2}.

To look into the cause, we investigated parameter correlations,
$\beta-T$ and $\beta-r_c$.  We found that the $\beta-T$ correlation is
not significant (the correlation coefficient is 0.24) and at most
$\beta\propto T^{0.17^{+0.08}_{-0.09}}$.  On the other hand, there is
a trend to find larger $r_c$ for larger $\beta$ noticeably at
$r_c>100$~kpc though such a trend is not clear for $r_c<100$~kpc.  In
addition, several clusters show highly inhomogeneous surface
brightness distributions and very large cores of $r_c>400$~kpc, which
form a small peak at around $r_c\sim 500$~kpc in the $r_c$ histogram
(Fig.~\ref{fig5}b).  Considering that a large fraction of large-core
clusters (24 out of 34) exhibit irregular X-ray surface brightness
distribution, we may say that the above change in the dispersion
likely reflects the X-ray morphology (see \S~3.2 and Table~2 of
\cite{Ota_Mitsuda_2004} for classification of the X-ray morphology).

Particularly, all the five clusters that are located far below the
mean relation (i.e. \#30 \object{A1895}, \#50 \object{AC118}, \#18
\object{MS0451.5+0250}, \#46 \object{1E0657-56}, and \#44
\object{A483}) have multiple peaks or a signature of cold front in the
X-ray images, which can be attributed to cluster merging.  In
addition, they have very large core radii of $r_c\gtrsim 400$~kpc and
large $\beta\sim 1$ in comparison to the average value for the distant
sample, $\langle \beta \rangle = 0.64$ \citep{Ota_Mitsuda_2004}.  In
Fig.~\ref{fig13}b, they appear at the largest end of $r_c$.  If we
exclude the five clusters and estimate the width of the distribution
of $L_{\rm 1keV}(=L_{\rm X}/(kT\beta)^{2.67})$, it results in a
smaller value of $\sigma/\mu=(6.4\pm5.5)\times10^{-3}$ for 29 large
core clusters though it is associated with the large statistical
uncertainty.  To show the relevance to the X-ray morphology in an
alternative way, we divide the sample into the regular and irregular
clusters and plot them by using different colors in Figs.~\ref{fig13}c
and d.  We roughly see three phases: i) irregular clusters with very
large core radius and smaller $L_{\rm 1keV}$ at $\log{r_c} > -0.4$,
ii) a coexistence of regular and irregular clusters with large core
radius at $-1< \log{r_c} < -0.4$, and iii) regular small-core clusters
with higher $L_{\rm 1keV}$ at $\log{r_c} < -1$.  Note that for iii),
there is only one exception, which is \#17 \object{A115}, an irregular
cluster lying at the smallest $r_c$. \footnote{\object{A115}
  ($z=0.1971$) exhibits two peaks in the surface brightness
  distribution and the temperature variation measured with {\it ASCA}
  implied that the cluster is indeed a merger system
  \citep{Shibata_etal_1999}. \cite{Gutierrez_Krawczynski_2005}
  reported based on the {\it Chandra} observation that the northern
  subcluster is highly nonuniform and there is a bright central core
  of about $10\arcsec$ in diameter, which might be explained by the
  ISM of the cD galaxy or a cooling core sloshing in the cluster. Our
  $\beta$-model analysis with {\it ROSAT} yielded $r_c =
  4.9^{+1.0}_{-0.6}\arcsec$, consistently with the {\it Chandra}
  result, and thus is very likely to have picked up the dense bright
  core of the northern subcluster. } Therefore we suggest that the
above results show a trend that merging clusters {particularly in the
  phase i)} segregate from more relaxed clusters on the $L_{\rm
  X}-T\beta$ plane.

\subsection{$L_{\rm X}-T'\beta$ Relation}\label{subsec:lx-tbeta3}
In Fig.~\ref{fig14}, we finally show a plot of $L_{\rm X}-T'\beta$
relation for $T'$ corrected in the same manner as in
\S~\ref{subsec:lx-t-tcool} and Fig.~\ref{fig13}.  The results of the
power-law fitting and the dispersions around the mean relation are
also listed in Tables~\ref{tab1} and \ref{tab2}. The power-law index
is obtained to be $2.54^{+0.29}_{-0.26}$ from the fitting to 69
clusters.  This is smaller in comparison with the $L_{\rm X}-T$ and
other three relations derived above, yet significantly steeper than
the self-similar model. In Fig.~\ref{fig14}b, we find that the
distribution of the data points is fairly constant against the cooling
time: in particular for 43 clusters with $t_{\rm cool}\le t_{\rm
  age}$, the scatter is small: $\sigma/\mu=(2.7\pm0.5)\times10^{-3}$
(Table~\ref{tab2}).  As suggested in the previous subsection, we see
again the marginal trend of segregating the merging cluster from the
regular clusters in Fig.~\ref{fig14}a.

In the next step, we further divide the sample into three subgroups
according to the X-ray morphology: the regular single-$\beta$ cluster,
the irregular single-$\beta$ cluster, and the regular double-$\beta$
cluster, and distinguish them by using different colors in
Fig.~\ref{fig14}c and d.  From Fig.~\ref{fig14}d, we may recognize a
trend of the morphological change along the $t_{\rm cool }$ axis.
Here it is possible to define three phases as follows: from larger to
smaller $t_{\rm cool}$, i) the irregular large-core clusters at
$\log(t_{\rm cool}/t_{\rm age})\gtrsim 0$, where a large scatter of
$L_{\rm 1keV}$ is seen, ii) the regular large-core clusters at
$-0.3\lesssim \log(t_{\rm cool}/t_{\rm age})< 0$, and iii) the regular
small-core clusters at $\log(t_{\rm cool}/t_{\rm age})<-0.3$.  The
regular and irregular clusters are no more mixed up in the $L_{\rm
  1keV}-t_{\rm cool}$ plot apart from a few exceptions (they are
irregular clusters with small $r_c$; \#17 \object{A115}, \#38
\object{MS1910+6736}, and \#11 \object{MS0906.5+1110}).  It also
should be noted that the boundary at $\log{t_{\rm cool}/t_{\rm
    age}}=0$ in Fig.~\ref{fig14}d, i.e. the boundary between i) and
ii), approximately corresponds to the crossover point at $\log{t_{\rm
    cool}}\sim\log{t_{\rm H}}$ and $\log{r_c}\sim-0.4$ in
Fig.~\ref{fig9}a, and thus also to the boundary between i) and ii) in
Fig.~\ref{fig13}d.  Unlike the $L_{\rm 1keV}-r_c$ plot
(Fig.~\ref{fig13}d) where the regular and irregular clusters are mixed
in the intermediate range of $r_c$, the three groups are now clearly
separated from each other in Fig.~\ref{fig14}d.
  
Furthermore, the double-$\beta$ clusters are located in iii)
$\log(t_{\rm cool}/t_{\rm age})\lesssim -0.3$, the same region as the
regular small-core clusters. As the inner-/outer-core dominant
double-$\beta$ clusters are at a shorter/longer side of the range,
they are suggested to be closely related to the small-/large-core
regular clusters.  We thus consider that this may be connected with
the evolution of the X-ray surface brightness distribution after the
onset of radiative cooling. We will discuss on this point in more
detail in the next section.

\section{DISCUSSION}
\label{sec:discussion}

\subsection{Summary of the Results}
From the statistical analysis of 69 distant clusters in $0.1<z<0.56$
with the {\it ROSAT} and {\it ASCA} X-ray catalog
\citep{Ota_Mitsuda_2004}, we have obtained the following observational
results.

\begin{enumerate}

\item Regarding the cluster outer region, the density profiles are
  nearly universal and the scatter around the $L_{\rm
    X}(>0.2r_{500})-T$ relation is surprisingly small.  However, the
  slope of the relation is significantly larger than the self-similar
  expectation of 2.

\item For the central region, the gas density exhibits a significant
  scatter and the self-similar condition is not satisfied particularly
  in the small core clusters.
  
\item We studied the redshift evolution of the $L_{\rm X}-T$ relation.
  If restricted the luminosity integration area to the central
  $0.2r_{500}$ region, there might be a weak, negative evolution,
  however, the normalization factor of the relation does not
  significantly change with redshift and is consistent with the nearby
  sample within the data scatter.

\item We investigated the parameter correlations focusing on the
  $L_{\rm X}-T$ relation and the connections to $r_c$ and $t_{\rm
    cool}$, and suggested based on the X-ray fundamental plane
  analysis that $t_{\rm cool}$ is likely to be a control parameter for
  the ICM structure of the cluster core region.

\item For all small core clusters, $t_{\rm cool}<t_{\rm age}$ and the
  central electron density is systematically higher as $n_{e0}\propto
  r_c^{-1.9}$, which is suggestive of a central concentration of the
  gas as cooling progresses. On the other hand, we confirmed that the
  temperature decrease is mild in the small core clusters; it is
  estimated as $T=1.3T'$ using the universal temperature profile.
  Then we derived the luminosity-`ambient temperature' relation and
  showed that the dispersion of $L_{\rm 1keV}$ significantly decreases
  and can be regarded as constant against $t_{\rm cool}$ within the
  data scatter.

\item From the $L_{\rm X}-T\beta$ and $L_{\rm X}-T'\beta$ relations,
  we showed that $L_{\rm 1keV}$ is related to the X-ray morphology,
  and that there is a marginal trend for the merging clusters
  segregating from the regular clusters on their planes. We indicated
  the possible three phases of the gas property along the $t_{\rm
    cool}$ axis in Fig.~\ref{fig14}d, according to the different level
  of cooling and ICM density structure.
\end{enumerate}

\subsection{Impact of Radiative Cooling on the $L_{\rm X}-T$ Relation
and Comparison to the Quasi-hydrostatic Model}
\label{subsec:discuss_quasi_static}

In the context of isobaric cooling flow model, the local density of
the gas increases against decreasing the temperature so that the
thermal pressure $P(r)$ at given $r$ is kept constant within the
cooling radius, i.e. $P(r) = n_e(r) kT(r) = \mbox{const.}$, and then
$n_e\propto T^{-1}$ would follow.  If this is the case, since $L_{\rm
X} \propto n_e^2 T^{1/2}$ for the gas hotter than $kT\sim2$~keV, the
$L_X-T$ relation would exhibit $L_{\rm X}\propto T^{-3/2}$.  For a
typical cooling radius of $r_{\rm cool}\sim 0.1r_{500}$ ($\sim
100$~kpc), $L_{\rm X}(<r_{\rm cool})$ contributes about 50\% of the
total luminosity. When we derived the $L_{\rm X}(<0.1r_{\rm 500})-T$
relation, however, we did not find any tendency for such an inverse
correlation (see also the $L_{\rm X}(<0.2r_{500})-T$ relation in
Fig.~\ref{fig4}a).

The observational results indicate that the gas in the core region of
small core clusters is undergoing radiative cooling but the
temperature decline towards the center seems mild.  The gas may be
considered close to a steady state.  We will discuss such a possible
state of the gas, quasi-hydrostatic cooling, proposed by
\cite{Masai_Kitayama_2004}.  The model predicts a moderate and smooth
gas inflow with hydrostatic balancing.  In the context, unlike
isobaric cooling flows that increase the local density so $P(r)$ is
kept against local cooling, quasi-hydrostatic cooling allows the gas
to modify its profile or core size so $\nabla P(r)$ matches with the
force by gravitational potential.

In \S~\ref{subsec:lx-t-tcool} we derived the $L_{\rm X}-T'$
relation by correcting the emission-weighted temperature for
small-core clusters with short $t_{\rm cool}$ based on the idea of the
universal temperature profile, and showed that the bolometric
luminosity normalized by the temperature dependence of $(kT')^{2.8}$,
$L_{\rm 1keV}$ is constant within the data scatter over a wide range
of $t_{\rm cool}/t_{\rm age}$.  This strongly suggests that the rate
of thermal energy loss is kept nearly constant after the onset of
cooling in the cluster core.

According to the quasi-hydrostatic model, the temperature starts to
decrease at the cooling radius and approaches a constant of $\sim 1/3$
the ambient temperature towards the center.  The temperature profile
does not depend on its absolute value.  This picture may explain the
observations of nearby cooling-flow clusters.  The mass inflow rate
from the outer region is controlled by cooling so as to maintain the
quasi-hydrostatic balance and is expected to vary little through the
flow.  The condition is well satisfied in the case that the
temperature is $kT \ga 2$~keV where bremsstrahlung dominates cooling
more than line emission.  The present sample meets this condition
since it has a temperature range of $2<kT~{\rm [keV]}<12$.  Therefore
our results can be consistently understood within a framework of the
quasi-hydrostatic cooling model.

At the same time, we have found that the gas density profile for the
outer part of the cluster does not vary very much from cluster to
cluster, and is almost independent of the gas cooling at the center.
This would also imply that the energy loss due to radiative cooling at
the center is only a portion of the total energy of the thermal gas,
and the outer region is likely to offer a great {\it reservoir} of the
heat, which is carried a fraction into the inner cooling region
advectively by the gas inflow under quasi-hydrostatic balance.

As described above, cooling with quasi-hydrostatic balancing modifies
the core structure of the gas.  This may give a clue to the origin of
the small core of $\sim 50$~kpc or the inner core component of the
double-$\beta$ cluster.  In addition the study would also have an
impact on the present understanding of the origin of the dispersions
around the $L_{\rm X}-T$ relation because it provided the
observational evidence that they are directly linked to the thermal
evolution of the ICM and showed how the relation would be modified
when the effect of the cooling is considered.

\subsection{Evolution of Density Structure of X-ray Clusters}
\label{subsec:discuss_evolution}

In this subsection we attempt to present an overall picture of the ICM
evolution on the basis of the observational results.  First we
consider about the $r_c$ distribution and the origin of the
double-$\beta$ nature of the clusters from the phenomenological point
of views.

We have reported that the cluster core radius has a large
cluster-to-cluster dispersion of more than an order of magnitude, even
if restricted the sample to the regular clusters
\citep{Ota_2001}. This is partly due to the presence of two peaks
separated by a factor of $\sim4$. For 9 (7) out of the 35 regular
clusters, the surface brightness is significantly (marginally) better
fitted with the double-$\beta$ model, and the average values of two
core radii for the double-$\beta$ clusters coincide well with the two
$r_c$ peaks of the single-$\beta$ clusters.  Then a question may
arise: are there three subclasses of clusters, small-core
single-$\beta$, large core single-$\beta$, and double-$\beta$
clusters, or are we looking at some different aspects of a single
evolutionary process?  From the imaging analysis, we noticed that the
double-$\beta$ clusters are further subdivided into inner-core
dominant and outer-core dominant ones; if we are to perform a single
$\beta$ model fitting for these clusters, we would pick up either of
the core radii as the best-fit result.  Thus the similarity in the
$r_c$ distributions for the single-$\beta$ and double-$\beta$ clusters
can be interpreted in the following way: the (regular) clusters have
more or less double-$\beta$ nature in their structure, and when X-ray
emission from one of the two cores dominates the whole cluster
emission, we recognize it as the single-$\beta$ case and pick up
either the inner or outer core as the core radius. This will account
for an apparent large dispersion of core radius. Accordingly the
single-$\beta$ clusters with a small/large core radius may
respectively correspond to extreme cases of inner/outer core dominant
double-$\beta$ clusters.

Then why are there inner-core dominant and outer-core dominant
clusters?  Such double structures have often been attributed to the
cooling-flow phenomenon at the cluster center. However, the present
analysis showed that the standard cooling-flow model with dynamic
inflow is not preferable to account for the results, in particular the
$T-n_{e0}$ and $L_{\rm 1keV}-t_{\rm cool}/t_{\rm age}$ relations.  On
the other hand, in \S~\ref{sec:lx-tbeta} we suggested the possible
three phases of the ICM property on the $L_{\rm 1keV}-t_{\rm
  cool}/t_{\rm age}$ plane in the light of the radiative cooling and
the relevance to a variety of density structure.  It is likely that
the double-$\beta$ clusters are those in the transient phase of their
cooling cores, i.e., core size transition from outer-core dominated to
inner-core dominated, mentioned in the quasi-hydrostatic cooling
model.  As the interpretation of the double core nature, we consider
in the following way: when some time passed after the collapse and the
cooling becomes important, the core is radiatively cooled and the
central density gradually becomes higher to evolve from an
outer-core-dominant cluster toward an inner-core-dominant cluster.

The above discussion on the evolution of gas structure focused on the
effect of radiative cooling and the possibility of quasi-hydrostatic
state since the X-ray analysis showed that a majority of small core
clusters are distributed in a narrow range of $L_{\rm 1keV}$ once the
temperature bias is corrected (Fig.~\ref{fig12}) and thus supports the
existence of a typical value. Moreover, a long tail at the smaller
$L_{\rm 1keV}$ is seen in the histogram for $r_c < 100$~kpc
(Fig.~\ref{fig12}c). This indicates existence of some peculiar
clusters or extra physical mechanism that produces such a
variation. Recently, \cite{O'hara_etal_2005} examined a relation
between substructure and cool cores and the scatter about cluster
scaling relations using 45 nearby clusters. They found that cool core
clusters, which usually have less morphological substructure, exhibit
higher intrinsic scatter about scaling relations even after correcting
for cooling core effect, suggesting a possibility that a more global
process is at work. Then the variation in the $L_{\rm 1keV}$ histogram
for the small $r_c$ clusters may be related with the scatter found by
\cite{O'hara_etal_2005}. To further clarify both the structural
evolution and the scatter of scaling relations, a systematic study
with higher sensitivities will be important.

\subsection{Related Issues and Future Observations}

We will remark on the effect of the cold front detected in nearby
clusters with the {\it Chandra} observations in regard to the present
interpretation, and the assumption of the universal temperature
profile with consideration on the recent progress of {\it Chandra} and
{\it XMM-Newton} observations.

The cold front structure is characterized by a compact core with the
density and temperature jumps in contrast to the surroundings, and
interpreted as a remnant of subcluster merger
\citep{Markevitch_etal_2000}.  For higher redshifts, it is more
difficult to directly identify such small structures due to the
limitation of the sensitivity to measure the temperature maps, and
some of them might possibly be fitted with the double-$\beta$
model. However, as the cold front is a signature of recent merger, it
is expected to be accompanied with the irregular X-ray
morphology. Thus for relatively bright clusters, a small core of the
cold front in an irregular cluster may be distinguishable from that of
a regular cluster by referring to the morphological information.  For
example, \object{A115} and \object{1E0657-56}, for which a possible
remnant of cooling core of a merging subcluster was pointed out
\citep{Gutierrez_Krawczynski_2005, Markevitch_etal_2002}, were
classified into irregular clusters from our image analysis, though
they have different $r_c$ and are located in the different phases in
Fig.~\ref{fig14}d ( i) and iii), respectively).  For more accurate
identification of the cold front in distant clusters, it is necessary
to measure the 2 dimensional density and temperature maps with higher
sensitivity observations.

We have utilized the universal temperature profile to estimate the
effect of temperature decrease and relate the emission-weighted
temperature to the ambient temperature for clusters with short cooling
time and apparent luminosity enhancement in Fig.~\ref{fig10}b. Due to
the limited spatial resolution of {\it ASCA}, the temperature profiles
were not directly constrained for the distant sample. However for some
of them, the results have been already published with the {\it
  Chandra} and {\it XMM-Newton} data. As for the double-$\beta$
clusters or the small core cluster with very short cooling time of
$-1.3 \lesssim \log{(t_{\rm cool}/t_{\rm age})} \lesssim -0.7 $,
\object{A2204}, \object{PKS0745-191}, \object{A2390}, \object{A1835},
the gradual temperature decline within the central region of roughly
$r<100 - 200$~kpc were reported \citep{Sanders_etal_2005,
  Hicks_etal_2002, Allen_etal_2001b, Majerowicz_etal_2002}. For the
other double-$\beta$ cluster, A1689, with slightly longer $\log{t_{\rm
    cool}/t_{\rm age}}=-0.56$, the nearly isothermal profile was found
with {\it Chandra} \citep{Xue_Wu_2002}, and a complex temperature
structure was further reported with {\it XMM-Newton}
\citep{Andersson_Madejski_2004}.  Therefore, from the overall
standpoint, the assumption of the universal temperature profile for
$\log{t_{\rm cool}/t_{\rm age}} \lesssim -0.5$ seems reasonable for
many of the small core and double-$\beta$ clusters.  To directly
confirm the temperature structure in relation to the gas cooling, we
need to analyze the temperature profile of individual clusters in
various phases with using higher resolutions.

In \S~\ref{subsection:lx-t_outside}, we derived the $L_{\rm
  X}(>0.2r_{500})-T$ relation without correcting for the temperature
profile.  Adopting the corrected temperature $T' =1.3 T$ (Eq.
\ref{eq:temp_correction}), we obtain a similar slope, $L_{\rm
  X}(>0.2r_{500})=2.34^{+2.35}_{-1.13}\times10^{42}
(kT')^{3.05^{+0.33}_{-0.35}}$ for 69 clusters ($\chi^2/{\rm
  d.o.f.}=610/67$) (Table~\ref{tab1}) and a smaller standard deviation
for $L_{\rm 1keV}$, $\sigma/\mu=(4.5\pm0.5)\times10^{-3}$.  Once the
ambient temperature $T'$ is directly measured for a large sample of
clusters by future observations, we would suggest to use it instead of
$T$ in the analysis of cluster scaling relations.

\acknowledgments N.O. acknowledges support from the Special
Postdoctral Researchers Program of RIKEN. The authors thank J. J. Mohr
for valuable comments and discussions, and T. Mihara and T. Oshima for
useful suggestions. 
This work is supported in part by the Grants-in-Aid by the Ministry
  of Education, Culture, Sports, Science and Technology of Japan
  (14740133:TK).
  
{\it Facilities:} \facility{ASCA ()}, \facility{ROSAT ()}

\appendix
\section{A SIMPLE $L_{\rm X}-T-r_c$ MODEL}\label{appendix:model}

\begin{figure}
\epsscale{0.8}
\plottwo{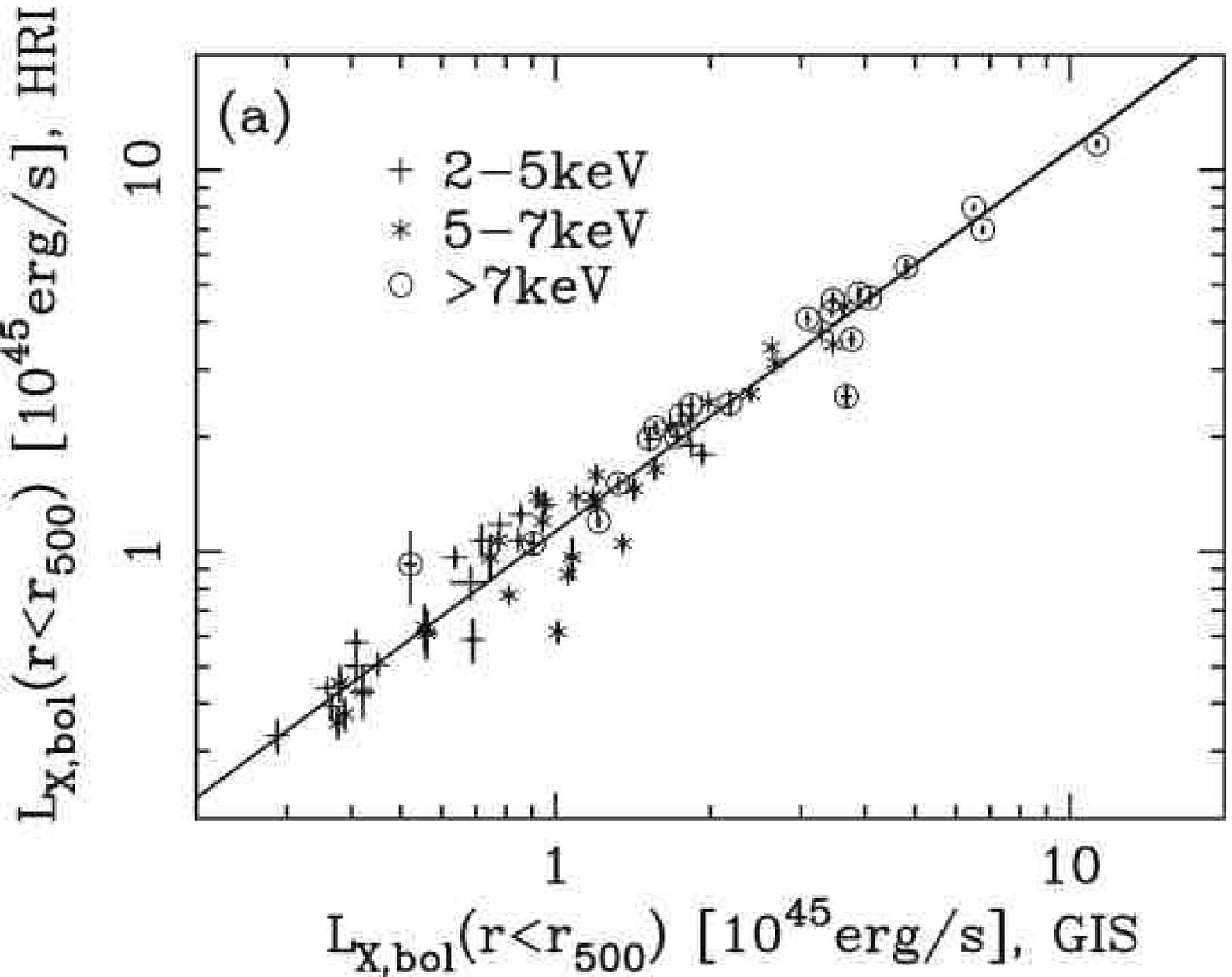}{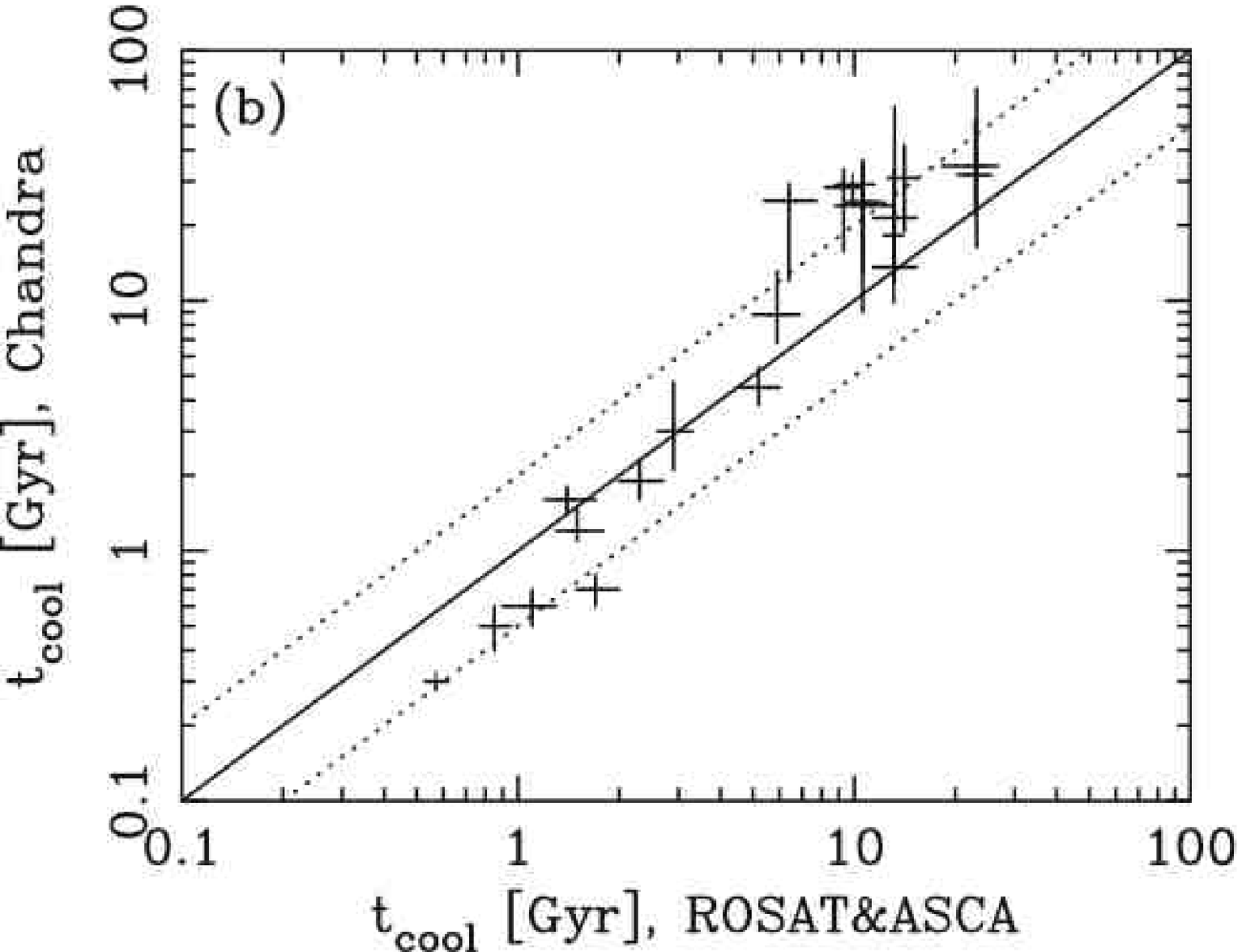}
\figcaption{(a) Comparison of two luminosity measurements.  The
  bolometric luminosities within $r_{500}$ estimated from the {\it
    ROSAT} HRI are plotted against those from the {\it ASCA} GIS. The
  clusters are divided into three subgroups according to the best-fit
  temperature from the {\it ASCA} spectroscopy, 18 clusters with
  $2<kT~{\rm [keV]}<5$, 30 clusters with $5<kT~{\rm [keV]}<7$ and 21
  clusters with $7<kT~{\rm [keV]}<12$ and denoted with the crosses,
  the asterisks, and the open circles, respectively. The solid line
  corresponds to $L_{\rm X, HRI} = L_{\rm X, GIS}$. (b)
    Comparison of the cooling time, $t_{\rm cool}$ estimated
    with the {\it ROSAT} and {\it ASCA} analysis \citep{Ota_Mitsuda_2004} and
    {\it Chandra} \citep{Bauer_etal_2005} regarding 21 clusters with
    $0.15<z<0.35$. 
          \label{fig1}}
\end{figure}

\begin{figure}
\epsscale{0.4}
\plotone{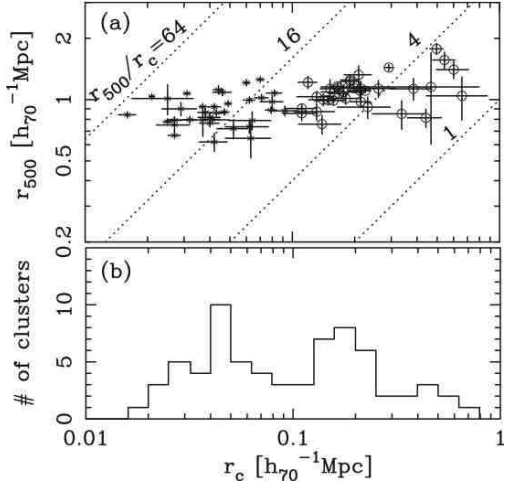}
        \figcaption{$r_{500}-r_c$ relation (a) and histogram of $r_c$ (b)
          for 69 clusters. In the panel (a), 35 clusters with small
          core of $r_c<100$~kpc and 34 clusters with larger core of
          $r_c>100$~kpc are shown with the green asterisks and the magenta
          circles, respectively. The dotted lines indicate the
          self-similar condition corresponding to four different
          constant values of $r_{500}/r_c$.  \label{fig2}}
\end{figure}

\begin{figure}
\epsscale{0.4}
\plotone{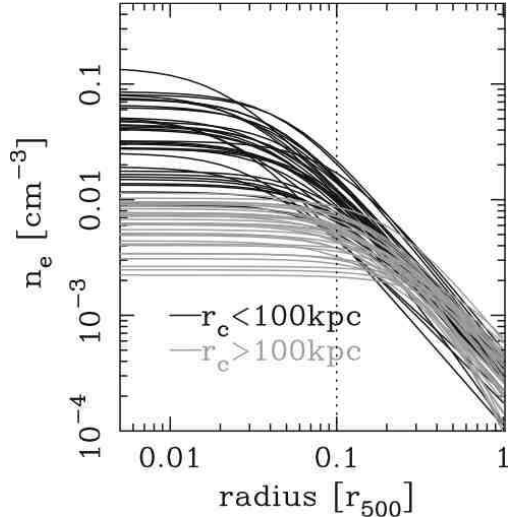}
        \figcaption{Electron density profiles for 69 clusters. The
          best-fit density profiles derived with the single
          $\beta$-model are plotted, where the radius is normalized
          with $r_{500}$. $0.1r_{500}$ is indicated with the vertical
          dotted line, inside which the scatter is the most
          prominent. \label{fig3}}
\end{figure}

\begin{figure}
\epsscale{0.8}
        \plottwo{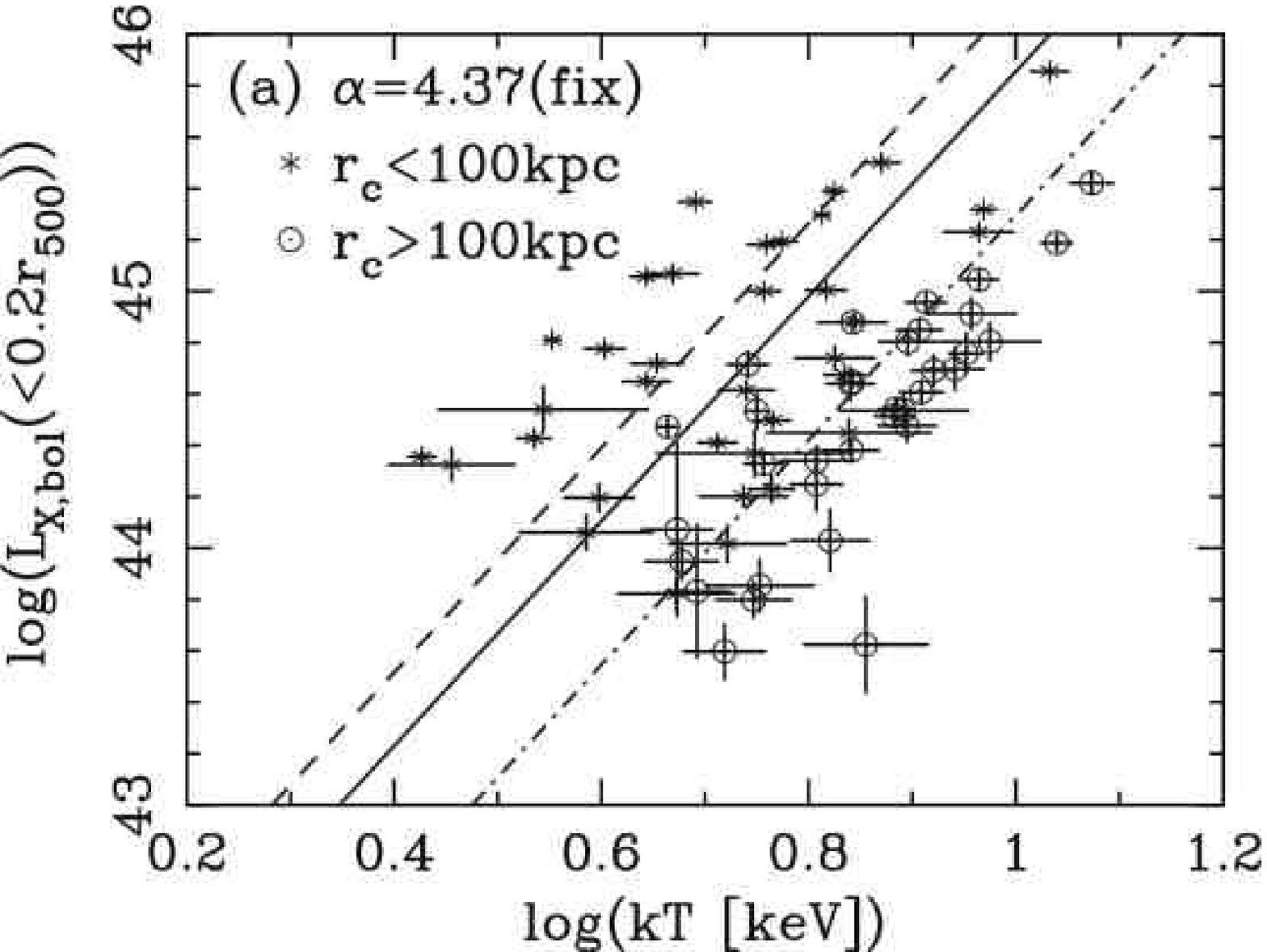}{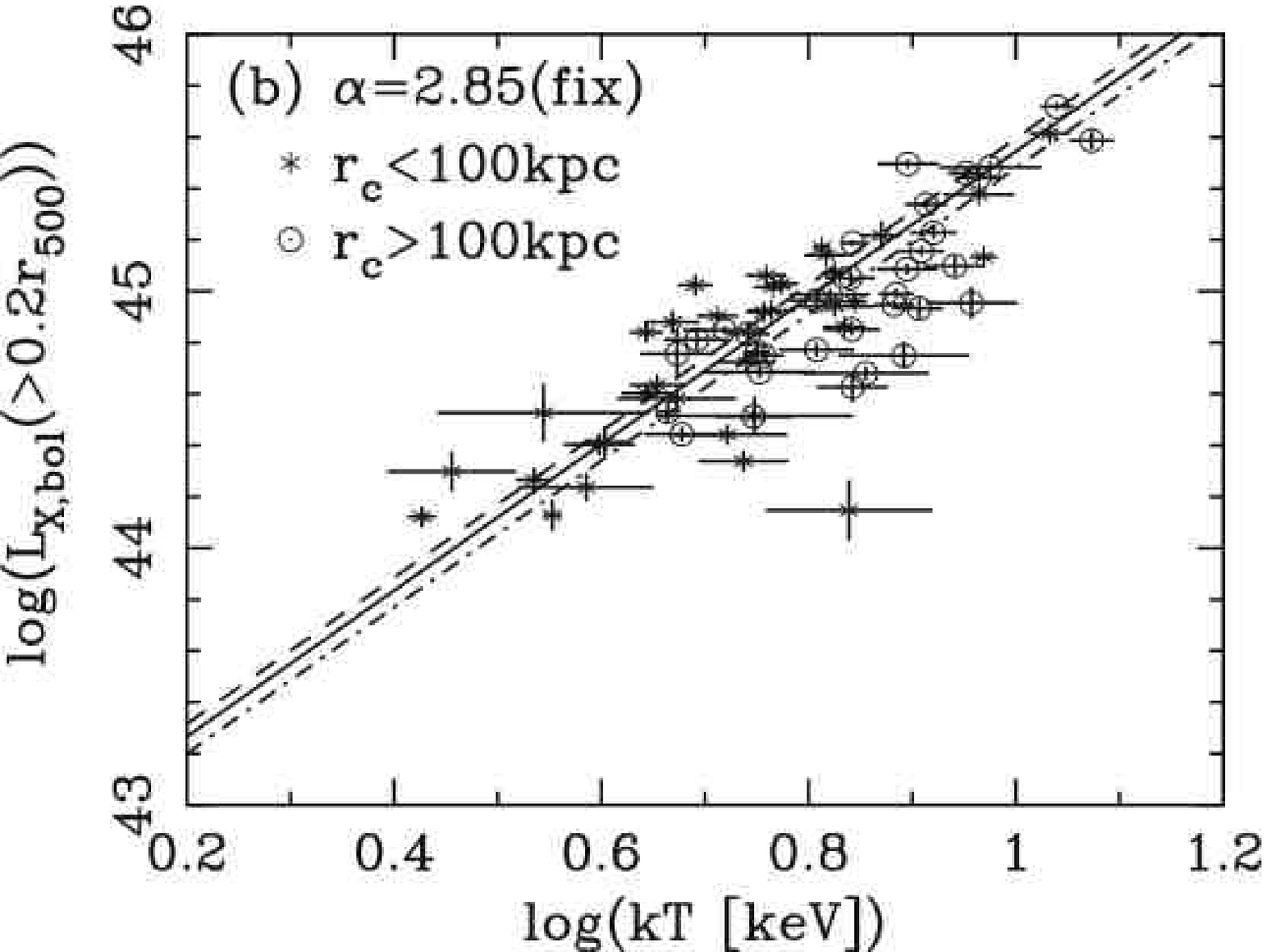}
        \figcaption{$L_{\rm X}(<0.2r_{500})-T$ relation (a) and $L_{\rm
            X}(>0.2r_{500})-T$ relation (b). 
          The sample is divided into two $r_c$ groups and thus the
            meaning of the symbols are the same as
            Fig.~\ref{fig2}. The solid line represent the best-fit
            power-law for 69 clusters in each panel. The best-fit
            models obtained for $r_c<100$~kpc and $>100$~kpc are
            also shown with the dashed and dot-dash lines,
            respectively. 
            \label{fig4}}
\end{figure}

\begin{figure}
\epsscale{0.8}
\centering
        \plottwo{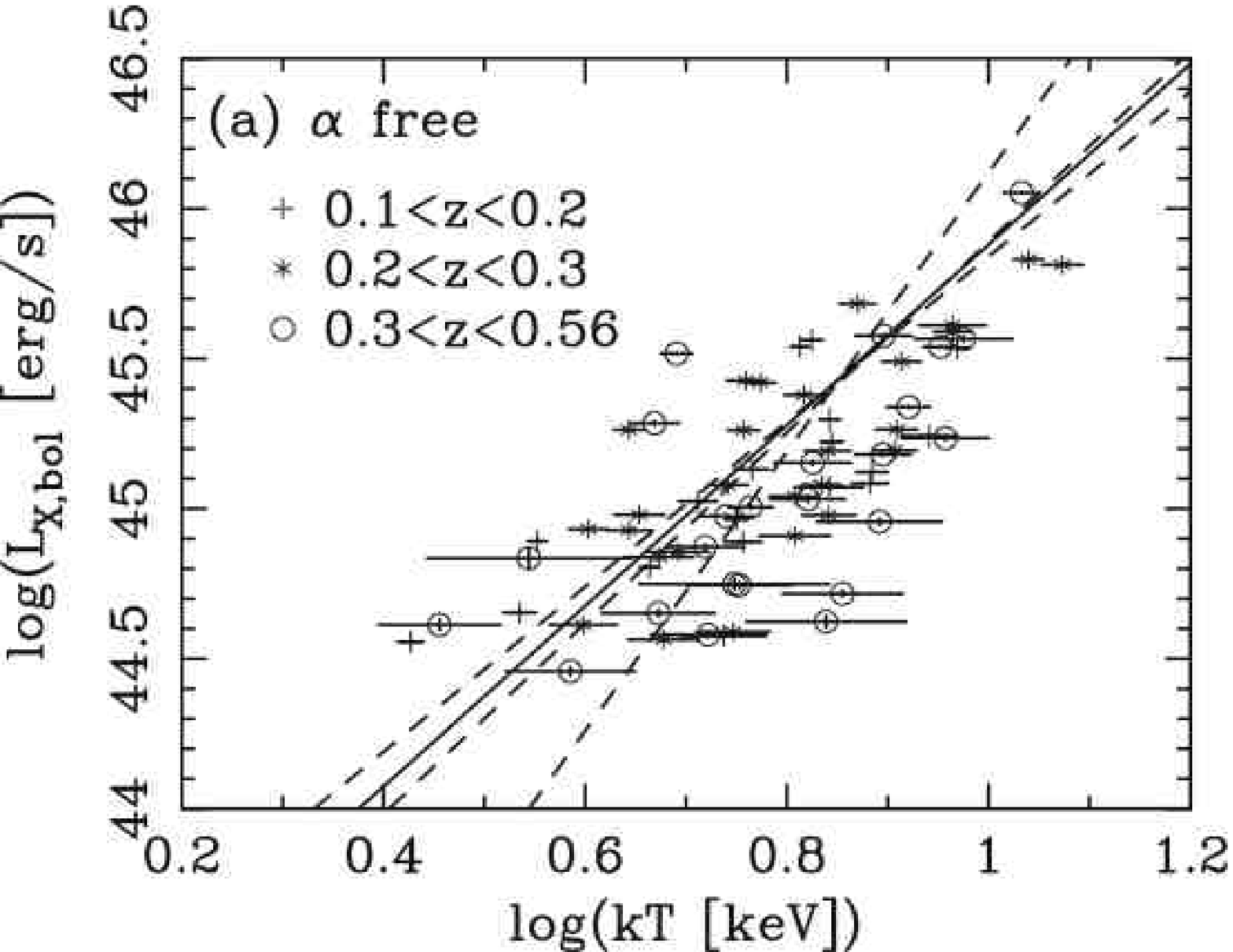}{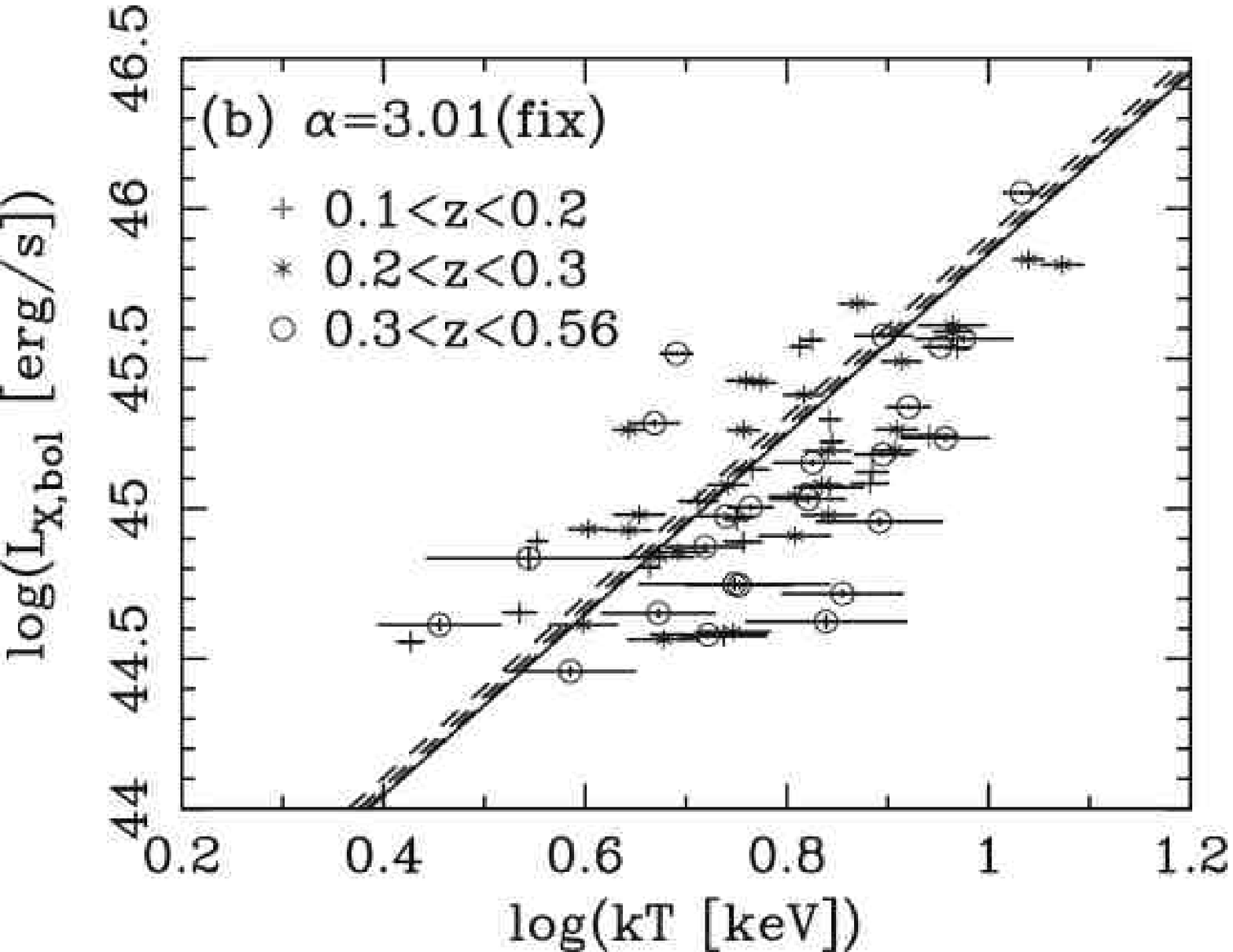}
        \plottwo{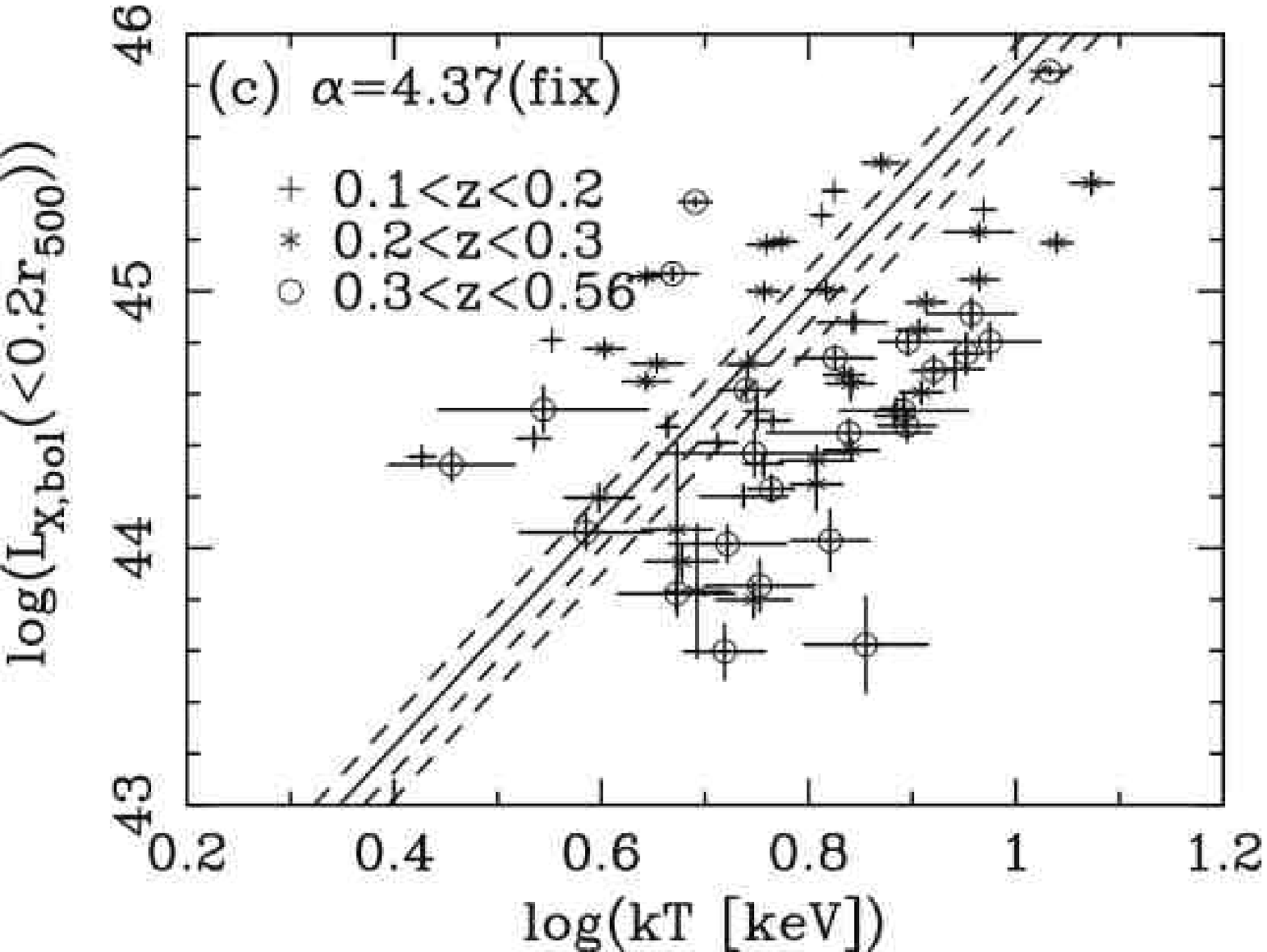}{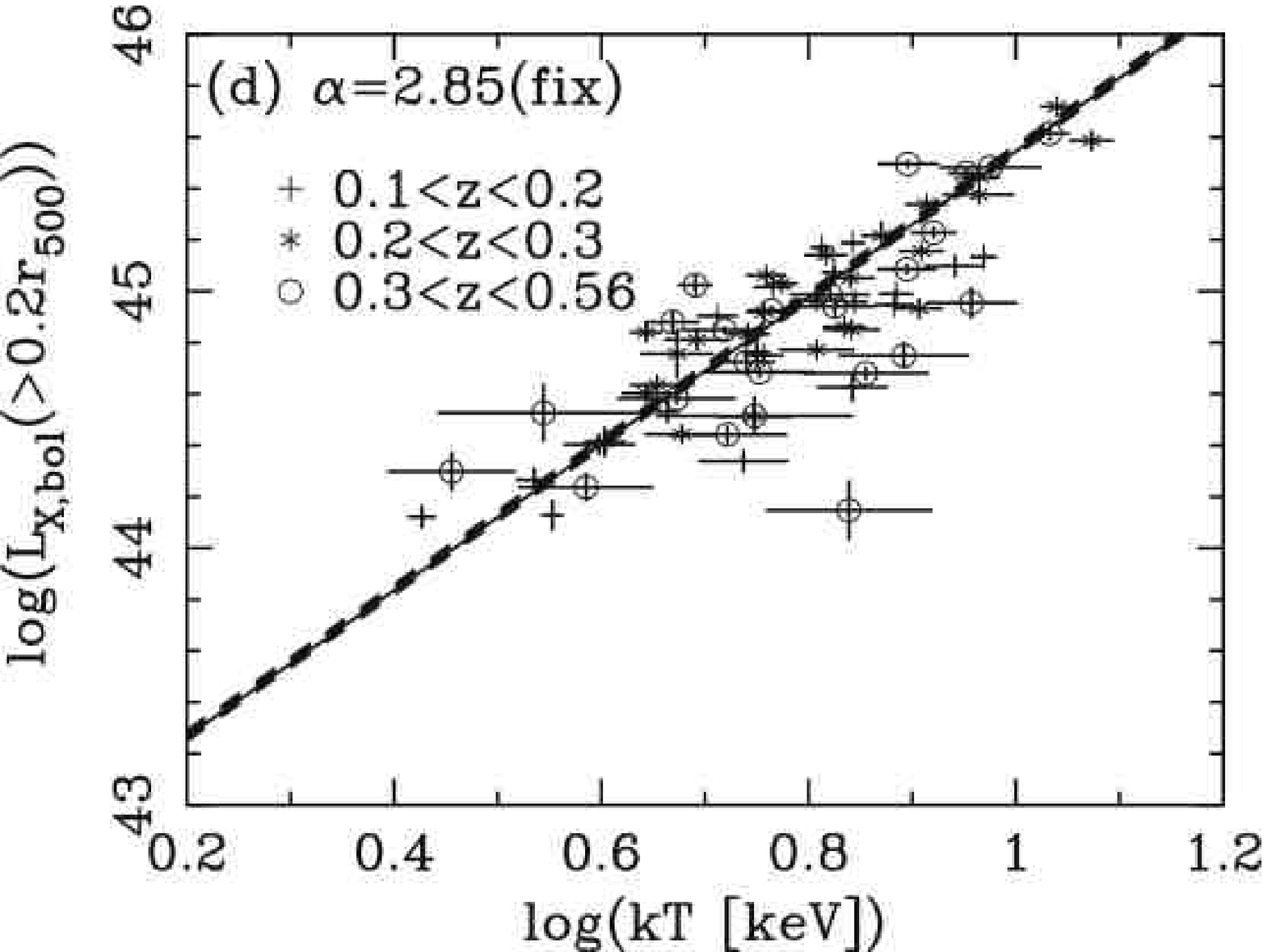}
        \plottwo{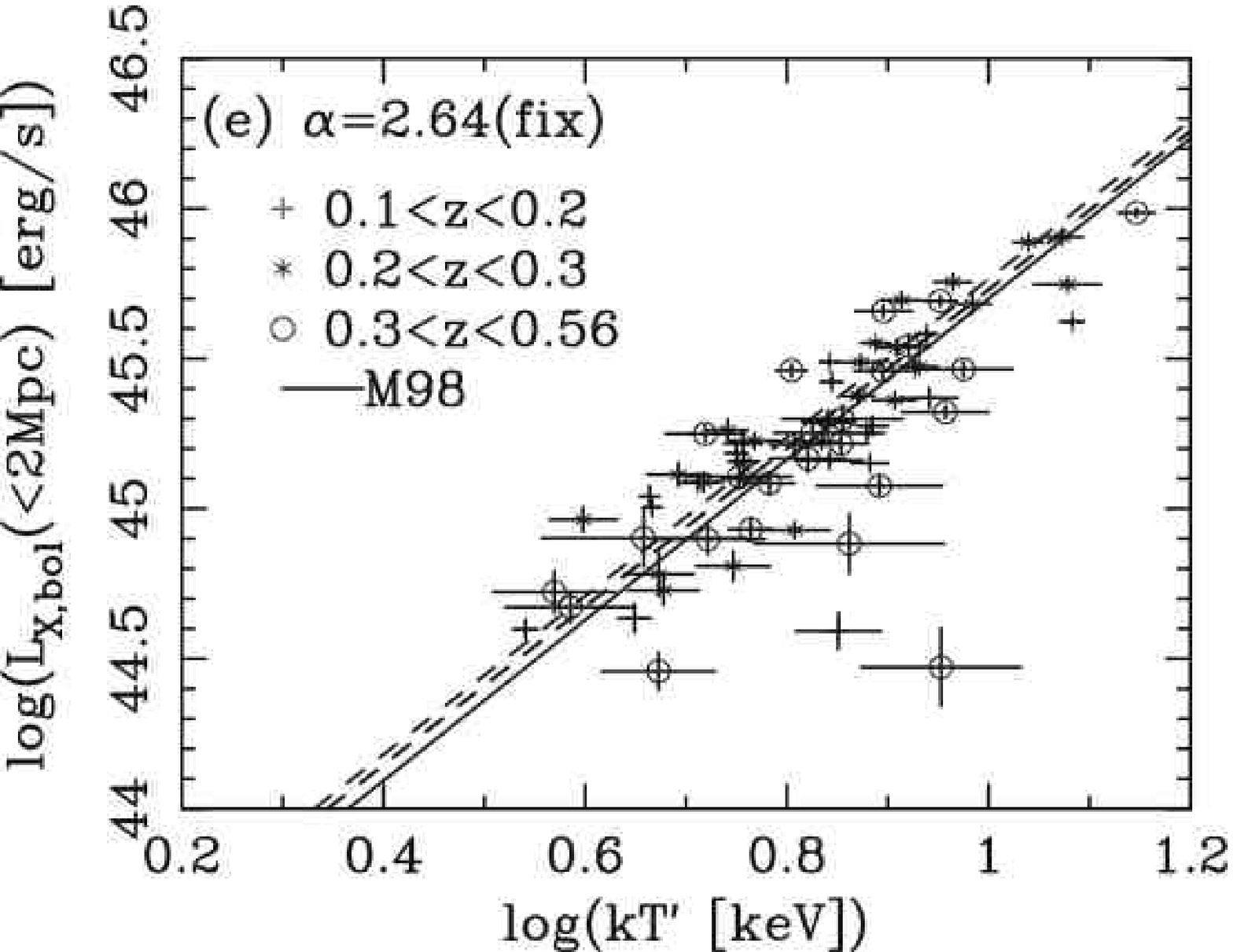}{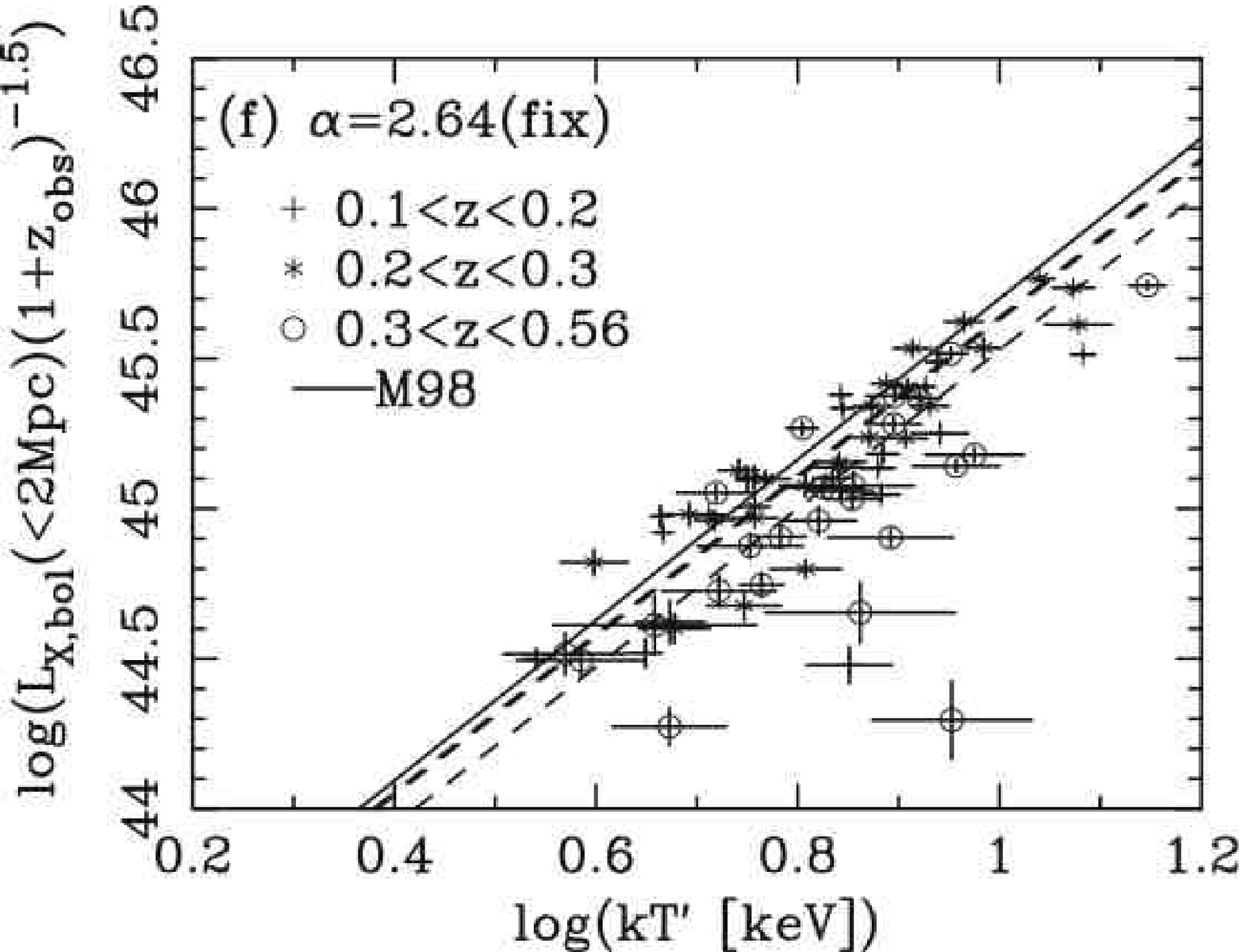}
        \figcaption{\small $L_{\rm X}-T$ relations for three subsets
          of the data: 18 clusters with $0.1<z<0.2$ (blue crosses), 27
          clusters with $0.2<z<0.3$ (green asterisks), and 24 clusters
          with $0.3<z<0.56$ (red circles). In the panels (a) and (b),
          the luminosity was calculated within the overdensity radius
          $r_{500}$.  The solid line shows the best-fit power-law
          model for 69 distant clusters.  In the panels (c) and (d),
          the $L_{\rm X}(<0.2r_{500})-T$ and $L_{\rm
            X}(>0.2r_{500})-T$ relations are shown, respectively.  In
          the panels (e) and (f), the luminosity was calculated for a
          fixed radius of 2~Mpc. For 26 clusters with a short cooling
          time, we corrected the effect of possible cool emission
          regarding both the luminosity and temperature (see text for
          details). The solid line indicates the local relation
          obtained by \cite{Markevitch_1998}.  In the panel (f), the
          luminosity was further scaled with redshift as $L_{\rm
            X}(1+z)^{-1.5}$.  In each panel, the blue, green, and red
          dashed lines show the best-fit power-laws for the above
          three redshift ranges, respectively. In the panel (a), the
          slope of the relation was included as a free parameter in
          the fits, while in the panels (b)--(f), the slope was fixed
          at $\alpha=3.01$, $4.37$, $2.85$, $2.64$, and $2.64$,
          respectively.
            \label{fig5}}
\end{figure}

\begin{figure}
\epsscale{0.8}
\centering
       \plottwo{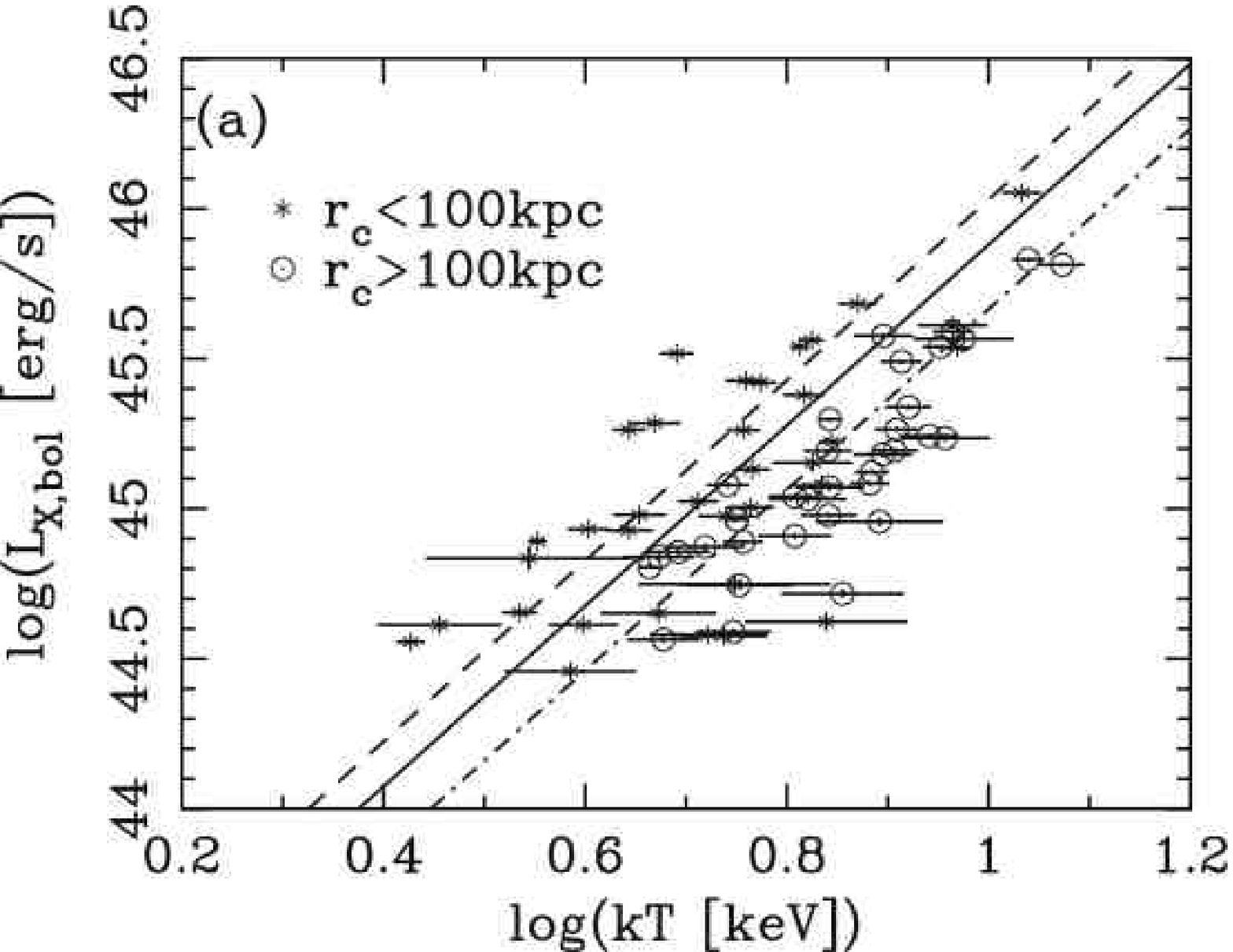}{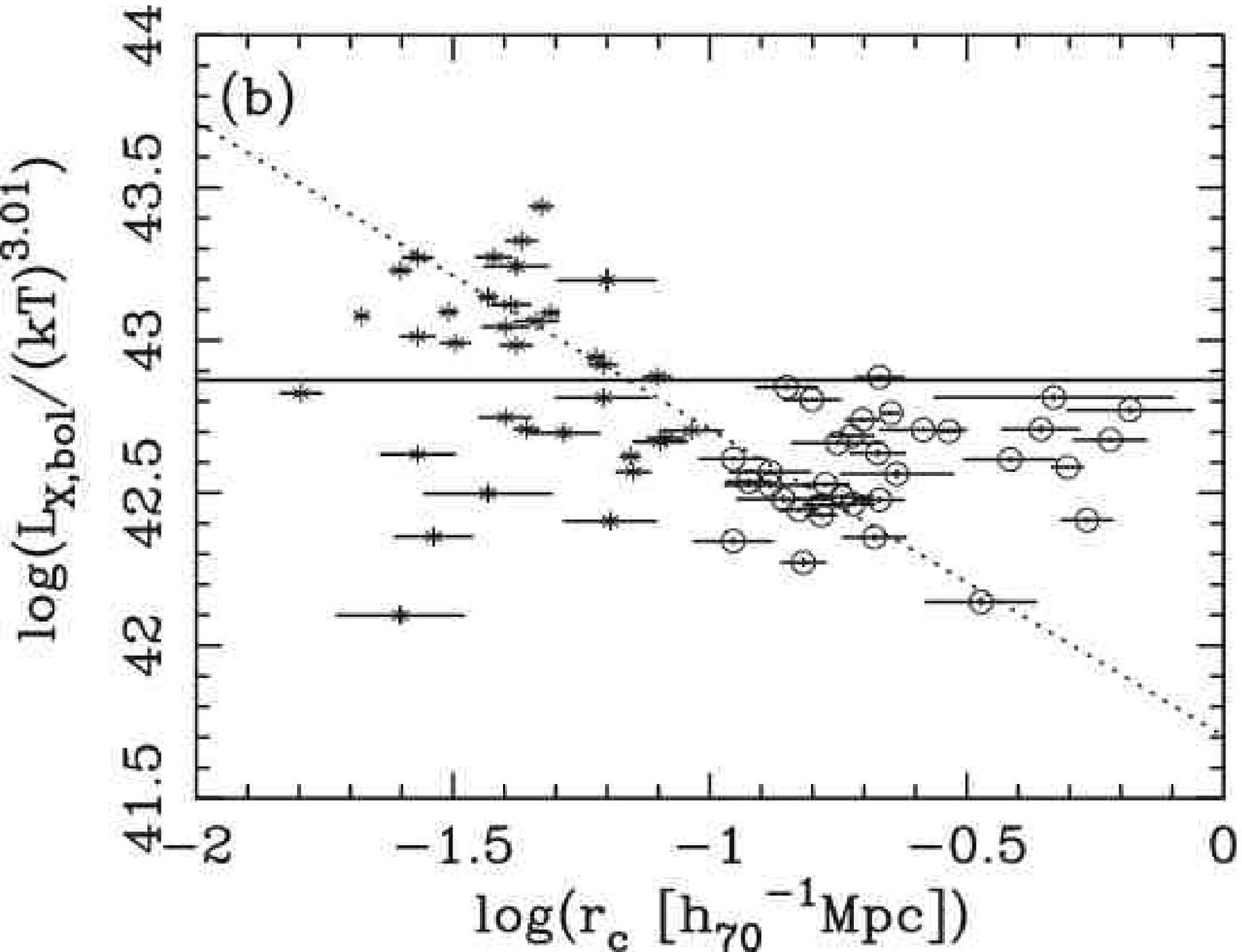}
        \figcaption{$L_{\rm X}-T$ relation (a) and $L_{\rm 1keV}-r_c$
          relation (b).  In the panel (a), we find the smaller core
          clusters are brighter on average compared to the large core
          ones of the same temperature. In the panel (b), $L_{\rm
          1keV}(=L_{\rm X}/(kT)^{3.01})$ is the normalization factor
          of the $L_{\rm X}-T$ relation derived for each cluster. The
          solid line indicates the best-fit normalization factor of
          the $L_{\rm X}-T$ relation for 69 clusters in the panel
          (b). The negative correlation between $L_{\rm 1keV}$ and
          $r_c$ corresponds to the luminosity difference between the
          two subgroups, which is roughly approximated as $L_{\rm
          1keV}\propto r_c^{-1}$ (the dotted line).
\label{fig6}}
\end{figure}

In order to evaluate the $L_{\rm X}-T$ relation quantitatively, we
assume the $\beta$ profile for the gas density;
\begin{equation}
\rho_{\rm gas} =  \rho_{\rm gas,0} \left[ 1 +
\left( \frac{r}{r_c} \right)^2 \right]^{-3\beta/2}.
\end{equation}
Then the total gas mass, $M_{\rm gas}$, within $r_{\rm vir}$ is
evaluated by
\begin{equation}
M_{\rm gas} =  4\pi r_c^3 \rho_{\rm gas,0} 
       \int_0^{x_{\rm v}} x^2(1+x^2)^{-3\beta/2} dx,
\label{eq:Mg}
\end{equation}
where $x_{\rm v}=r_{\rm vir}/r_c$.  For the observed range of $x_{\rm
  v}$, i.e. $4< x_{\rm v} <64$ (see Fig.~\ref{fig1}), the integral
in Eq. \ref{eq:Mg} can be well approximated by a single power law
function, thus is $\propto x_{\rm v}^a$ for a fixed value of $\beta$
\citep{Evrard_Henry_1991}.  Under the assumptions that the cluster
mean density is $\Delta_c$ times the critical density of the universe,
$\bar{\rho} = 3M_{\rm vir}/4\pi r_{\rm vir}^3=\Delta_c \rho_{\rm
  crit}$, and the gas-mass fraction, $f_{\rm gas}=M_{\rm gas}/M_{\rm
  vir}$ is constant among clusters, Eq.~\ref{eq:Mg} yields
\begin{equation}
M_{\rm gas} \simeq \frac{4\pi}{3} r_{\rm vir}^3 \Delta_c\rho_{\rm
  crit} f_{\rm gas}.
\label{eq:Mg_rhoc}
\end{equation}
Eqs. \ref{eq:Mg} and \ref{eq:Mg_rhoc} relate the central gas
density, $\rho_{\rm gas,0}$, with $r_c$ and $r_{\rm vir}$ as follows.
\begin{equation}
\rho_{\rm gas,0} = \frac{1}{3}x_{\rm v}^{3-a}\Delta_c \rho_{\rm
  crit}f_{\rm gas}.
\label{eq:rho_gas0}
\end{equation}
The X-ray luminosity, $L_{\rm X}$, is expressed as
\begin{eqnarray}
L_{\rm X}  &=& \Lambda(T_{\rm gas}) 4\pi r_c^3 
\left(\frac{\rho_{\rm gas,0}}{\bar{m}}\right)^2
\int^{x_{\rm v}}_0 x^2(1+x^2)^{-3\beta}dx, \\ 
&\simeq & \Lambda(T_{\rm gas}) 4\pi r_c^3
\left(\frac{\rho_{\rm gas,0}}{\bar{m}}\right)^2 x_{\rm v}^b,\\
&\simeq & 1.435\times10^{-27}\frac{4\pi}{9\bar{m}^2}\left(\frac{9\beta  k}{4\pi\bar{m}G}\right)^{3-a+b/2} \\
& & \cdot f_{\rm gas}^2  (\Delta_c\rho_{\rm crit})^{-1+a-b/2}
T_{\rm gas}^{7/2-a+b/2} r_c^{-3+2a-b}, \label{eq:Lx}
\end{eqnarray}
where $\Lambda$ is the emissivity of the free-free emission and
$\bar{m}=\mu m_p$ ($\mu$ : the mean plasma weight, 0.63).  In the case
of $\beta=0.7$, $a=0.96$, and $b=0.025$ for $x_{\rm v}=32$ ($x_{\rm v}$
dependence is small), which leads
\begin{equation}
L_{\rm X} \propto  f_{\rm gas}^2 (\Delta_c\rho_{\rm crit})^{-0.05}
T_{\rm gas}^{2.6}r_c^{-1.1}. \label{eq:LT_model}
\end{equation}
We define the virial temperature as 
\begin{equation}
kT_{\rm vir} \equiv \frac{\bar{m}GM_{\rm vir}}{3 r_{\rm vir}} 
= \frac{4\pi\bar{m}G\Delta_c \rho_{\rm crit}r_{\rm vir}^2}{9}, 
\end{equation}
and assume that it is related to the gas temperature through $kT_{\rm
  vir}\equiv \beta' kT_{\rm gas}$.

\begin{figure}
\centering
        \epsscale{0.4}
        \plotone{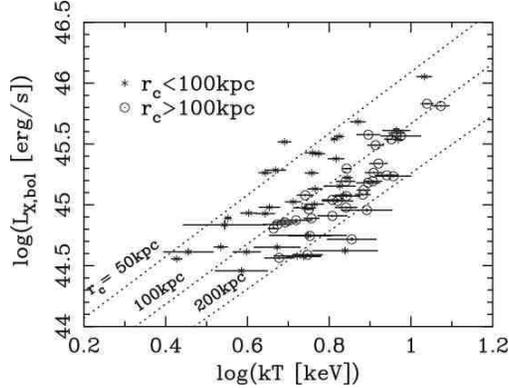}
        \figcaption{$L_{\rm X}-T$ relations expected for clusters with a
          constant core radius, $r_c=50$ or 100 or
          200~kpc (dotted lines). $\beta=0.7$ is assumed in the model
          calculation. See Appendix~\ref{appendix:model} for details
          of the model. \label{fig7}}
\end{figure}

\begin{figure}
\centering
        \epsscale{0.8}
        \plottwo{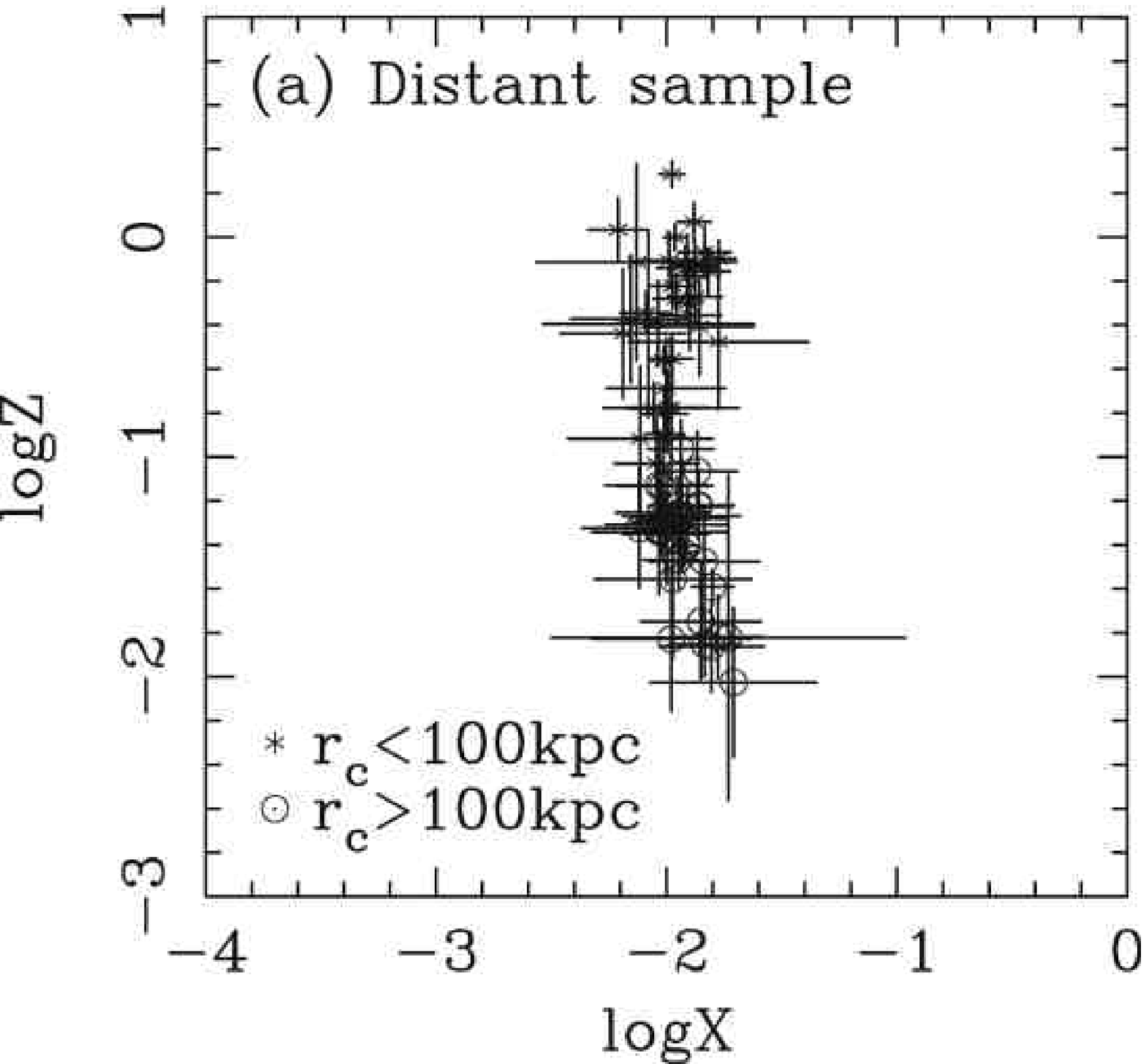}{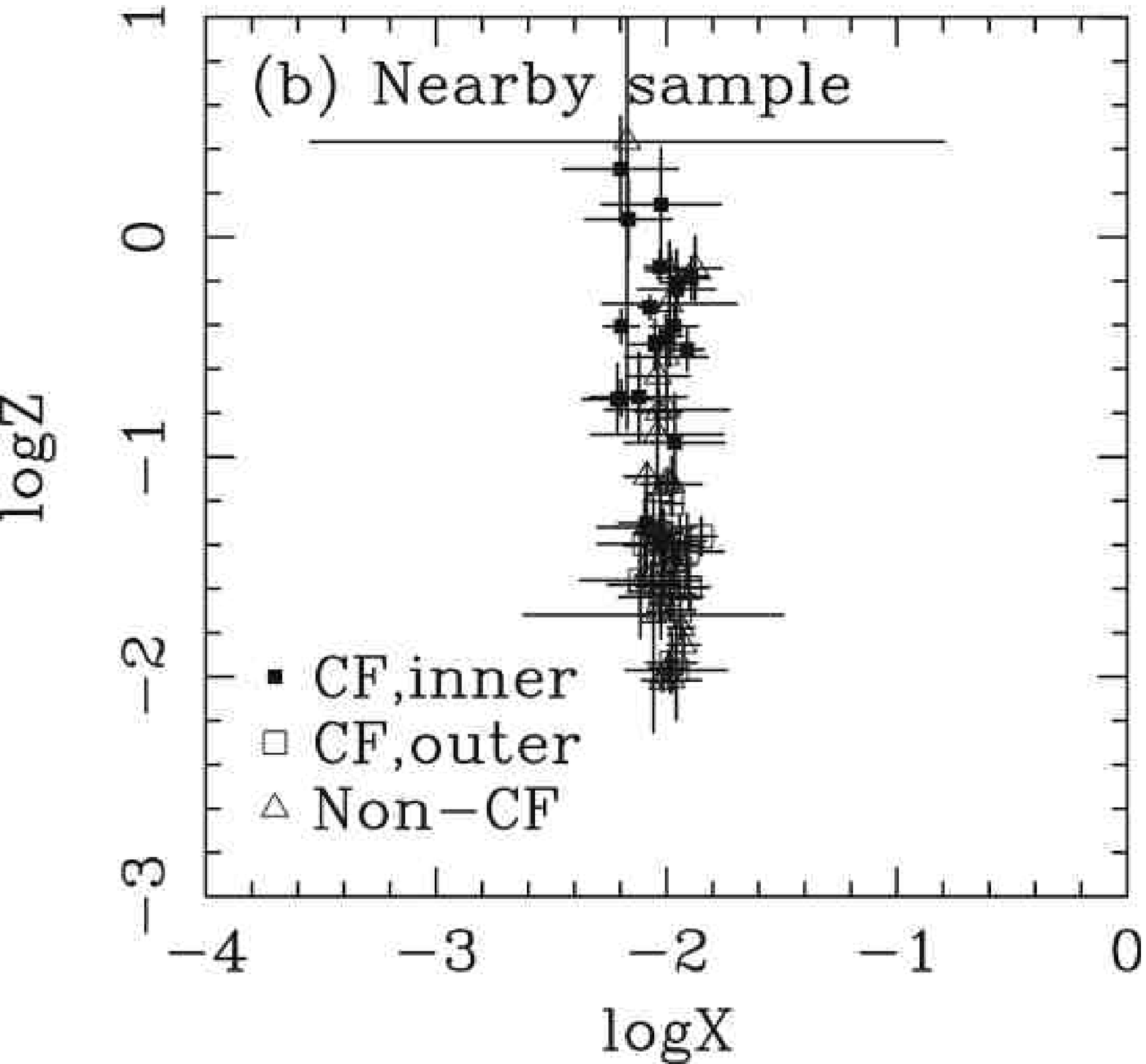}
        \figcaption{Results of X-ray fundamental plane analysis. In
          panel (a), the best-fit $\log{X}-\log{Z}$ plane obtained for
          the distant clusters is shown. The meaning of the symbols
          are the same as Fig.~\ref{fig2}. In the panel (b), 45 nearby
          clusters taken from \cite{Mohr_etal_1999} were projected
          onto the same fundamental plane as panel (a). According to
          Table 2 of \cite{Mohr_etal_1999} non cooling-flow clusters
          are shown with the red triangles, and inner-core and
          outer-core components of cooling-flow clusters are
          separately shown with the solid blue boxes and open black
          boxes. \label{fig8}}
\end{figure}

\begin{figure}
\centering
        \plottwo{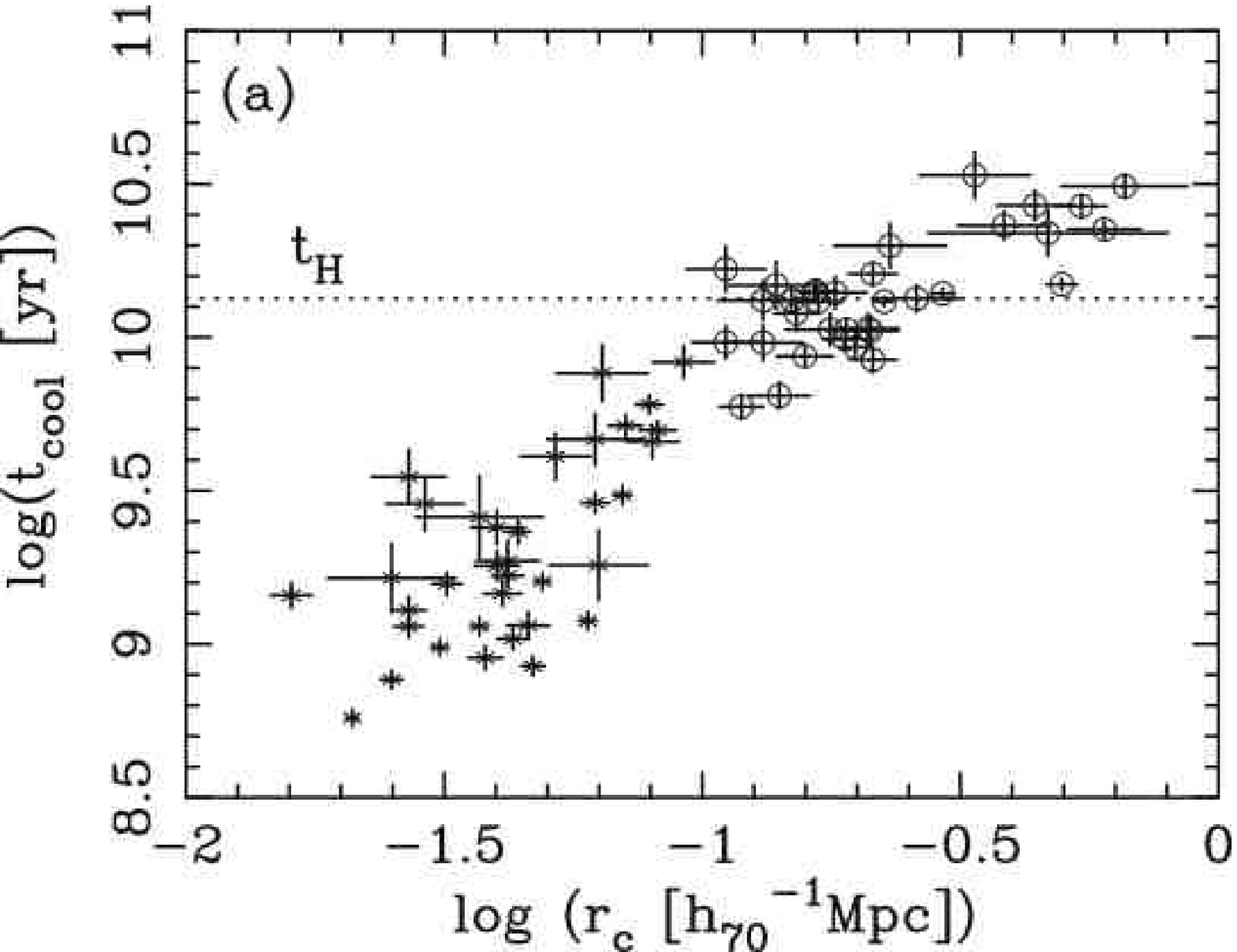}{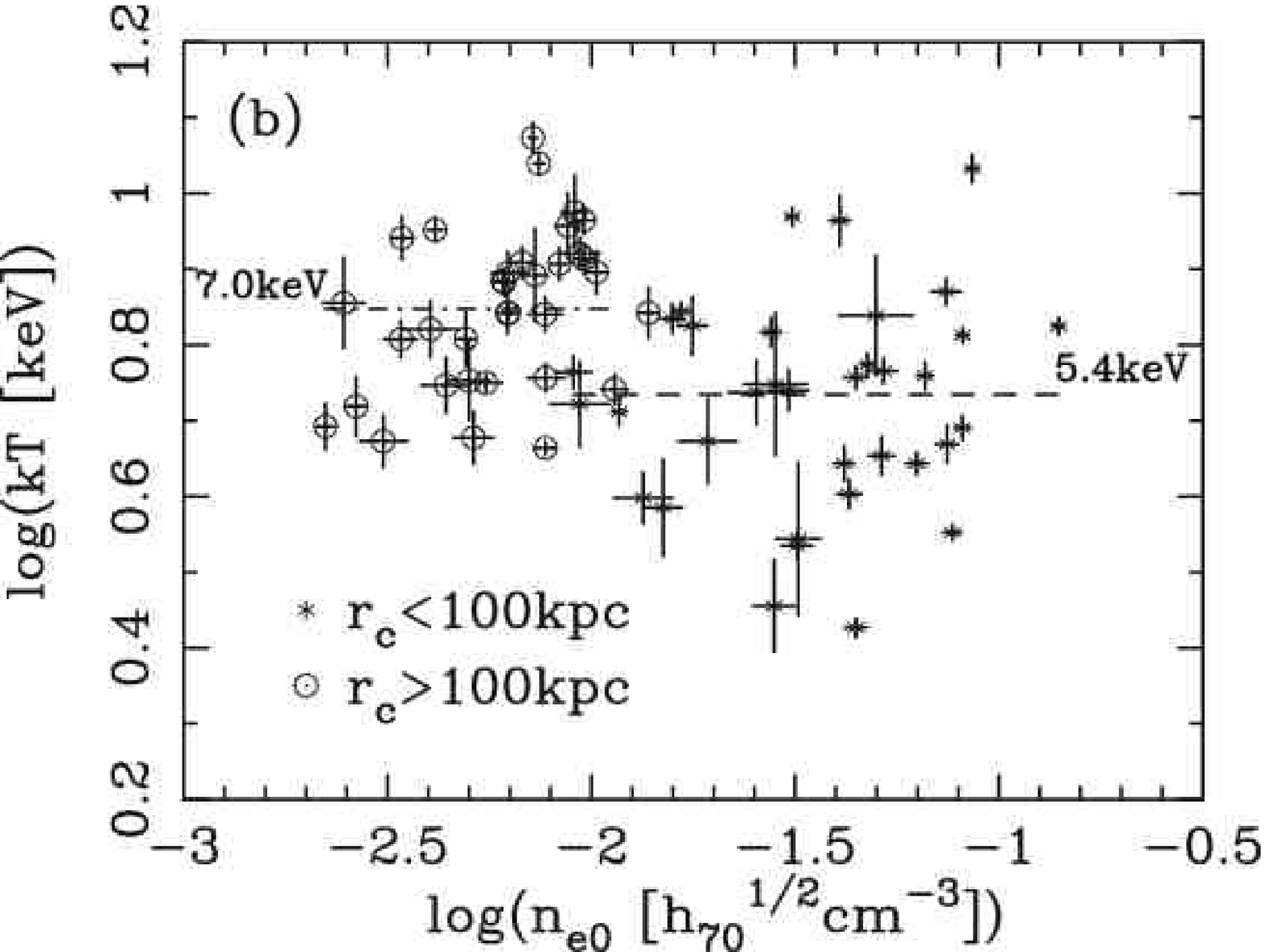}
        \figcaption{$t_{\rm cool}-r_c$ (a) and $T-n_{e0}$ (b) relations.
          For panel (a), $t_{\rm cool}$ is the radiative cooling timescale (see text).
          There is a strong
          correlation with $r_c$ which follows $t_{\rm cool}\propto
          r_c^{1.7}$. The Hubble time, $t_{\rm H}=13.4$~Gyr
          \citep{Spergel_etal_2003} is indicated with the horizontal
          dotted line.  Then $t_{\rm cool} < t_{\rm H}$ for all
          clusters belonging to the smaller core group.  For panel
          (b), $n_{e0}$ is the central electron density obtained from
          the $\beta$ model analysis. There is no clear
          difference in the temperature range between the two
          subgroups though $n_{e0}$ is scattered nearly over two
          orders of magnitudes. Their average temperatures of 5.4 keV
          and 7.0 keV are shown with the dashed line and dot-dash
          line, respectively.
\label{fig9}}
\end{figure}

\begin{figure}
\centering
       \plottwo{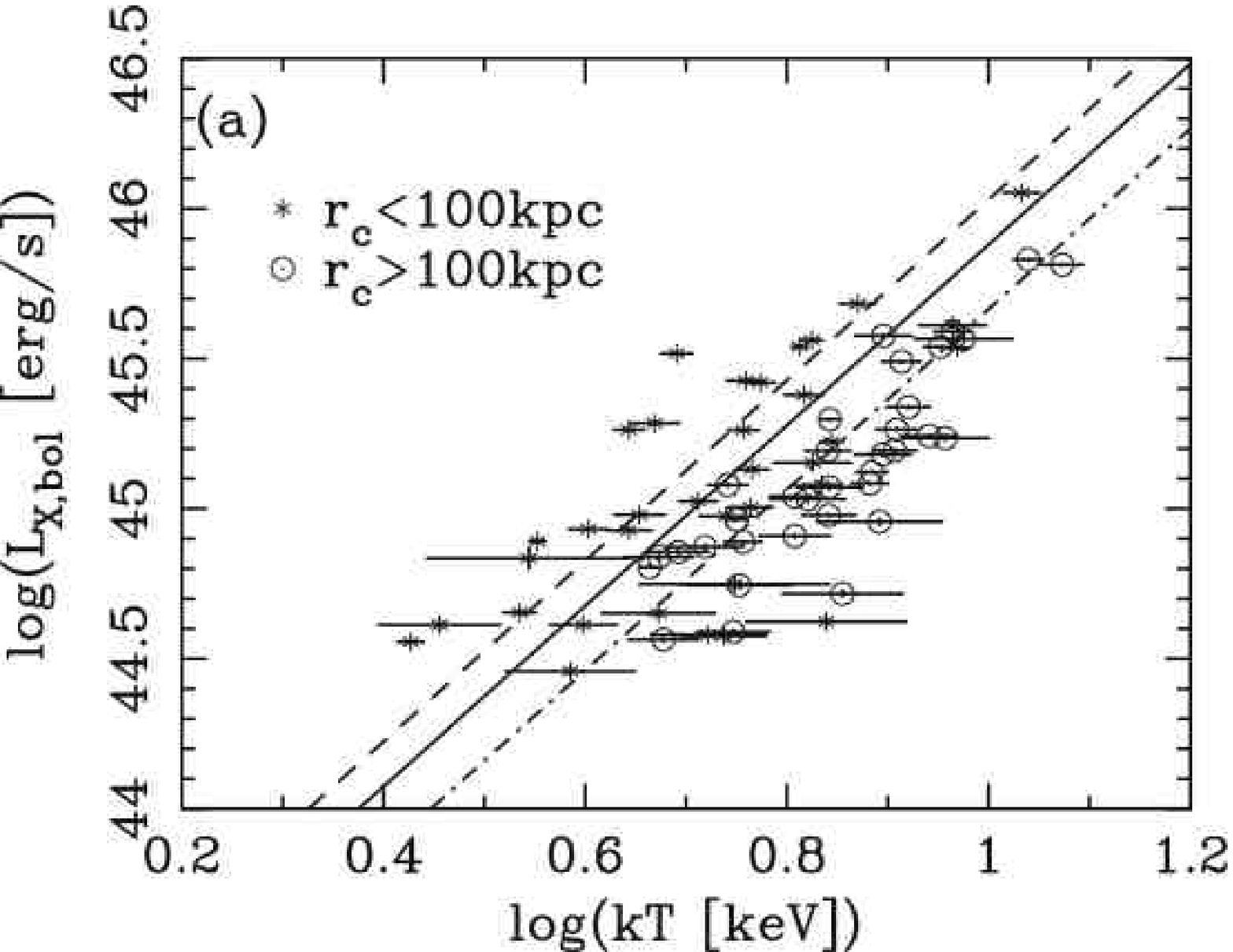}{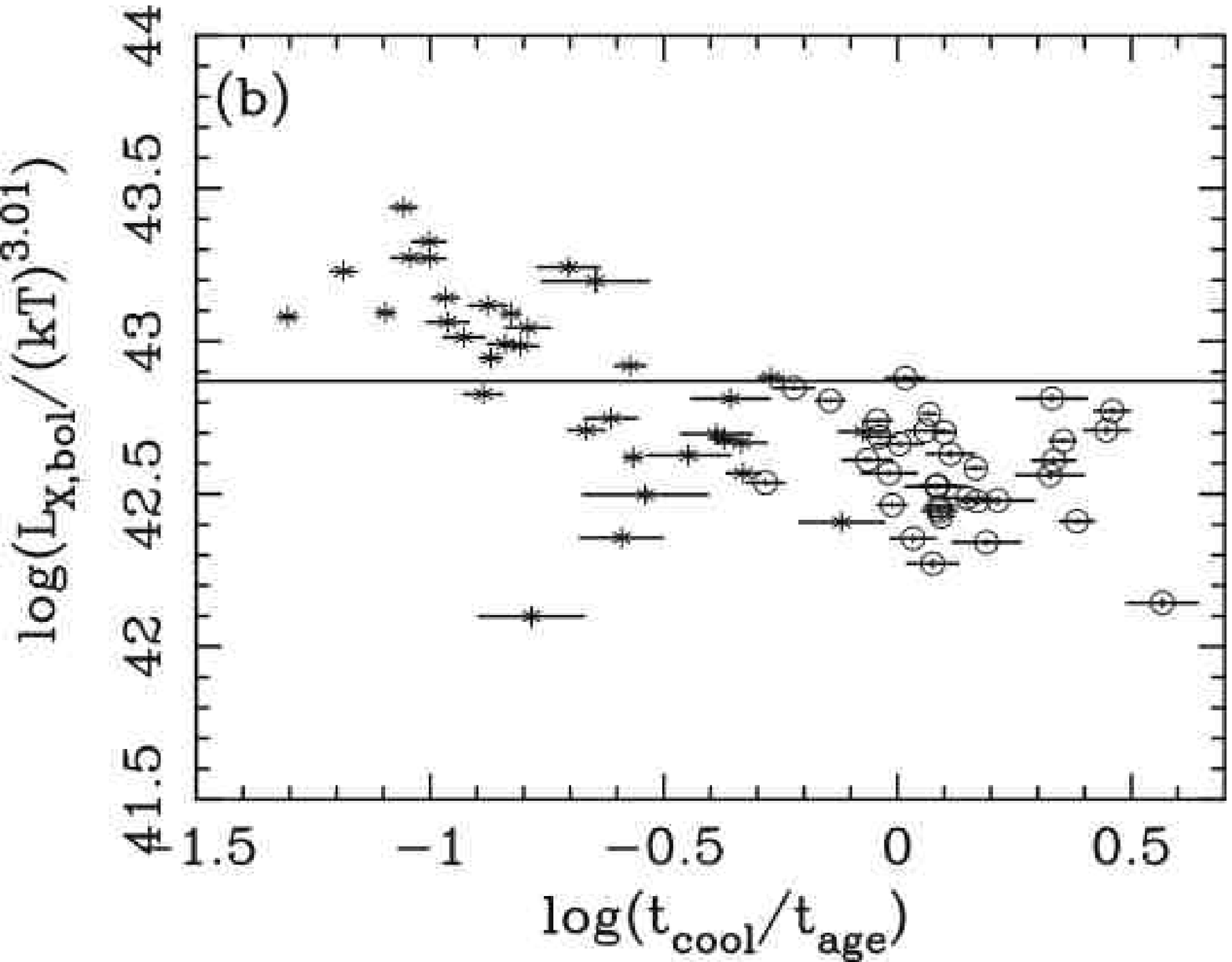}
        \plottwo{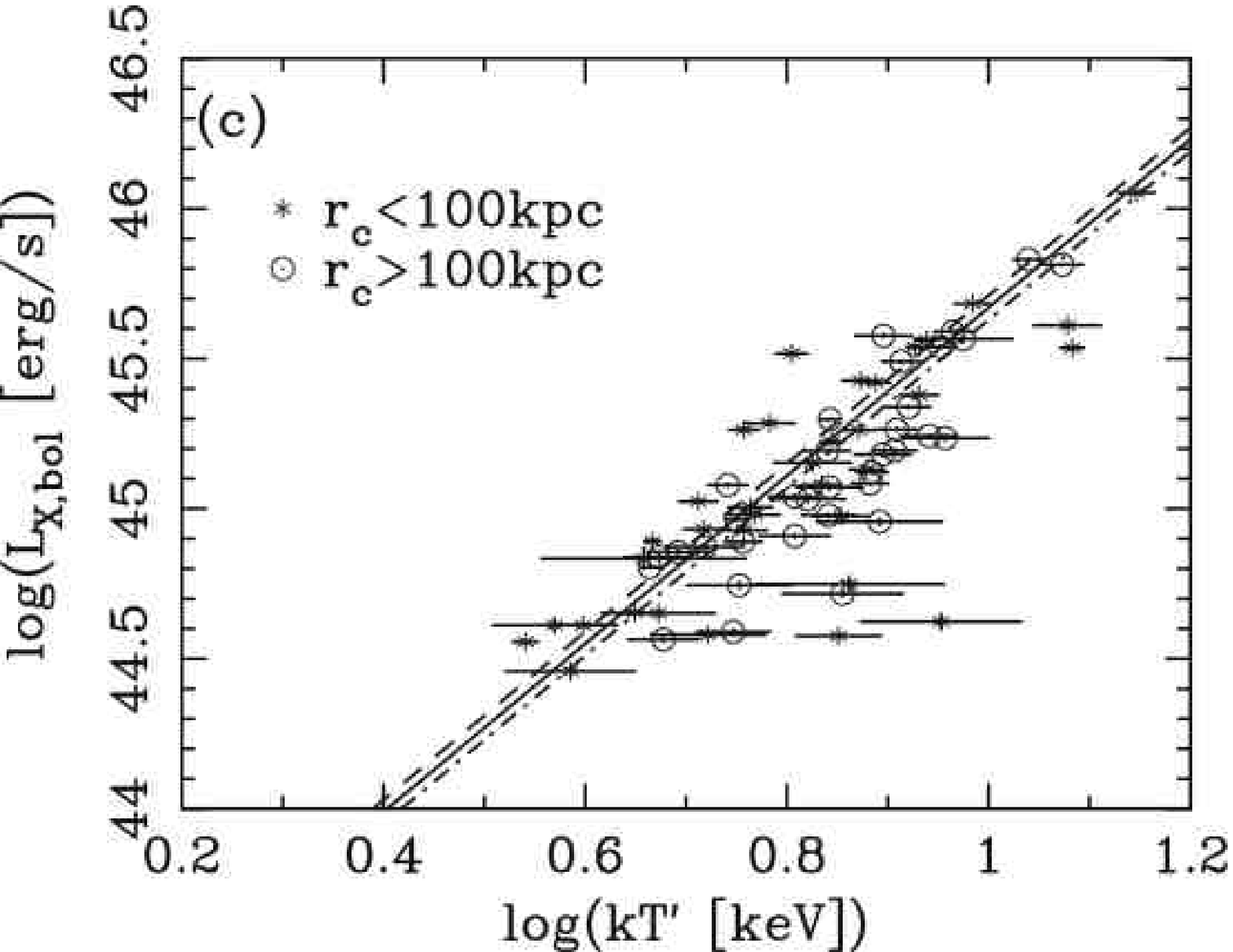}{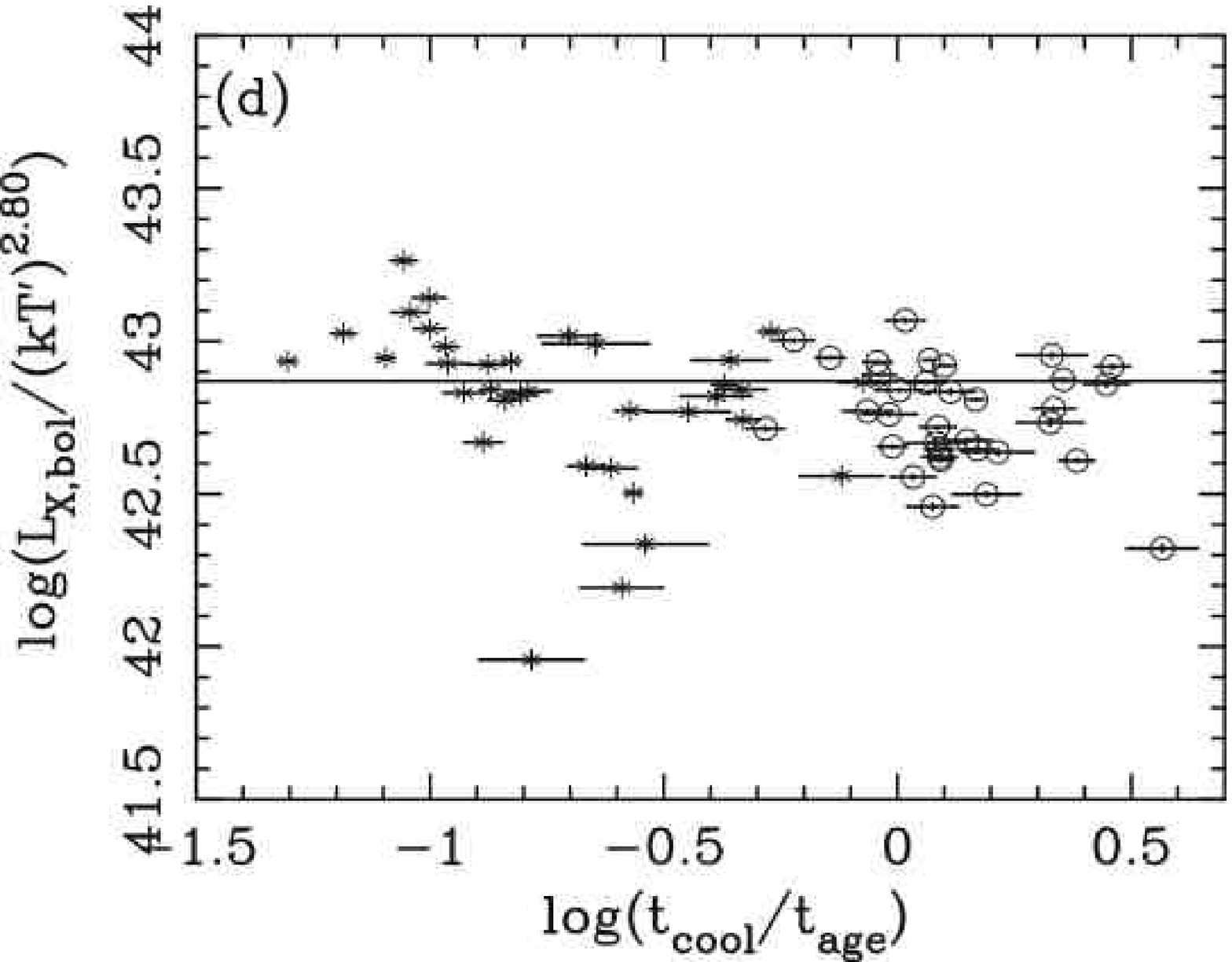}
        \figcaption{$L_{\rm X}-T$ relation and $t_{\rm cool}$.  Upper
          panels, $L_{\rm X}-T$ relation of clusters (a) (the same as
          Fig.~\ref{fig6}a), and $L_{\rm 1keV}$ as a function of
          $t_{\rm cool}/t_{\rm age}$ (b).  Bottom panels, the
          luminosity to the `ambient temperature' relation, $L_{\rm
          X}-T'$ (c) and $L_{\rm 1keV} = L_{\rm X}/(kT')^{2.80}$ as a
          function of $t_{\rm cool}/t_{\rm age}$ (d). In the panels (c)
          and (d), for clusters with short cooling timescale of
          $\log{t_{\rm cool}/t_{\rm age}}<-0.5$, the 
          temperature decrease is corrected with $T'=1.3T$, 
          otherwise $T'=T$ (see \S~\ref{subsec:lx-t-tcool}). We find
          a decrease of the dispersion around the mean $L_{\rm X}-T'$
          relation in comparison to the $L_{\rm X}-T$, which we show
         in a more quantitative manner in Fig.~\ref{fig12}.
\label{fig10}}
\end{figure}

\begin{figure}
\centering
        \plottwo{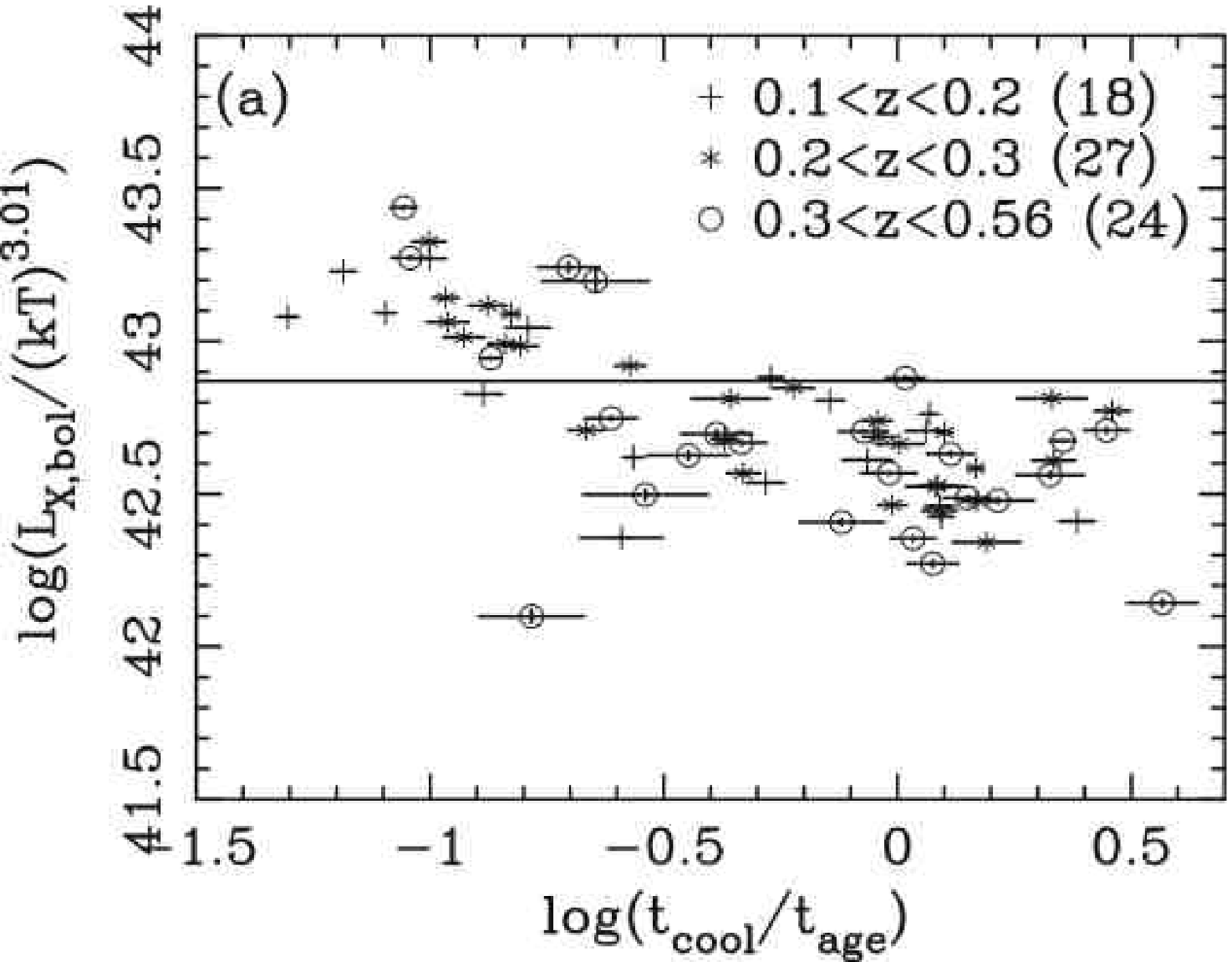}{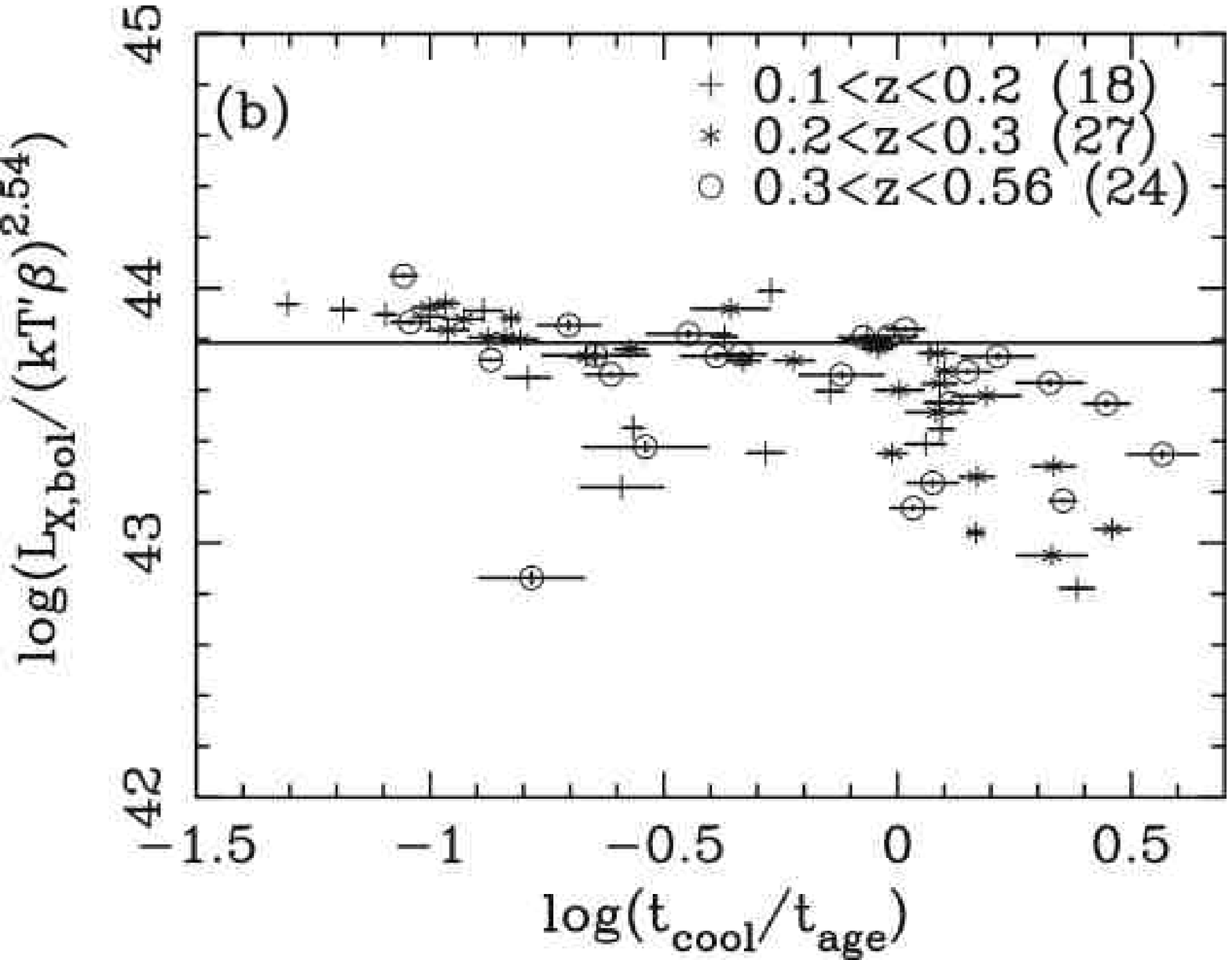}
        \figcaption{$L_{\rm 1keV}$ as a function of $t_{\rm cool}/t_{\rm
        age}$ in three redshift ranges.  In the panels (a) and (b),
        $L_{\rm 1keV}=L_{\rm X}/(kT)^{3.01}$
        (\S~\ref{subsec:lx-t-tcool}) and $L_{\rm
        X}/(kT'\beta)^{2.54}$ (\S~\ref{subsec:lx-tbeta3}) are
        plotted, respectively.
\label{fig11}}
\end{figure}

\begin{figure}
\centering
        \plottwo{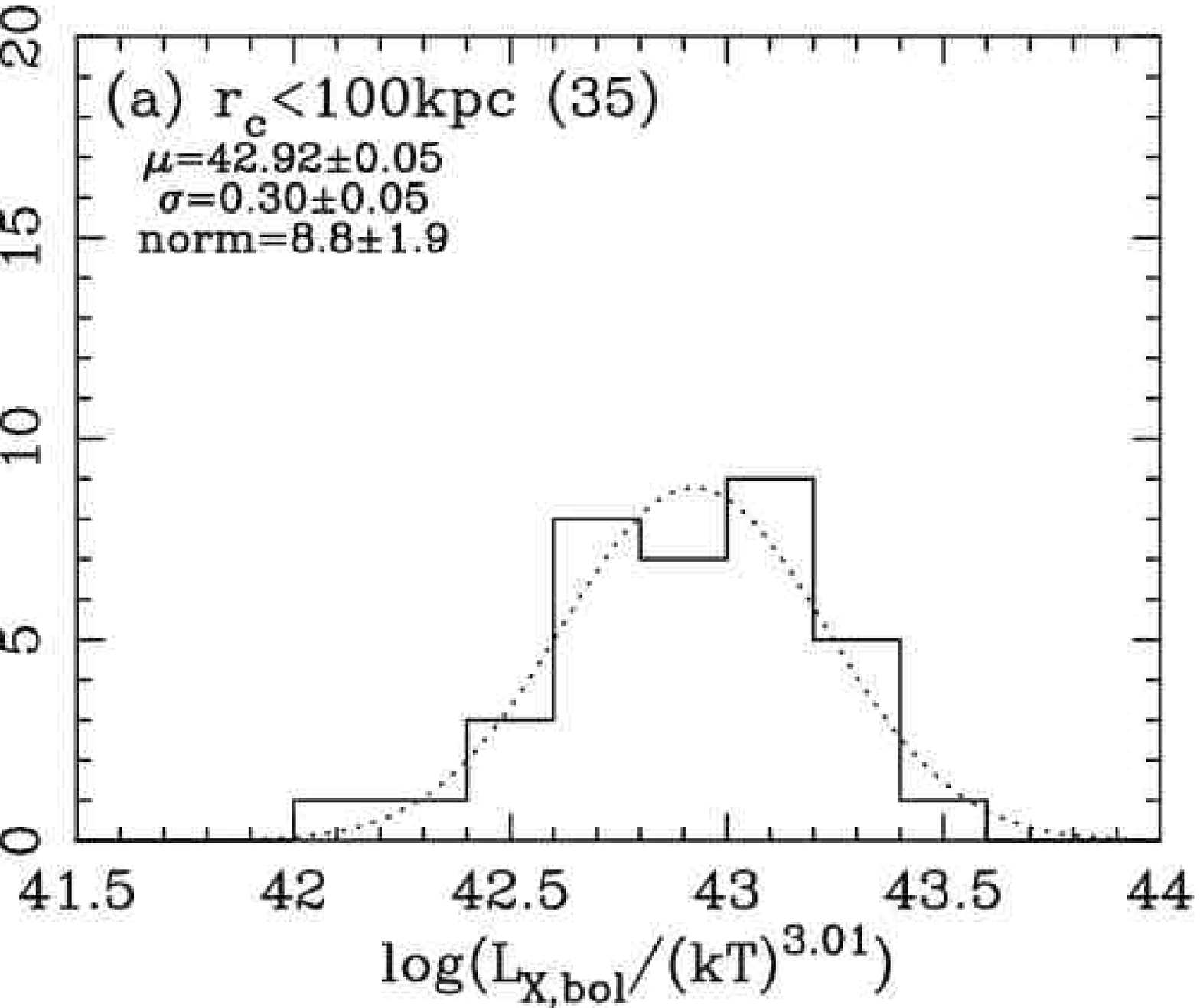}{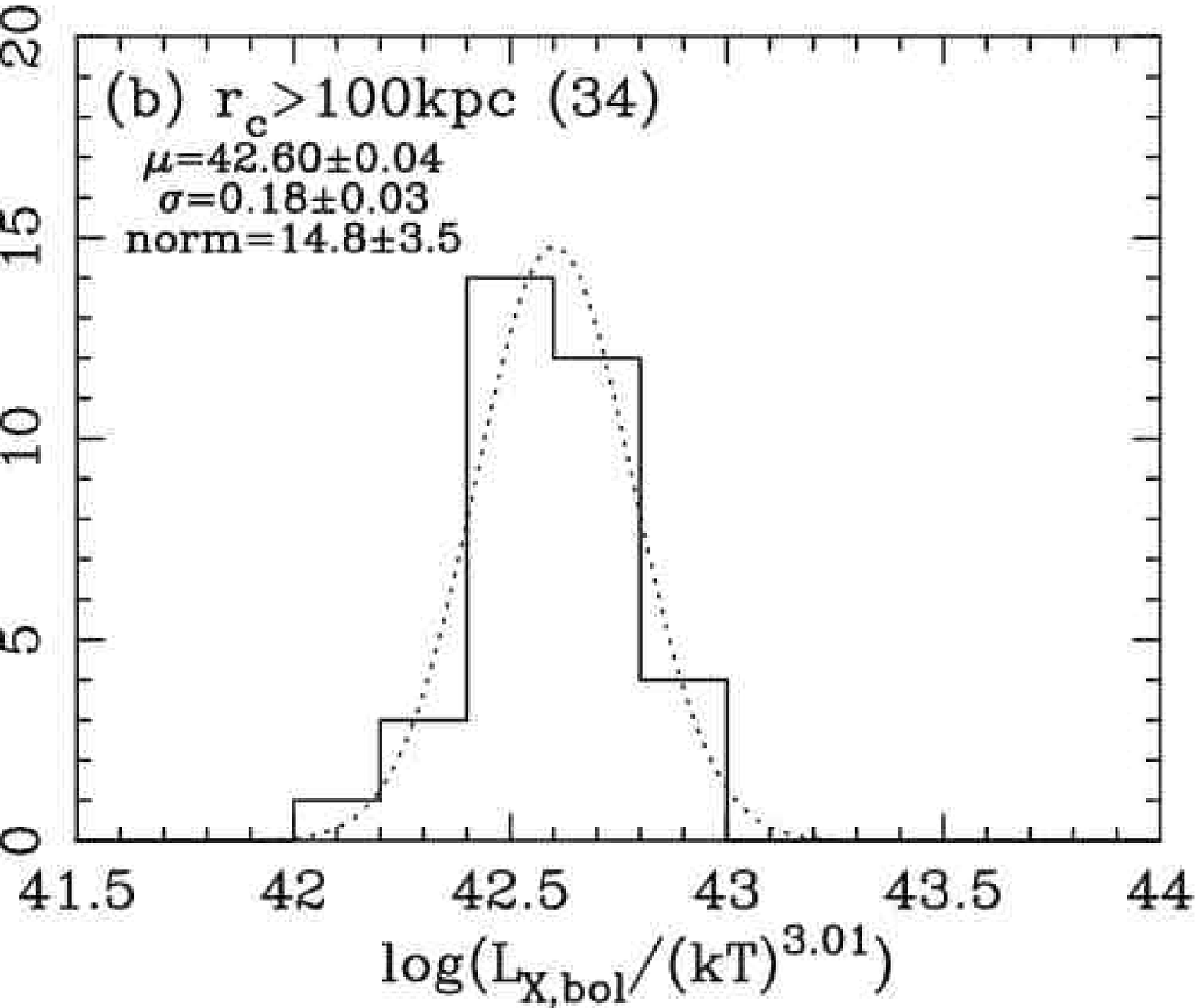}
        \plottwo{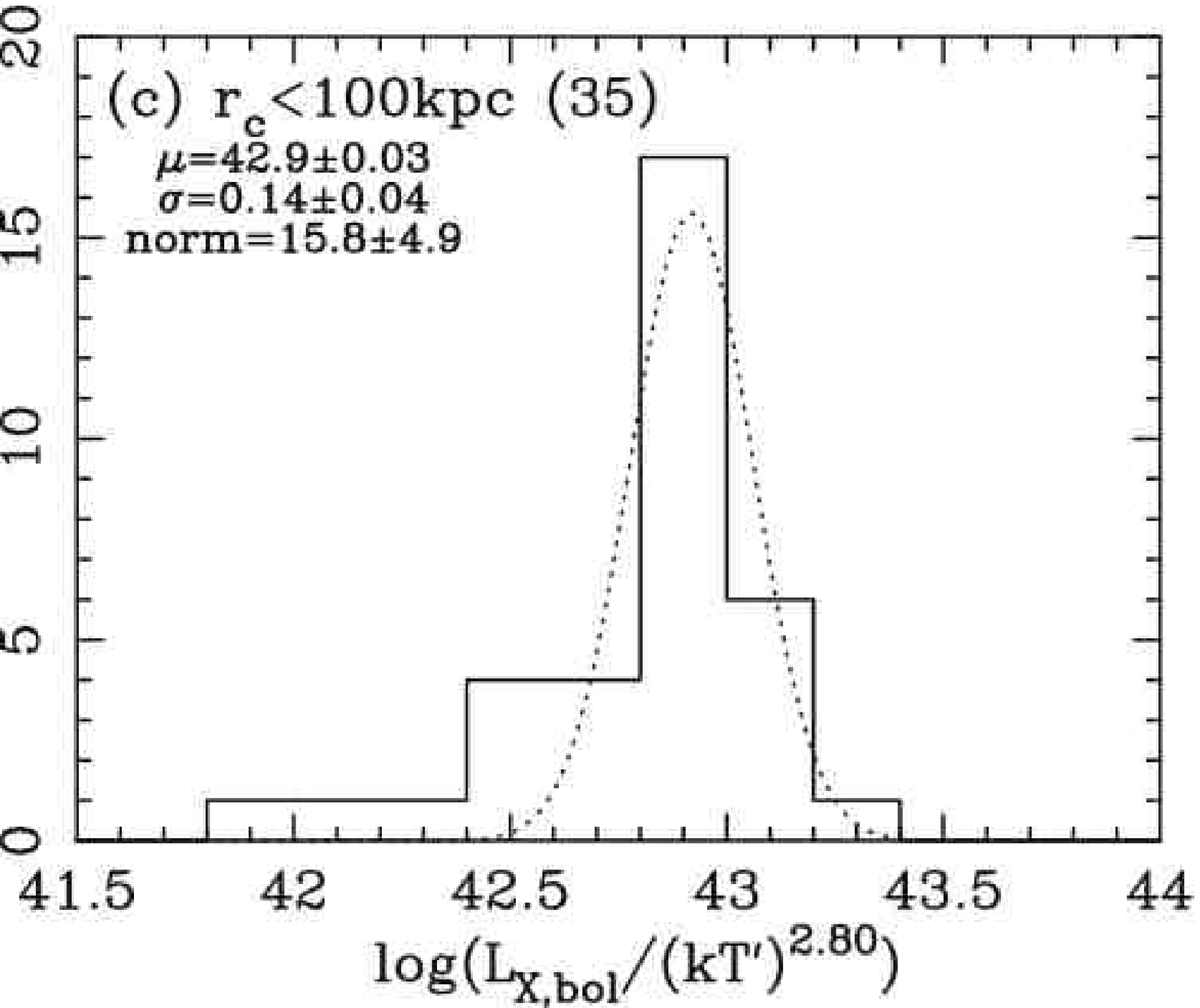}{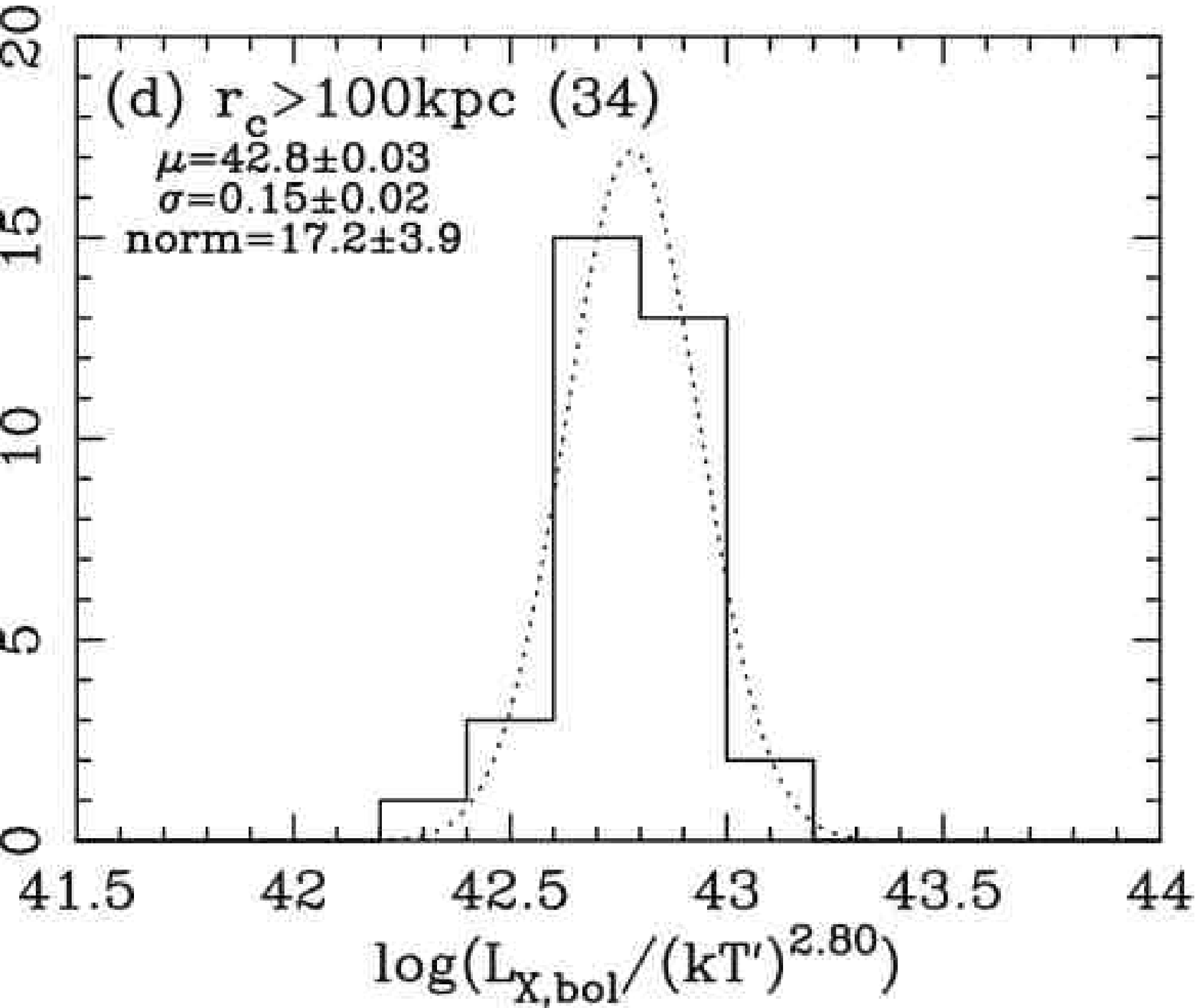}
        \figcaption{Histograms of $L_{\rm 1keV}$.  The distributions of
          $L_{\rm 1keV}(=L_{\rm X}/(kT)^{3.01})$ for 35 small core
          clusters and 34 large core clusters are shown with the green
          and magenta stepped lines in the panels (a) and (b),
          respectively.  The results of $L_{\rm 1keV}(=L_{\rm
          X}/(kT')^{2.80})$ are shown in the panels (c) and (d). The
          best-fit Gaussian model is shown with the dotted curve
          in each panel. The Poisson error of each bin was considered
          in the fitting though it is not displayed in the panels.
          The fitting results are summarized in Table~\ref{tab2}.
\label{fig12}}
\end{figure}

\begin{figure}
\centering
        \plottwo{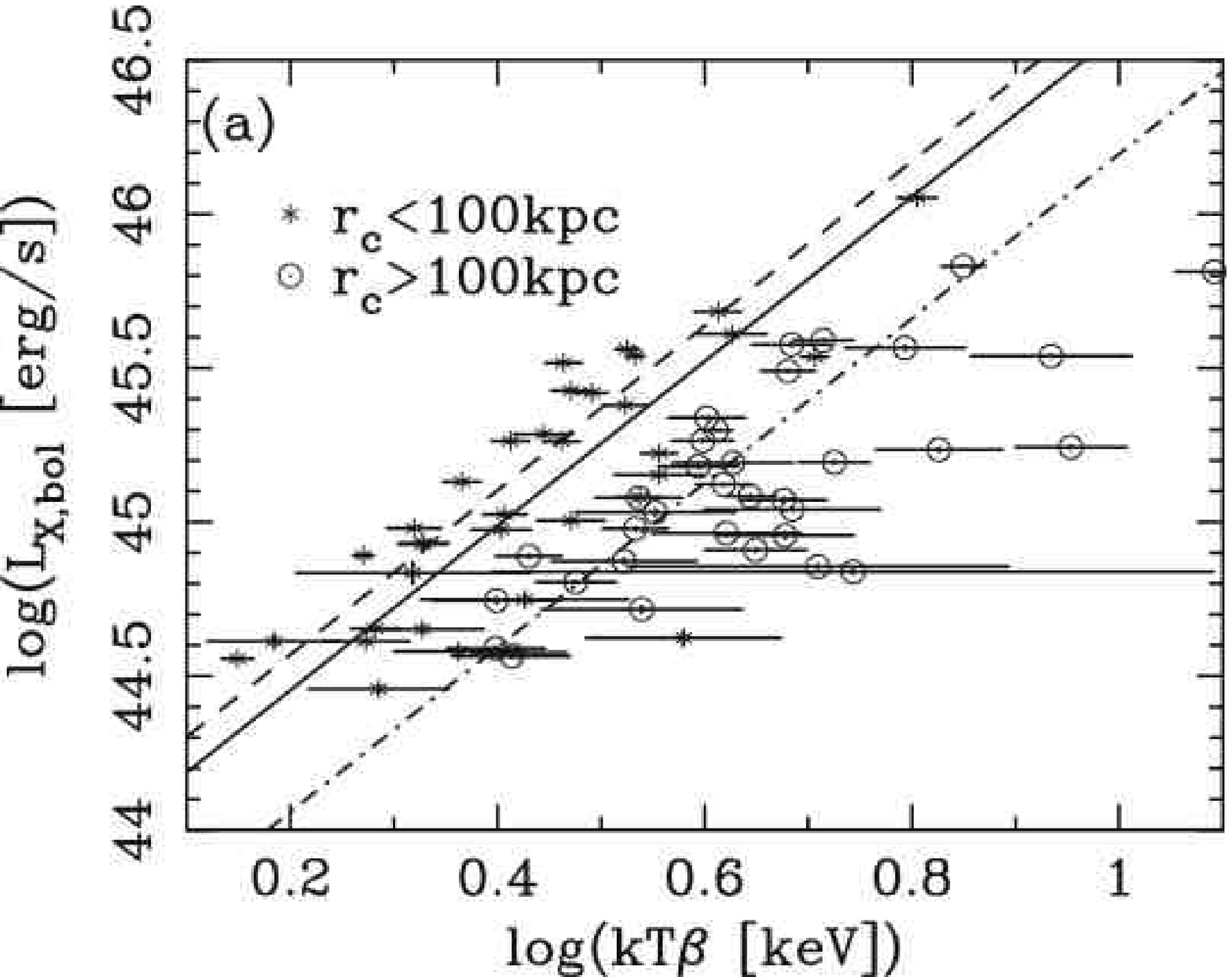}{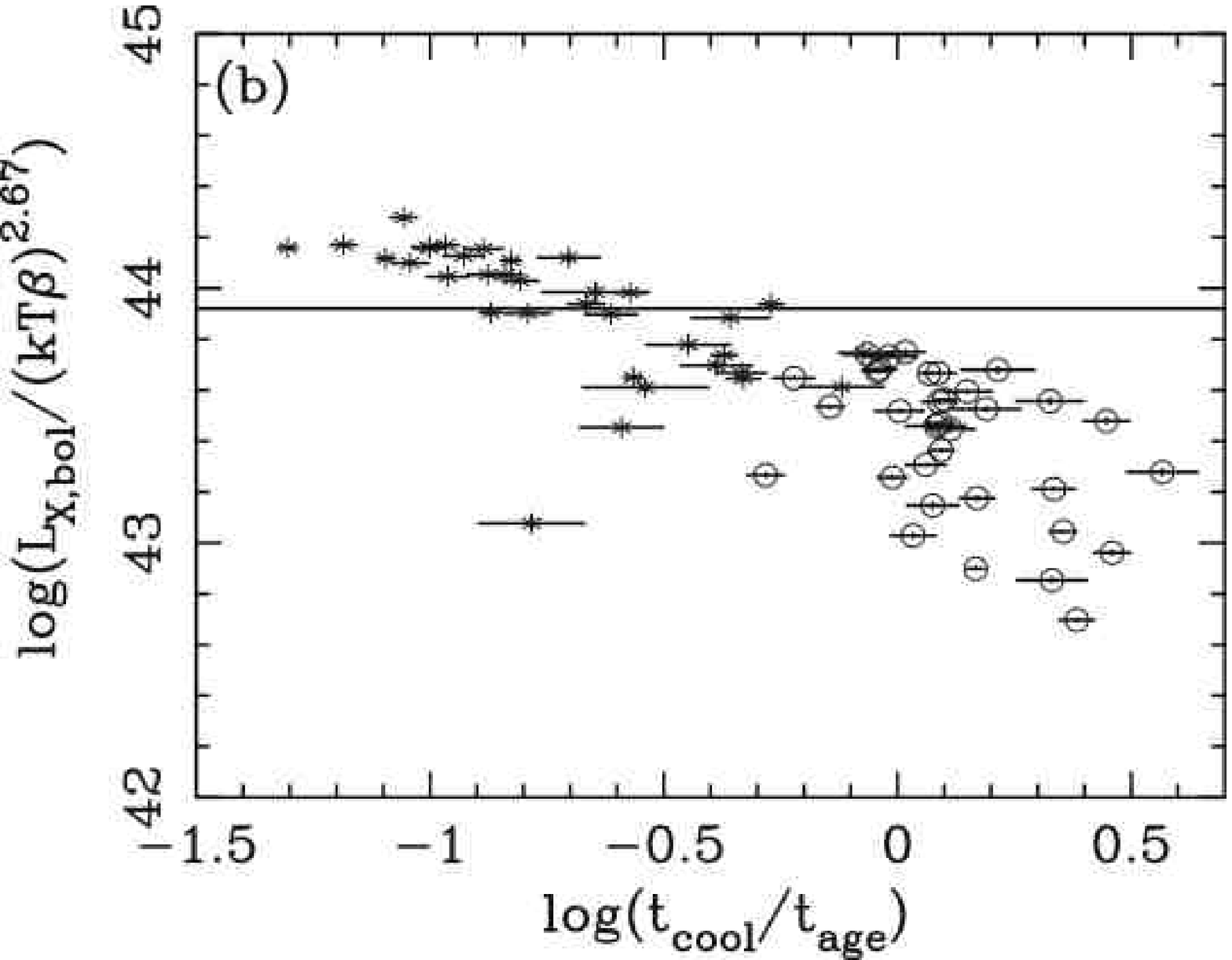}
        \plottwo{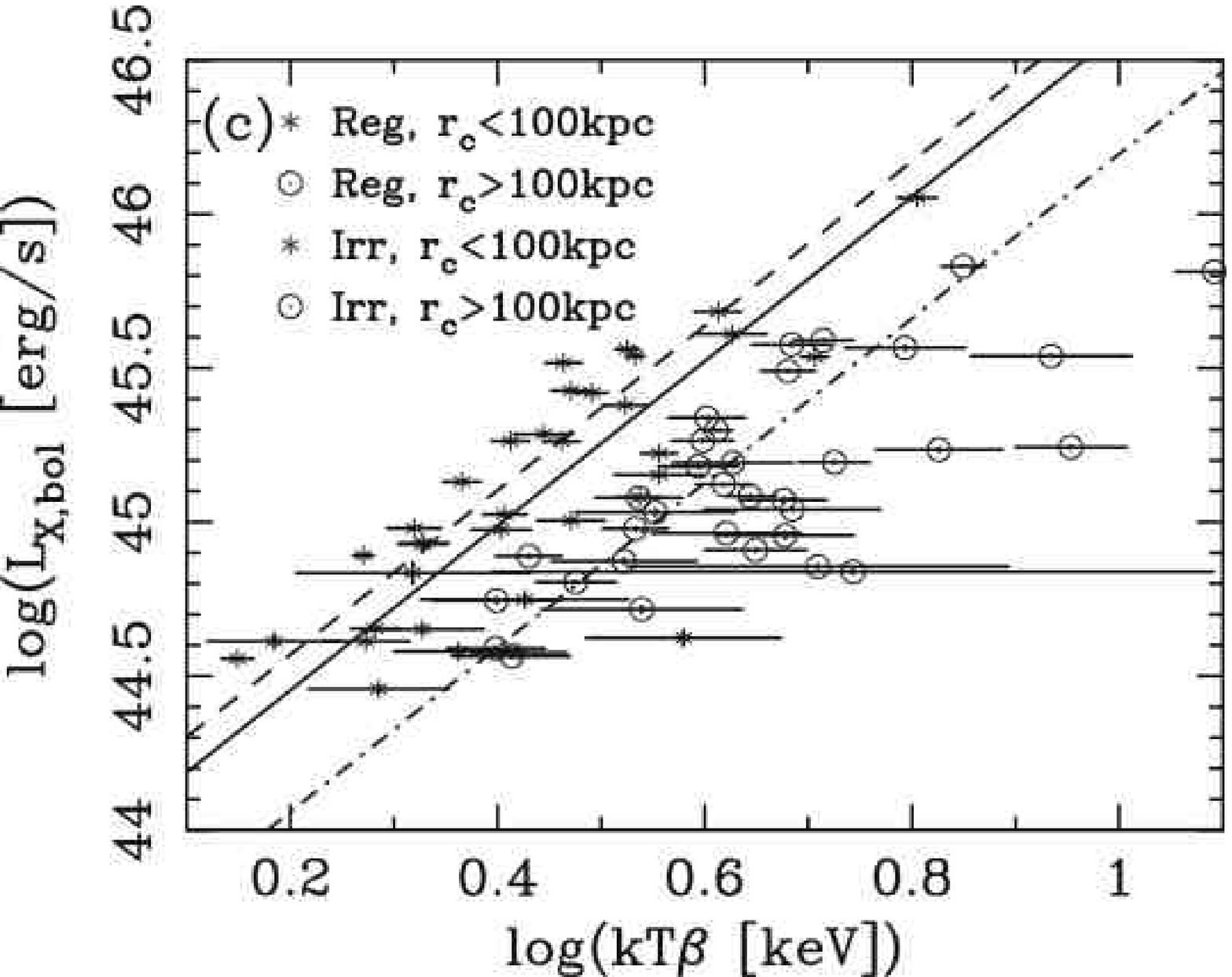}{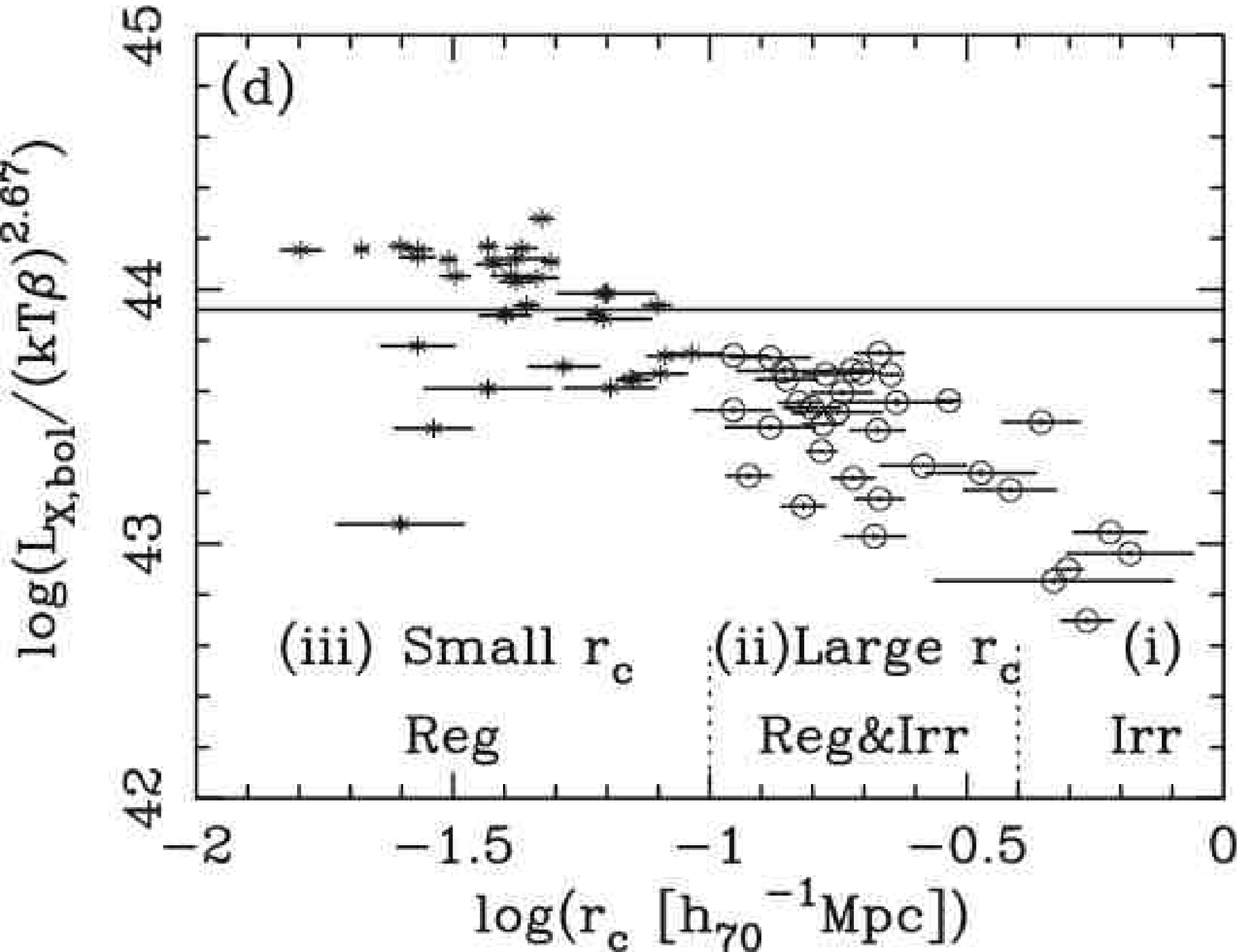}
        \figcaption{$L_{\rm X}-T\beta$ relation and X-ray morphology.
          Upper panels, $L_{\rm X}-T\beta$ relation of clusters (a)
          and $L_{\rm 1keV}=L_{\rm X}/(kT\beta)^{2.67}$ as a function
          of $t_{\rm cool}/t_{\rm age}$ (b).  Bottom panels, the same
          as panel (a) (c), and $L_{\rm 1keV}$ as a function of $r_c$
          (d).  The meaning of the symbols for two $r_c$
          classes are the same as (a) and (b), but the regular and
          irregular clusters are shown in blue and red
          respectively. In the panel (d), we indicated the three
          phases corresponding to i) the irregular clusters with a
          very large $r_c$ and a signature of merger, ii) the
          coexistence of the regular and irregular clusters with a
          large $r_c$, and iii) the regular clusters with a small
          $r_c$ (see \S~\ref{subsec:lx-tbeta}).
          \label{fig13}}
\end{figure}

\begin{figure}
\centering
        \plottwo{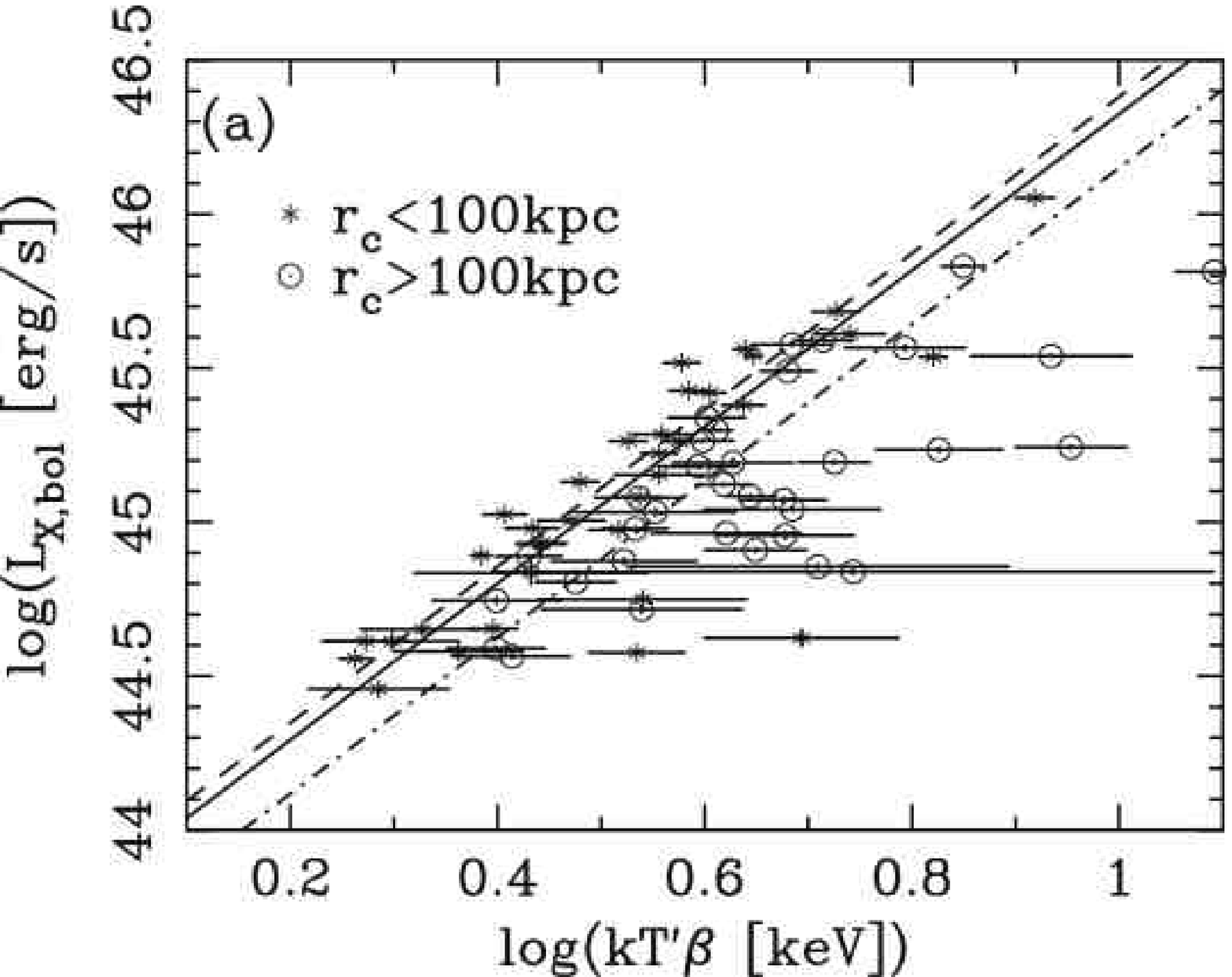}{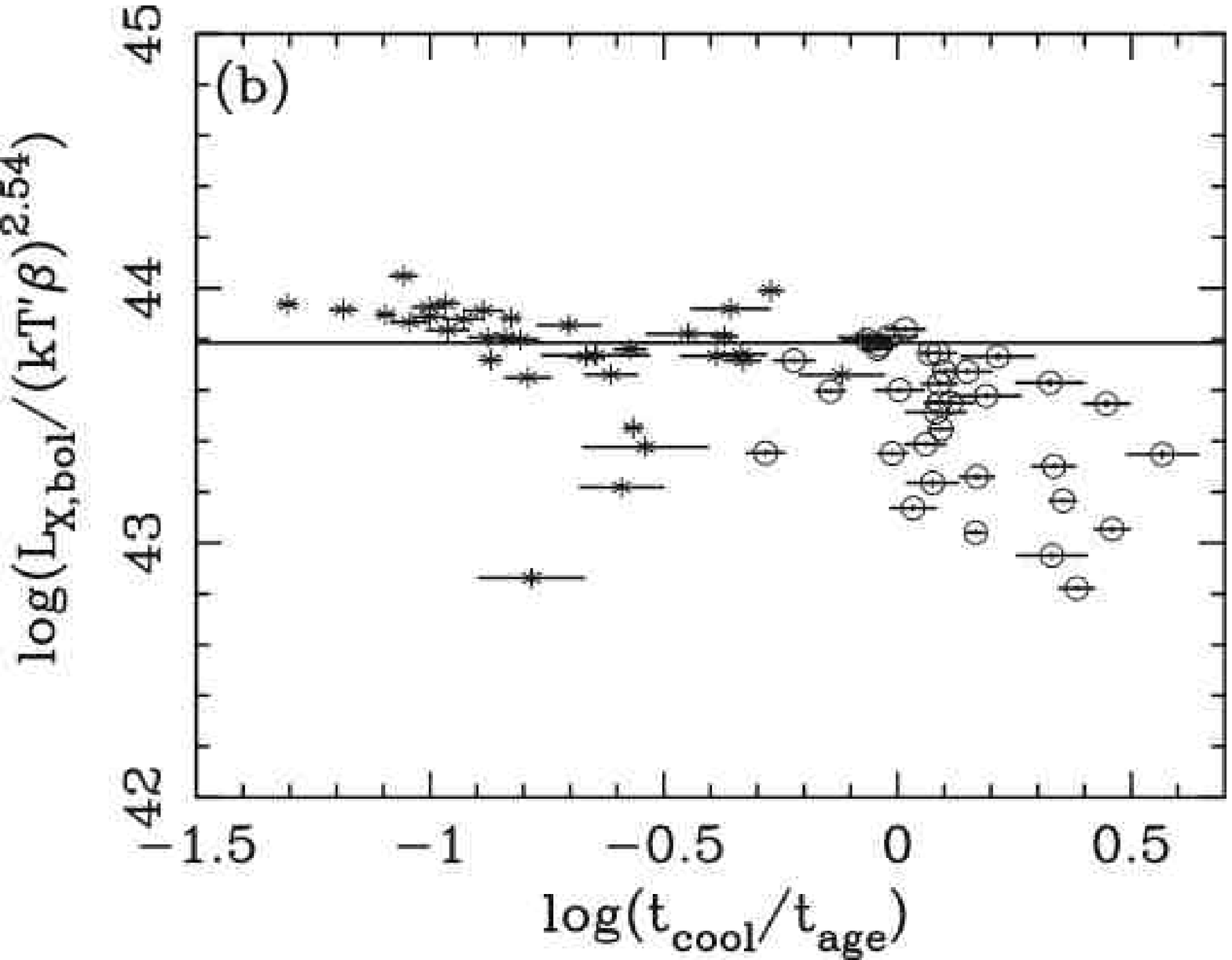}
        \plottwo{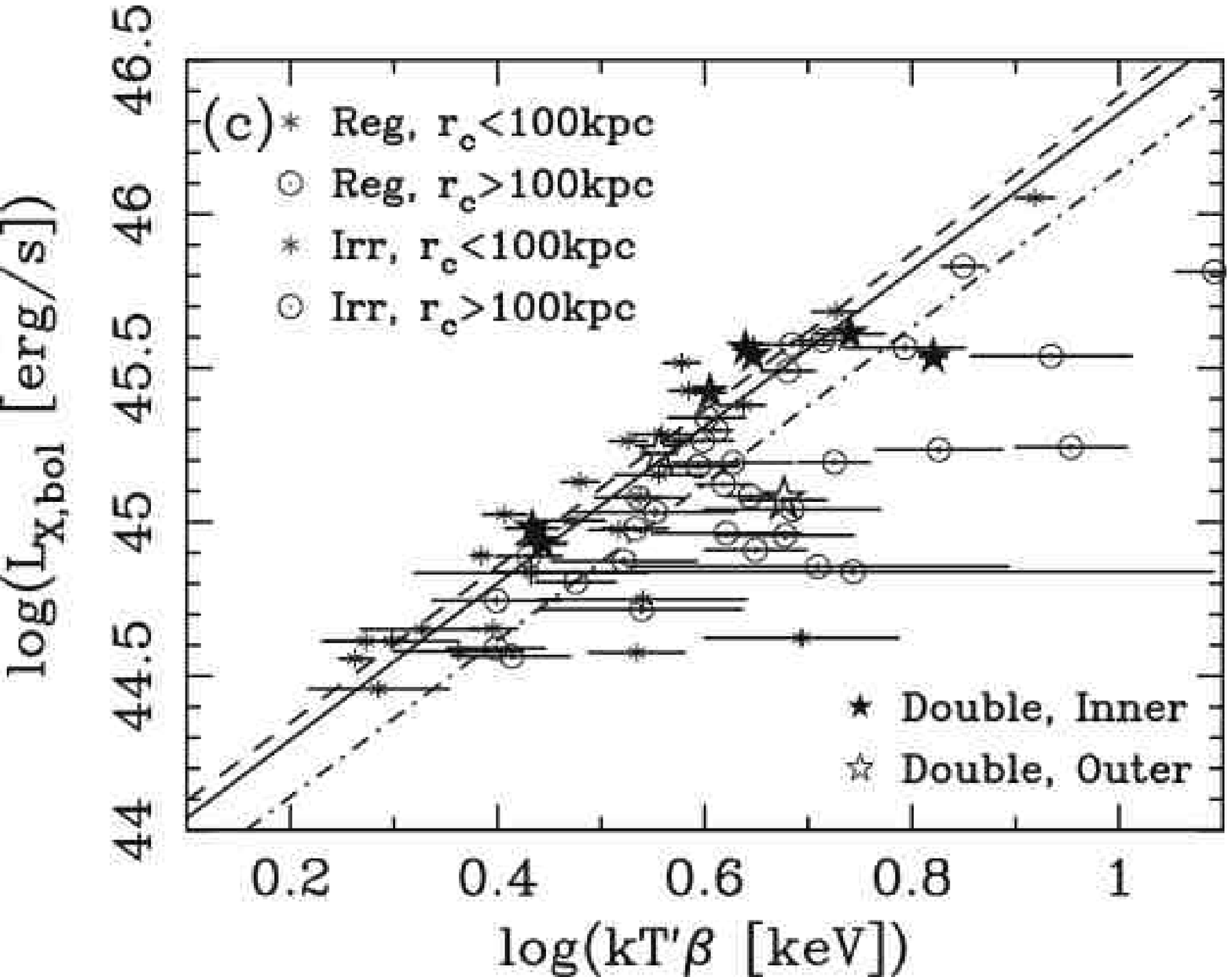}{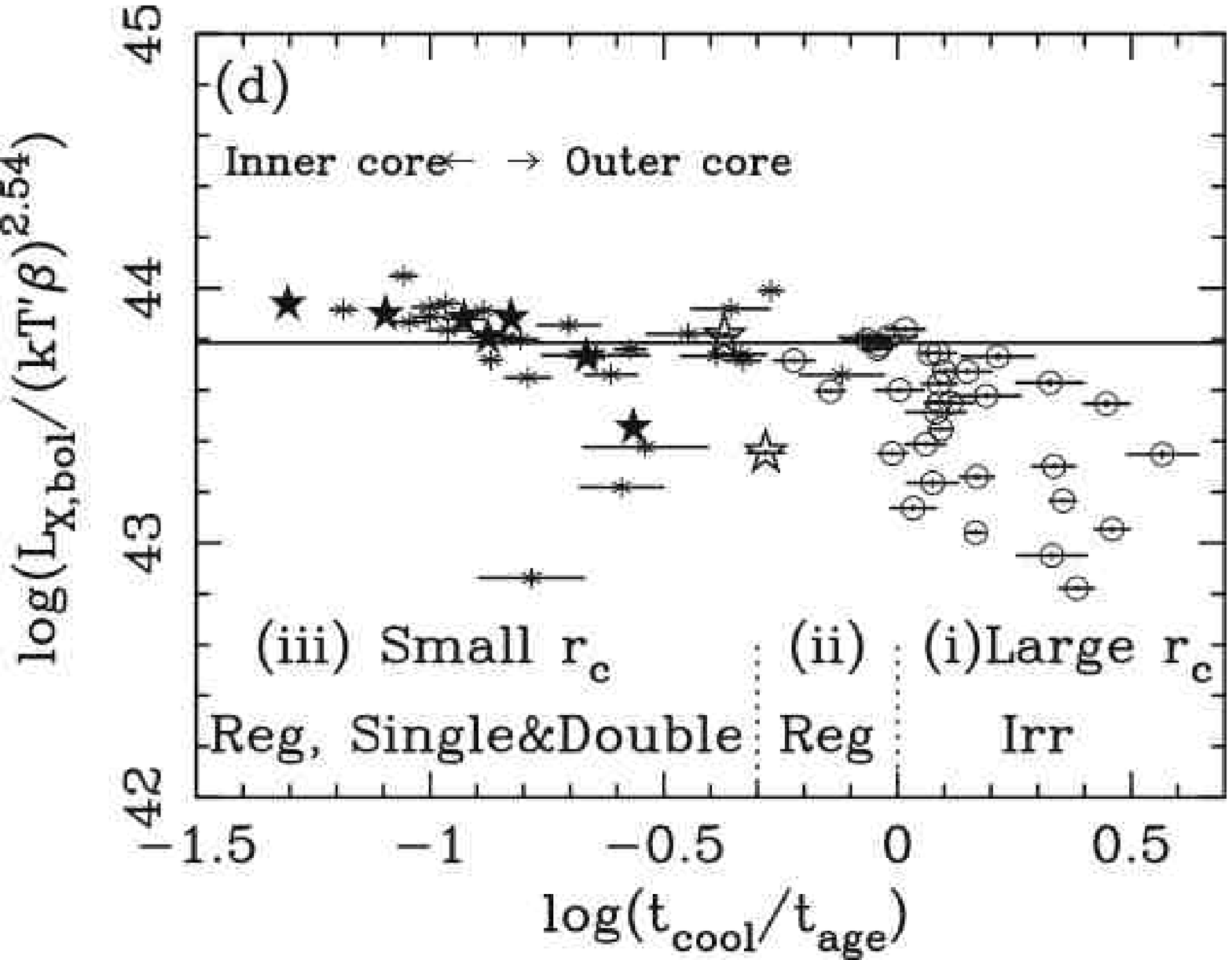}
        \figcaption{$L_{\rm X}-T'\beta$ relation and X-ray
          morphology. The double-$\beta$ clusters are denoted with the
          light-blue stars. The filled (open) stars correspond to the
          inner-core (outer-core) dominant double-$\beta$ clusters. In
          the panel (d), we indicated the three phases according to the
          different level of cooling and density structure: i) the
          large core, irregular clusters with long $t_{\rm cool}$
          relative to $t_{\rm age}$, ii) the large core, regular
          clusters with moderate $t_{\rm cool}$, and iii) the small
          core, regular clusters with short $t_{\rm cool}$.  In the
          phase iii), 9 of the sample shows the significant
          double-$\beta$ structure, and the inner-core/outer-core
          dominant double-$\beta$ clusters are located at the
          shorter/longer cooling time (see
          \S~\ref{subsec:lx-tbeta3}). \label{fig14}}
\end{figure}

In Fig.~\ref{fig7}, we show the $L_{\rm X}-T$ relation numerically
calculated according to Eq. \ref{eq:Lx} for $r_c = 50, 100,
200~h_{70}^{-1}{\rm kpc}$, $\beta =0.7$, and $\Delta_c=500$.  The
gas-mass fraction within $r_{\rm vir}(=r_{\rm 500})$ is fixed at the
mean value of $f_{\rm gas}=0.27$ for the distant sample
\citep{Ota_Mitsuda_2004}.  We should notice that dependence of $L_{\rm
  X}-T$ relation on $\rho_{\rm crit}\Delta_c$ is very small; the
power-law index is $-0.05$, and is almost independent of the
cluster-collapse epoch.  Then the collapse redshift is assumed to be
$z_{\rm col}=0.3$, which is an average observed redshfit of the sample.
We adjusted the value of $\beta'$ to $\beta'=0.4\beta$ in
the calculation so as to make the absolute values
of $L_{\rm X}$ consistent with the observations.

\clearpage



\clearpage

\begin{table}
\footnotesize
\begin{center}
\caption{X-ray luminosity-temperature relations 
for the distant sample. \label{tab1}}
\begin{tabular}{llll} \tableline \tableline
	& Sample & Relation\tablenotemark{a} & $\chi^2/{\rm d.o.f.}\tablenotemark{b}$ \\ \tableline
$L_{\rm X}-T$ & all (69) & $7.41^{+8.89}_{-4.40}\times10^{42} (kT)^{3.01^{+0.49}_{-0.44}}$ & 1603/67\\
 & $r_c\le0.1$~Mpc (35) & $1.05^{+0.12}_{-0.11}\times10^{43} (kT)^{3.01}$ & 722/34\\
 & $r_c>0.1$~Mpc (34) & $ 4.52^{+0.37}_{-0.34}\times10^{42} (kT)^{3.01}$ & 172/33\\ 
 & $r_c\le0.1$~Mpc (35) & $1.45^{+1.62}_{-0.89}\times10^{43} (kT)^{2.82^{+0.56}_{-0.44}}$ & 716/33\\
 & $r_c>0.1$~Mpc (34) & $ 4.37^{+4.84}_{-2.30}\times10^{42} (kT)^{3.03^{+0.38}_{-0.38}}$ & 172/32\\ \tableline
$L_{\rm X}-T'$ & all (69) & $7.41^{+4.49}_{-3.12}\times10^{42} (kT')^{2.80^{+0.28}_{-0.24}}$ & 669/67\\
 & $r_c\le0.1$~Mpc (35) &  $8.04^{+0.71}_{-0.65}\times10^{42} (kT')^{2.80}$ & 461/34\\
 & $r_c>0.1$~Mpc (34) & $ 6.76^{+0.56}_{-0.35}\times10^{42} (kT')^{2.80}$ & 177/33\\ 
 & $r_c\le0.1$~Mpc (35) &  $1.04^{+0.85}_{-0.54}\times10^{43} (kT')^{2.67^{+0.37}_{-0.32}}$ & 457/33\\
 & $r_c>0.1$~Mpc (34) & $ 4.37^{+4.84}_{-2.30}\times10^{42} (kT')^{3.03^{+0.38}_{-0.38}}$ & 171/32 \\ \tableline
$L_{\rm X}-T\beta$ & all (69) & $ 8.32^{+5.52}_{-3.32}\times10^{43} (kT\beta)^{2.67^{+0.44}_{-0.44}}$ & 1310/67\\
 & $r_c\le0.1$~Mpc (35) &  $1.08^{+0.09}_{-0.12}\times10^{44} (kT\beta)^{2.67}$ & 420/34\\
 & $r_c>0.1$~Mpc (34) & $3.35^{+0.47}_{-0.49}\times10^{43} (kT\beta)^{2.67}$ & 169/33 \\ 
 & $r_c\le0.1$~Mpc (35) &  $1.10^{+0.49}_{-0.40}\times10^{44} (kT\beta)^{2.66^{+0.41}_{-0.36}}$ & 420/33\\
 & $r_c>0.1$~Mpc (34) & $2.85^{+3.62}_{-1.78}\times10^{43} (kT\beta)^{2.77^{+0.65}_{-0.54}}$ & 169/32 \\ \tableline
$L_{\rm X}-T'\beta$ & all (69) & $6.10^{+2.57}_{-1.96}\times10^{43} (kT'\beta)^{2.54^{+0.29}_{-0.26}}$ & 628/67\\
 & $r_c\le0.1$~Mpc (35) &  $6.92^{+0.48}_{-0.67}\times10^{43} (kT'\beta)^{2.54}$ & 294/34 \\
 & $r_c>0.1$~Mpc (34) & $ 4.03^{+0.56}_{-0.40}\times10^{43} (kT'\beta)^{2.54}$ & 172/33\\ 
 & $r_c\le0.1$~Mpc (35) &  $7.16^{+3.29}_{-2.25}\times10^{43} (kT'\beta)^{2.51^{+0.30}_{-0.28}}$ & 293/33\\
 & $r_c>0.1$~Mpc (34) & $2.85^{+3.62}_{-1.78}\times10^{43} (kT'\beta)^{2.77^{+0.65}_{-0.54}}$ & 169/32 \\ \tableline
 $L_{\rm X}(<0.2r_{500})-T$ & all (69) & $3.02^{+45.22}_{-2.83}\times10^{41} (kT)^{4.37^{+1.54}_{-1.50}}$ & 2995/67\\
 & $r_c\le0.1$~Mpc (35) &  $5.82^{+1.91}_{-1.12}\times10^{41} (kT)^{4.37}$ & 1289/34 \\
 & $r_c>0.1$~Mpc (34) & $8.32^{+1.58}_{-1.57}\times10^{40} (kT)^{4.37}$ & 301/33 \\ 
 & $r_c\le0.1$~Mpc (35) &  $2.88^{+12.70}_{-2.35}\times10^{42} (kT)^{3.43^{+0.98}_{-0.98}}$ & 1231/33 \\
 & $r_c>0.1$~Mpc (34) & $9.23^{+20.29}_{-6.77}\times10^{41} (kT)^{3.13^{+0.67}_{-0.59}}$ & 252/32 \\ \tableline
$L_{\rm X}(>0.2r_{500})-T$ & all (69) & $4.95^{+3.45}_{-2.13}\times10^{42} (kT)^{2.85^{+0.30}_{-0.27}}$ & 551/67\\
  & $r_c\le0.1$~Mpc (35) &  $5.56^{+0.62}_{-0.56}\times10^{42} (kT)^{2.85}$ & 321/34 \\
 & $r_c>0.1$~Mpc (34) & $4.27^{+0.48}_{-0.32}\times10^{42} (kT)^{2.85}$ & 174/33 \\ 
& $r_c\le0.1$~Mpc (35) &  $7.41^{+7.19}_{-3.91}\times10^{42} (kT)^{2.68^{+0.43}_{-0.37}}$ & 317/33 \\
 & $r_c>0.1$~Mpc (34) & $1.45^{+1.75}_{-0.90}\times10^{42} (kT)^{3.40^{+0.48}_{-0.40}}$ & 148/32 \\ \tableline
 $L_{\rm X}(>0.2r_{500})-T'$ 
  & all (69) & $2.34^{+2.35}_{-1.13}\times10^{42} (kT')^{3.05^{+0.33}_{-0.35}}$ & 610/67\\
& $r_c\le0.1$~Mpc (35) &  $2.04^{+0.26}_{-0.24}\times10^{42} (kT')^{3.05}$ & 362/34 \\
 & $r_c>0.1$~Mpc (34) & $2.88^{+0.29}_{-0.26}\times10^{42} (kT')^{3.05}$ & 158/33 \\ 
  & $r_c\le0.1$~Mpc (35) &  $5.07^{+5.89}_{-2.97}\times10^{42} (kT')^{2.59^{+0.45}_{-0.38}}$ & 336/33 \\
 & $r_c>0.1$~Mpc (34) & $1.45^{+1.75}_{-0.59}\times10^{42} (kT')^{3.40^{+0.22}_{-0.40}}$ & 148/33 \\ \tableline
\end{tabular}
\tablenotetext{a}{The power-law relations derived for the
  distant sample. The results for two $r_c$ subgroups with/without
  fixing the slope at the best-fit value obtained for 69 clusters are
  also shown. The measurement uncertainties of both the $x$ and $y$
  axes are took into account in the fit.} 
\tablenotetext{b}{$\chi^2$ and the degree of freedom of the power-law fitting.}
\end{center}
\end{table}

\begin{table}
\begin{center}
\caption{Means and standard deviations of $\log{L_{\rm 1keV}}$.
\label{tab2}}
\begin{tabular}{llllll} \tableline\tableline
$\log{L_{\rm 1keV}}$	& Sample & $\mu$\tablenotemark{a}
	& $\sigma$\tablenotemark{a}	& $\chi^2/{\rm d.o.f.}\tablenotemark{b}$ & $\sigma/\mu~[10^{-3}]$\tablenotemark{c} \\ \tableline
$\log{L_{\rm X}/(kT)^{3.01}}$	& all (69)   &$42.75\pm0.04$&$0.31\pm0.04$	&4.73/5 & $7.3\pm0.9$ \\
	&  $r_c \leq 0.1$~Mpc (35)& $42.92\pm0.05$ & $0.30\pm0.05$	& 1.95/5 & $7.0\pm1.2$ \\
	& $r_c >0.1$~Mpc	(34) & $42.60\pm0.04$& $0.18\pm0.03$	& 0.82/2 & $4.2\pm0.7$ \\ 
	&  $t_{\rm cool} \leq t_{\rm age}$ (43)& $42.88\pm0.05$& $0.30\pm0.04$	& 2.07/5 & $7.0\pm0.9$ \\
	& $t_{\rm cool} > t_{\rm age}$ (26) & $42.57\pm0.04$& $0.18\pm0.04$	& 0.56/2 & $4.2\pm0.9$ \\ \tableline
$\log{L_{\rm X}/(kT')^{2.80}}$ & all (69)& $42.82\pm0.03$ & $0.19\pm0.02$& 4.53/5 & $4.4\pm0.5$ \\
	& $r_c \leq 0.1$~Mpc (35)	& $42.92\pm0.03$	& $0.14\pm0.04$	& 7.35/5& $3.3\pm0.9$ \\
	& $r_c > 0.1$~Mpc	(34) & $42.78\pm0.03$	& $0.15\pm0.02$	& 0.78/2 & $3.5\pm0.5$ \\ 
	& $t_{\rm cool} \leq t_{\rm age}$ (43)	& $42.87\pm0.04$	& $0.19\pm0.04$	& 5.23/5 & $4.4\pm0.9$ \\
	& $t_{\rm cool} > t_{\rm age}$ (26) & $42.77\pm0.04$	& $0.16\pm0.03$	& 0.99/2 & $3.7\pm0.7$ \\ \tableline 
$\log{L_{\rm X}/(kT\beta)^{2.67}}$ & all (69)& $43.63\pm0.05$ & $0.35\pm0.05$& 11.53/6 & $8.0\pm1.1$\\
	& $r_c \leq 0.1$~Mpc (35)	& $43.92\pm0.04$	& $0.20\pm0.02$	& 7.67/3 & $4.6\pm0.7$\\
	& $r_c > 0.1$~Mpc	(34) & $43.69\pm0.76$	& $0.47\pm0.34$	& 0.75/3 & $10.8\pm7.8$ \\  
	& $t_{\rm cool} \leq t_{\rm age}$ (43)	& $43.86\pm0.04$	& $0.22\pm0.03$	& 11.22/4 & $5.0\pm0.7$\\
	& $t_{\rm cool} > t_{\rm age}$ (26) & $43.44\pm0.27$	& $0.41\pm0.20$	& 2.62/3 & $9.4\pm4.6$\\ \tableline 	
$\log{L_{\rm X}/(kT'\beta)^{2.54}}$ & all (69)& $43.60\pm0.27$ & $0.26\pm0.20$& 18.98/4 & $6.0\pm4.6$\\
	& $r_c \leq 0.1$~Mpc (35)	& $43.83\pm0.02$	& $0.11\pm0.02$	& 3.58/3 & $2.5\pm0.5$\\
	& $r_c > 0.1$~Mpc	(34) & $43.81\pm0.02$	& $0.29\pm0.06$	& 2.66/3 & $6.7\pm1.4$\\ 
	& $t_{\rm cool} \leq t_{\rm age}$ (43)	& $43.81\pm0.02$	& $0.12\pm0.02$	&5.92/3 & $2.7\pm0.5$ \\
	& $t_{\rm cool} > t_{\rm age}$ (26) & $43.41\pm0.07$	& $0.29\pm0.07$	& 2.13/3 & $6.7\pm1.6$\\ \tableline
\end{tabular}
\tablenotetext{a}{The Gaussian mean, $\mu$ and width, $\sigma$
  obtained from the $\chi^2$ fitting to the $L_{\rm 1keV}$
  distribution. The quoted errors are the $1\sigma$.}
\tablenotetext{b}{$\chi^2$ and d.o.f. of the Gaussian fitting.} 
\tablenotetext{c}{$\sigma/\mu$ and the $1\sigma$ error in $10^{-3}$.}
\end{center}
\end{table}

\end{document}